\documentclass[prd,aps,twocolumn,amssymb,
amsmath,
amsfonts,eqsecnum,a4paper,floatfix,showpacs,nofootinbib,superscriptaddress,nolongbibliography,twocolumn,twoside]{revtex4-2}

\usepackage{graphicx,psfrag}
\usepackage{dcolumn}
\usepackage{bm}
\usepackage{tabularx}
\usepackage{multirow}
\usepackage{capt-of}
\usepackage{color}
\usepackage{url}
\usepackage[utf8]{inputenc}
\usepackage[T1]{fontenc}
\usepackage{accents}
\usepackage[normalem]{ulem}

\usepackage[nolist,nohyperlinks]{acronym}
\usepackage{xspace,xcolor}
\usepackage[colorlinks]{hyperref}
\usepackage{float}

\usepackage[labelformat=empty,caption=false]{subfig}

\usepackage[capitalise]{cleveref}

\usepackage{gensymb, upgreek}
\usepackage{commath, braket}

\newcommand{\codename}[1]{\textmd{#1}}

\newcommand{\modelname}[1]{\textsc{#1}}

\newcommand{\phX}{\modelname{IMRPhenomX}\xspace}

\newcommand{\phXHM}{\modelname{IMRPhenomXHM}\xspace}

\newcommand{\phXPHM}{\modelname{IMRPhenomXPHM}\xspace}
\newcommand{\phXPHMMSA}{\modelname{IMRPhenomXPHM-MSA}\xspace}
\newcommand{\phXPHMST}{\modelname{IMRPhenomXPHM-SpinTaylor}\xspace}
\newcommand{\phXOfourA}{\modelname{IMRPhenomXO4a}\xspace}

\newcommand{\phT}{\modelname{IMRPhenomT}\xspace}
\newcommand{\phTPHM}{\modelname{IMRPhenomTPHM}\xspace}

\newcommand{\parallelto}{{\mkern3mu\vphantom{\perp}\vrule depth 0pt\mkern2mu\vrule depth 0pt\mkern3mu}} 

\newcommand{\Mf}{\mathit{Mf}}
\newcommand{\fring}{\Mf_\mathrm{\!ring}}
\newcommand{\fdamp}{\Mf_\mathrm{\!damp}}

\newcommand{\fcutinsp}{\Mf_\mathrm{\!insp}^\mathrm{cut}}
\newcommand{\fcutrd}{\Mf_\mathrm{\!rd}^\mathrm{cut}}
\newcommand{\fcutrdtwo}{\Mf_\mathrm{\!rd,2}^\mathrm{cut}}
\newcommand{\frdaux}{\Mf_\mathrm{\!rdaux}^\mathrm{cut}}

\newcommand{\Msun}{\,\mathrm{M_\odot}}
\newcommand{\chieff}{\chi_\mathrm{eff}}

\definecolor{darkorange}{rgb}{1, 0.549,0}

\newcommand{\UIB}{Departament de F\'isica, Universitat de les Illes Balears, IAC3 -- IEEC, Crta. Valldemossa km 7.5, E-07122 Palma, Spain}

\newcommand{\UCD}{University College Dublin, Belfield, D4, Dublin, Ireland}

\allowdisplaybreaks[1]

\begin{document}

\title[PhenomXPHM-SpinTaylor]
{
	Fast frequency-domain gravitational waveforms for precessing binaries with a new twist
}

\author{Marta Colleoni}\email{marta.colleoni@uib.es}
\affiliation{\UIB}
\author{Felip A. Ramis Vidal}\email{f.ramis@uib.es}
\affiliation{\UIB}
\author{Cecilio García-Quirós }\email{cecilio.garciaquiros@uzh.ch}
\affiliation{Astrophysik-Institut, Universit\"{a}t Z\"{u}rich, Winterthurerstrasse 190, 8057 Z\"{u}rich, Switzerland}
\affiliation{Laboratoire Astroparticule et Cosmologie, 10 Rue Alice Domon et Léonie Duquet, 75013 Paris, France}
\author{Sarp Ak\c{c}ay}\email{sarp.akcay@ucd.ie}
\affiliation{\UCD}
\author{Sayantani Bera}\email{sayantani.bera@uib.eu}
\affiliation{\UIB}

\date{\today}

\begin{abstract}
	Gravitational waveform (GW) models are a core ingredient for the analysis of compact binary mergers observed by current ground-based interferometers. We focus here on a specific class of such models known as \modelname{PhenomX}, which has gained popularity in recent years thanks to its computational efficiency. We introduce a new description of the ``twisting-up'' mapping underpinning the construction of precessing waveforms within this family. The new description is an adaptation to the frequency domain of a technique previously implemented in time-domain models, where the orbit-averaged post-Newtonian spin-precession dynamics is numerically solved on the fly. We also present an improved version of the gravitational-wave strain amplitudes approximating the signal in the co-precessing frame. We demonstrate that the new description yields improved matches against numerical relativity simulations, with only a modest computational overhead. We also show that the new model can be reliably employed in parameter estimation follow-ups of GW events, returning equivalent or more stringent measurements of the source properties compared to its predecessor.
\end{abstract}


\maketitle

\section{Introduction}
\label{sec:Introduction}

The interferometers operated by the LIGO-Virgo-KAGRA (LVK) collaboration~\cite{LIGOScientific:2014pky,VIRGO:2014yos,KAGRA:2020tym} have thus far detected $\mathcal{O}(100)$ compact binary mergers during the first three observing runs~\cite{LIGOScientific:2021djp}, with the fourth observing run, O4, expected to deliver $\mathcal{O}(200)$ events~\cite{gracedb}.
More than 90\% of the hitherto detected sources have been binary black holes (BBHs) which comprise a pair of spinning (Kerr) black holes. As such, there should be imprints of spin-related effects on the gravitational waveforms emitted by such systems \cite{Kidder:1992fr,Kidder:1995zr,Cutler:1994ys, Poisson:1995ef}. At the leading order, these effects are manifest via the components of the spins parallel to the orbital angular momentum~\cite{Damour:2001tu, Racine:2008qv, Ajith:2011ec, Schmidt:2010it, Baird:2012cu, Hannam:2013oca}. Beyond the leading order, effects due to the precession of spins emerge~\cite{Arun:2008kb}, caused by non-zero planar spin components \cite{Apostolatos:1994mx}.

Though the data for a significant fraction of the detected BBHs is consistent with the components having non-zero spins  \cite{LIGOScientific:2018mvr, LIGOScientific:2020kqk, LIGOScientific:2021usb, KAGRA:2021vkt, KAGRA:2021duu}, the evidence for spin precession has been rare, with at least one, and at most a few events exhibiting spin precession. The most convincing of these events is GW200129 \cite{LIGOScientific:2021djp, Hannam:2021pit, Payne:2022spz, Macas:2023wiw}. Given that there exist theoretical formation mechanisms leading to precessing BBHs~\cite{Mandel:2018hfr, Mapelli:2021taw, Gerosa:2013laa,Vitale:2015tea, Rodriguez:2016vmx, Stevenson:2017dlk}, and that the sensitivity of the currently operating interferometers will improve, we can expect to detect one event like
GW200129 out of every $\lesssim 50$ GW detections by the ground-based interferometers \cite{Hoy:2024qpy}.

The existence of a potential subpopulation of BBHs with precessing spins necessitates the existence of waveform models that faithfully incorporate the effects of spin precession. Accordingly, significant effort has gone into the development of precessing waveforms and several competing waveform families have emerged with specific waveform ``approximants'' that are mature and robust enough to be employed in large-scale parameter estimation campaigns. The most accurate waveform model currently available is the numerical relativity surrogate \modelname{NRSur7dq4}~\cite{Varma:2019csw}. However, this model has a limited domain of validity, dictated by the dataset of numerical relativity simulations employed for its training, which was restricted to binaries with mass ratios ranging from $1:1$ to $4:1$ and dimensionless spin magnitudes less than 0.8. The model's fiducial extrapolation regime extends up to mass ratios of $6:1$ and spin magnitudes of 0.9. Another drawback of \modelname{NRSur7dq4} is that it can not generate waveforms with durations longer than $4300M$, where $M=1\Msun$ roughly corresponds to $5\times 10^{-6}$ seconds. Despite these limitations, the surrogate is the most faithful model in its training regime and was recently used in a re-analysis of most of the O3 events \cite{Islam:2023zzj}. It also proved to be the only waveform model with sufficient accuracy for the analysis of the significantly precessing high-SNR event GW200129~\cite{Hannam:2021pit}.

On a parallel front, there is a long pedigree of time-domain waveform models developed within the effective-one-body framework~\cite{Buonanno:1998gg, Buonanno:2000ef,Damour:2000we,Damour:2001tu}.
For quasi-circular precessing binaries, the state of the art among these are the numerical relativity calibrated models \modelname{SEOBNRv5PHM}~\cite{Ramos-Buades:2023ehm,Khalil:2023kep,Mihaylov:2023bkc,Pompili:2023tna} and \modelname{TEOBResumS-GIOTTO}~\cite{Nagar:2018zoe,Akcay:2020qrj,Gamba:2021ydi}.

Thus far, we have only listed time-domain models. However, frequency-domain waveforms are more convenient for data analysis applications based on matched filtering and they are roughly an order of magnitude faster to evaluate than time-domain waveforms.
Frequency-domain models also have a long history in gravitational-wave astronomy starting with \modelname{PhenomA}~\cite{Ajith:2007qp, Ajith:2007kx} and
going all the way up to the current state of the art models \phXPHM~\cite{Pratten:2020ceb} and \phXOfourA~\cite{Hamilton:2021pkf,Hamilton:2023znn,Ghosh:2023mhc,Thompson:2023ase}
with many versions developed along the way \cite{Hannam:2013oca, Schmidt:2014iyl, Khan:2018fmp, Khan:2019kot}.
In recent years, the suite of waveform models dubbed \phT~\cite{Estelles:2020osj,Estelles:2020twz,Estelles:2021gvs} have emerged, representing the first example of phenomenological models natively built in the time domain.

A common feature of precessing waveform models is the use of the twisting-up approximation, where the signal in a frame co-precessing with the binary is approximated by an aligned-spin waveform~\cite{Schmidt:2010it} and the mapping between co-precessing and detector frames is provided by time/frequency-dependent Euler angles. Currently, for a given spherical harmonic index $\ell$, the aligned-spin model \phXHM only includes a subset of the corresponding $m$ modes, e.g. only the $(2,1)$ and $(2,2)$ modes are available for $\ell=2$ subset. Nonetheless, in the precessing case all of the $\ell\leq4$ multipoles will be present in the waveform strain, through the mode mixing of all the $(\ell,m')$, $|m'|\leq\ell$ harmonics (see, e.g., App. E of~\cite{Pratten:2020ceb}). Another difference with respect to the aligned-spin case is the estimate of the remnant's spin, which is modified to account for the contribution of the in-plane spins (i.e., the spin components lying in the binary's orbital plane).

Though \phXPHM is known to be less faithful to numerical relativity than \modelname{SEOBNRv5PHM} in some parts of the binary black hole parameter space \cite{Ramos-Buades:2023ehm, MacUilliam:2024oif}, its speed and overall accuracy make it a valuable tool for computationally intensive data analysis applications. In recognition of these qualities, it is one of the fiducial waveform models routinely employed by the LVK Collaboration in their catalogs of gravitational-wave transients~\cite{LIGOScientific:2021djp}.
The modular construction of \phX also offers a relatively low threshold to extending the model beyond its original purpose, e.g. with the inclusion of extreme matter~\cite{Colleoni:2023czp,Abac:2023ujg} or enviromental effects~\cite{Roy:2024rhe}.

In this article, we present two improvements to \phX that are available through the \codename{LALSuite}~\cite{lalsuite} algorithm library and ready to use for parameter estimation studies. First, we have improved the amplitude model of \phXHM, making it more accurate and robust across parameter space. Second, we have added a new precession prescription to \phXPHM, where the Euler angles used in the twisting-up approximation are computed via the numerical integration of the post-Newtonian (PN) spin-precession equations, with SpinTaylorT4 phasing. This approach is adopted in many time-domain models~\cite{Estelles:2021gvs,Gamba:2021ydi,Ramos-Buades:2023ehm}: here, we obtain a frequency-domain representation of the solutions by applying the stationary phase approximation. In previous frequency-domain models~\cite{Khan:2018fmp,Pratten:2020ceb}, the Euler angles were derived in closed form by applying a multiple scale analysis (MSA) to the PN precession equations. This technique leverages the fact that the precession timescale is much shorter than the radiation-reaction one~\cite{Kesden:2014sla,Gerosa:2015tea}, allowing for a perturbative treatment of the dynamics. Within this scheme, the Euler angles are expanded in terms of a precession-averaged term plus a leading order correction ~\cite{Chatziioannou:2017tdw}. While computationally more efficient, we will show that this method is less accurate than the one described here.

This article is organized as follows. In Sec.\,\ref{sec:implementation}, we cover some technical details regarding the implementation of the updated amplitude model (Subsec.~\ref{subsec:amplitude_model}) and inspiral Euler angles (Subsec.~\ref{subsec:angles_model}). We then demonstrate how the updated model, \phXPHMST, achieves a higher faithfulness against numerical relativity (Subsec.~\ref{sec:nr_comparison}) and \modelname{NRSur7dq4} waveforms (Subsec.~\ref{sec:model_comparison}) than its predecessor \phXPHMMSA or simply \phXPHM. In Sec.~\ref{sec:benchmarks}, we present some timing tests comparing the efficiency of \phXPHMST to that of other state of the art models. Sec.~\ref{sec:pe} contains a reanalysis of several gravitational-wave events from the observing run O3b (Subsec.~\ref{subsec:real_gw}), as well an injection-recovery study. We summarise our findings in Sec.~\ref{sec:conclusions}.

\section{Implementation notes}
\label{sec:implementation}

In this section, we describe the implementation of the new amplitude and precession prescriptions added to \phXPHM.

Let us start by introducing the notation that we will be using throughout this paper. We denote the masses of the binary components by \( m_1 \) and \( m_2 \), assuming \( m_1 \geq m_2 \), the total mass by \( M = m_1 + m_2 \), the mass ratio by \( Q = m_1 / m_2 \geq 1 \) or \( q = m_2 / m_1 \leq 1 \), the symmetric mass ratio by \( \eta = m_1 m_2 / (m_1 + m_2)^2 \), and the chirp mass by \( \mathcal{M} = \eta^{3/5}M \). The spin vectors of each component are denoted by \( \vec{S}_i \) and their dimensionless forms by \( \vec{\chi}_i = \vec{S}_i / m_i^2 \). The magnitude of each dimensionless spin is \( \chi_i = \| \vec{\chi}_i \| \), with the spin component parallel to the Newtonian orbital angular momentum \(  \vec{L}_\mathrm{N}  \) given by \( \chi_{iz} = \vec{\chi}_i \cdot \hat{L}_\mathrm{N}  \), and the perpendicular component by \( \chi_i^\perp = \| \vec{\chi}_i - \chi_{iz} \hat{L}_\mathrm{N}  \| \). The effective spin parameter is defined as \( \chi_\mathrm{eff} = (m_1 \chi_{1z} + m_2 \chi_{2z}) / M \), and the effective precession parameter is \(\chi_\mathrm{p} = \max(A_1 S_1^\perp \!,\: A_2 S_2^\perp)/(A_1 m_1^2)\) where \( A_1 = 2 + \frac{3}{2Q} \) and \( A_2 = 2 + \frac{3Q}{2} \).

\subsection{New amplitude model}
\label{subsec:amplitude_model}

The modelling strategy has been redefined to achieve simplicity and uniformity across modes and parameter space. Some of the main improvements are a smoother connection of the extreme mass ratio and comparable mass regime, more robust ansätze for the merger-ringdown signal and updated fits of phenomenological coefficients across parameter space.

\subsubsection{Transition frequencies}

The three main frequency regions, \textit{inspiral}, \textit{intermediate} and \textit{ringdown}, are separated by two cutting frequencies $\fcutinsp$ and $\fcutrd$.

In the previous \phXHM version, these frequencies had different definitions for each mode and $\fcutinsp$ had a sharp transition to the phenomenological cutting frequency for the extreme-mass-ratio inpirals (EMRI) inspiral $\Mf_{\mathrm{EMRI}}^{\mathrm{cut}}$. In the new amplitude model, we apply a smooth transition function between the comparable mass and the EMRI inspiral transition:
\begin{equation}
	\fcutinsp=
	\begin{cases}
		\begin{aligned}
			 & \Mf_{\!\rm{MECO}}                                                  & \quad Q<20\,,    \\
			 & w \Mf_{\!\rm{MECO}} + (1 - w) \Mf_{\!\mathrm{EMRI}}^{\mathrm{cut}} & \quad Q\geq20\,,
		\end{aligned}
	\end{cases}
\end{equation}
where $w=1/2(1 +\tanh(\eta - 0.0192234)/0.004)$.

The ringdown transition frequency is defined as
\begin{equation}
	\fcutrd =
	\begin{cases}
		\begin{aligned}
			 & \fring - \fdamp              & \quad (\ell,m)\neq(3,2)\,, \\
			 & \fring^{22} - 0.5\fdamp^{22} & \quad (\ell,m)=(3,2)\,,
		\end{aligned}
	\end{cases}
\end{equation}
where $\fring,\fdamp$ denote the ringdown and damping frequencies of a given $(\ell,m)$ mode, and a $22$ superscript indicates the corresponding quantities refer to the $(2,2)$ mode.
We introduce an extra ringdown region for all the modes to ensure that the behavior at high frequencies is a purely exponential decay. This ``late-ringdown'' region starts at the frequency
\begin{equation}
	\fcutrdtwo{} =\fring + 2\fdamp\,.
\end{equation}

\subsubsection{\label{sec:inspiral}Inspiral region}
The inspiral region covers frequencies $\Mf\leq\fcutinsp{}$.
The ansatz is the same as for the old release, i.e., we add three pseudo-PN terms whose coefficients are obtained through three fitted collocation points defined at the frequency points:
\begin{align}
	\Mf_{\mathrm{insp},i}^\text{CP} = \{0.5 \fcutinsp{},\ 0.75\fcutinsp{},\ \fcutinsp{}\}\,.
\end{align}
This is now employed consistently over the full mass ratio range.

\subsubsection{\label{sec:inter}Intermediate region}
The intermediate region covers the frequency range:
\begin{align}
	\fcutinsp{} < \Mf < \fcutrd{}.
\end{align}

The ansatz presented in Eq. (5.3) of \cite{Garcia-Quiros:2020qpx} is an inverse polynomial, with poles in different regions of parameter space. In order to keep the reconstructed amplitude finite, the ansatz had to be regularized by iteratively decreasing the number of degrees of freedom until the polynomial did not cross zero. In the new model, we adopt instead the ansatz
\begin{align}{\label{eq:interansatz}}
	A_{\mathrm{inter}}(\Mf)=\frac{1}{\Mf^{7/6}} \sum_{i=0}^N a_{i} \Mf^i\,,
\end{align}
which entirely avoids this issue. At low frequency, the ansatz captures the leading order behavior predicted by PN ($\sim\!\Mf^{-7/6}$); $N$ will depend on the number of collocation points employed for the reconstruction. By default, the ansatz will use 4 fitted collocation points (\textit{inner} collocation points), as well as the amplitude values and their derivatives at the 2 boundaries of the intermediate region, giving a total of 8 degrees of freedom and an overall $C^1$ function. For the $(2,1)$ mode, however, we drop the left derivative and two of the fitted collocation points, leaving us with just 5 degrees of freedom. This helped to better describe the transition region between the amplitude decay observed in the inspiral and the plateau occurring in the intermediate region, while avoiding at the same time spurious oscillations in the reconstruction due to fitting errors, which are especially problematic for configurations where the amplitude is very small.

The four inner collocation points, which are fitted across parameter space, are defined at equispaced frequency points inside the intermediate frequency region
\begin{align}\label{eq:inter_cpoints}
	\Mf^{\mathrm{CP}}_{\!\mathrm{inter}, i} = \fcutinsp + \frac{i}{4} \del[0]{\fcutrd - \fcutinsp}\,,
	\quad i=1...4\,.
\end{align}
The coefficients of the ansatz (\ref{eq:interansatz}) are computed by numerically solving the linear system of equations obtained by requiring that the reconstructed amplitude takes on the values predicted by the fits at the frequency nodes given by Eq.~\eqref{eq:inter_cpoints}.

\subsubsection{\label{sec:ringdown}Ringdown region}
The ringdown region is defined in the interval
\begin{align}
	\Mf\geq\fcutrd{}.
\end{align}
As before, the frequency array is filled with zeros for $\Mf>0.30$ and for $\Mf>0.33$ for cases with $\chieff>0.99$ to fully include the (4,4) mode.

\subsubsection{No mode-mixing}\label{sec:nomodemixing}

We distinguish two regions separated by $\fcutrdtwo{}$. The first one describes the typical Lorentzian weighted with an exponential decay, while the second consists of a pure exponential decay whose two degrees of freedom are fixed by imposing the resulting piecewise amplitude is $C^1$ throughout.

The ringdown ansatz for the first region, $\fcutrd < \Mf < \fcutrdtwo$, is
\begin{equation}
	A_{\mathrm{ring,1}}(\Mf) = \frac{a_0 \exp\sbr{- \frac{a_1 (\Mf - \fring)}{a_2 \fdamp}}}{(\Mf - \fring)^2 + (a_2 \fdamp)^2} \,.
\end{equation}
We have dropped an overall factor going as a power of the frequency with respect to the previous ansatz, presented in Eq.~(6.2) of \cite{Garcia-Quiros:2020qpx}. In the general case, the coefficients \( a_i \) are determined by solving a system of equations using three colocation points; with the exception of the (3,2) mode, for which these parameters are instead obtained via a direct fit across the parameter space.

The collocation points values $v_i$ are defined at the frequencies
\begin{equation}
	\Mf^{\mathrm{CP}}_{\!\mathrm{rd},i} = \{ \fring - \fdamp,\ \fring,\ \fring + \fdamp \}\,.
\end{equation}
Solving the system $A_{\mathrm{ring,1}}(\Mf^{\mathrm{CP}}_{\!\mathrm{rd}, i}) = v_i$, we get:
\begin{align}\label{eq:ringcoeff}
	a_0 & = \frac{v_1 \fdamp^2}{\sqrt{\dfrac{v_1}{v_3}}-\dfrac{v_1}{v_2}} \,, \\
	a_1 & = \frac{a_2}{2} \log{\del{\frac{v_1}{v_3}}} \,,                     \\
	a_2 & = \pm \frac{1}{\fdamp}\sqrt{\frac{a_0}{v_2}} \,.
\end{align}



The ansatz for the second ringdown region, $\Mf > \fcutrdtwo $, is
\vspace{-2ex}
\begin{equation}
	A_{\mathrm{ring,2}}(\Mf) = b_0 e^{-b_1 (\Mf - \fcutrdtwo)}.
\end{equation}
By imposing the amplitude is $C^1$ throughout the ringdown, we get
\begin{align}
	\begin{split}
		b_0 & = A_{\mathrm{ring,1}}(\fcutrdtwo)\,,                         \\
		b_1 & = -\frac{1}{b_0} A^{\prime}_{\mathrm{ring,1}}(\fcutrdtwo)\,.
	\end{split}
\end{align}

\subsubsection{Mode-mixing}

The ringdown region for the (3,2) mode is described in a spheroidal harmonic basis, as discussed in the \phXHM model paper~\cite{Garcia-Quiros:2020qpx}. Due to the mixing with the (2,2) mode, the ringdown region of the (3,2) spans a larger frequency range than that covered by the ringdown of other modes. This is due to the fact the ringdown region needs to be started around the ringdown frequency of the (2,2) mode, which is lower than that of the (3,2) mode. To account for this, we introduce a new early-ringdown region, starting at $\fcutrd$ and ending at
\begin{equation}
	\frdaux = \fring - \fdamp\,.
\end{equation}
For the high-frequency region, $\Mf > \frdaux$, we employ the same method described in Sec.~\ref{sec:nomodemixing}, but we use fitted coefficients instead of collocation points.

For the low-frequency part, $\fcutrd < \Mf < \frdaux$, we employ a fourth-order polynomial that connects the intermediate region with the late-ringdown region. We impose continuity in the left boundary and continuity plus differentiability in the right one. The last degree of freedom is fixed with a collocation point defined at
\begin{equation}
	\Mf^\mathrm{CP}_\mathrm{\!rdaux} = \frac{1}{2}(\fcutrd + \frdaux) \,.
\end{equation}
The linear system of equations is solved numerically.

\subsection{SpinTaylor angles}

For our description of the twisting-up Euler angles, we rely on the orbit-averaged equations implemented in \codename{LALSuite} \cite{Sturani_note}, with 3.5PN  \textmd{SpinTaylorT4} phasing (see~\cite{Buonanno:2009zt} for a review) and 3PN spin effects in the phasing and spin-precession equations.\footnote{After our code was merged into the public master branch of \codename{LALSuite}, a further extension of \modelname{\phXPHM} along these lines was proposed in~\cite{Yu:2023lml}.} This is also the maximum PN order at which one can work without incurring inaccuracies due to the orbit-averaging of 2PN spin squared terms~\cite{Sturani_note}. By default, we do not include linear-in-spin corrections to the evolution of the orbital angular momentum, though these can be activated toggling a flag (\verb|Lscorr|) in the dictionary of optional waveform parameters.

Using conserved-norm spin vectors, and neglecting the radiation of angular momentum, the evolution of the binary's spin vectors, $\vec{S}_{i}$, and angular momentum unit vector, $\hat{L}_\mathrm{N} $, is described by a set of precession-type equations~\cite{Bohe:2012mr}
\begin{align}
	\dot{\vec{S}}_{i} & =\vec{\Omega}_{i}\times \vec{S}_{i},\ \ i\in\{1,2\} \label{eq:sdot} \\
	\dot{\hat{L}}_\mathrm{N}      & =\vec{\Omega}_{L}\times\hat{L}_\mathrm{N} ,\label{eq:ldot}
\end{align}
where a dot denotes differentiation with respect to time and $\hat{L}_\mathrm{N} $ is the unit vector aligned with the binary's Newtonian orbital angular momentum. In what follows, we will be restricting to quasi-circular inspirals. The orbital frequency and, hence, the magnitude of the orbital angular momentum, are assumed to vary by negligible amounts on the orbital timescale. Under the adiabatic approximation, the evolution of $\omega$ over the radiation reaction timescale can be computed by means of flux-balance equations:
\begin{equation}
	\dot{\omega}=-\frac{d\omega}{dv}\frac{\mathcal{F}}{d\mathcal{E}/dv},
	\label{eq:omegadot}
\end{equation}
where we introduced the PN velocity parameter $v=(M\omega)^{1/3}$ and $\mathcal{F}$ denotes the gravitational-wave energy flux $\mathcal{F}=-d\mathcal{E}/dt$. In the  \textmd{SpinTaylorT4} approximation, the RHS of Eq.~\eqref{eq:omegadot} is Taylor-expanded up to 3.5PN order.

\Crefrange{eq:sdot}{eq:omegadot} are numerically evolved using a fourth-order Runge-Kutta-Fehlberg method. The integration is performed until one of the stepping conditions of the integrator is triggered. These conditions have the purpose to ensure that the output of the integration is physically meaningful and does not break any of the assumptions underlying the calculation (e.g., the energy should be decreasing as the inspiral progresses, and the orbital frequency should remain finite; the PN velocity parameter $v$ needs to be smaller than the speed of light).

As in previous phenomenological model, the conversion to the frequency domain is carried out via the stationary phase approximation. More specifically, for each $(\ell,m)$ mode, we associate the PN velocity parameter, $v$ to the Fourier domain variable $f$ via the mapping $v=(2 M\pi f/m)^{1/3}$, which follows from applying the stationary phase approximation mode by mode. This mapping readily allows one to transform the result of the time-domain integration to the frequency domain in which \modelname{PhenomX} models operate. Following the \modelname{\phXPHM} implementation, we employ this rescaling throughout the coalescence.\footnote{An improved treatment of higher harmonics has been implemented in \phXOfourA~\cite{Thompson:2023ase}, where a different rescaling, based on the quasinormal mode spectrum of the remnant, is invoked in the late stages of the coalescence.}

\begin{figure}[h]
	\begin{center}
		\includegraphics[width=\columnwidth]{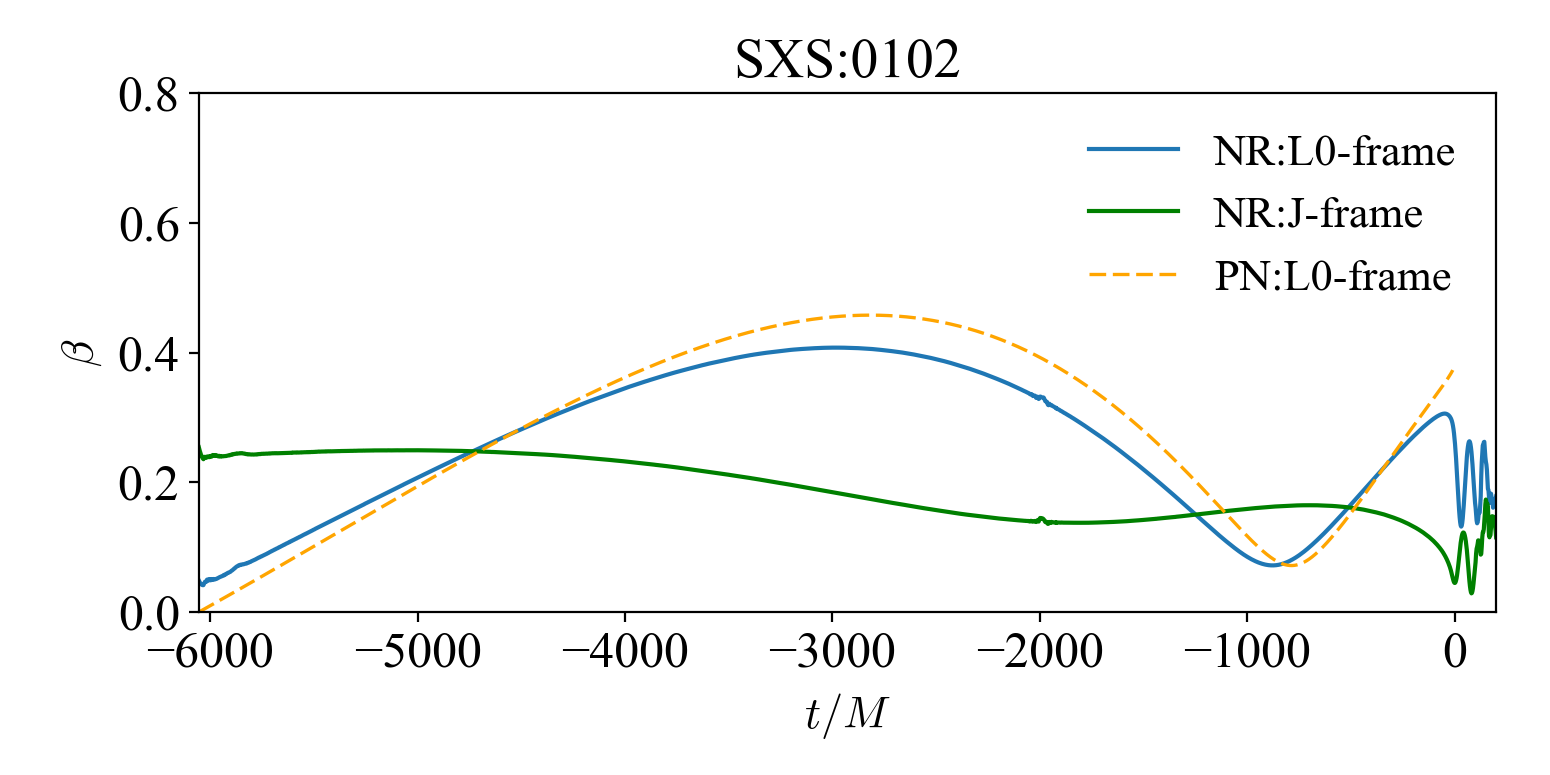}
		\caption{The plot shows the evolution of the angle between the $z$-axis of the co-precessing frame and either the $L_0:=\hat{L}_\mathrm{N}(f_{\mathrm{ref}})$ (blue solid curve) or the total angular momentum $J$ (dark green solid curve), as estimated from the full set of $\ell=2$ modes of $\psi_4$ for the NR simulation SXS:0102, taken from the public SXS gravitational waveform database~\cite{Boyle:2019kee,sxsdb}. The evolution of the first angle is well approximated by the evolution of $\arccos(\hat{L}_{\mathrm{N},z})$, as estimated through the numerical integration of the PN spin-precession equations (orange dashed curve). Transforming between co-precessing and $J$-frame results in a much simplified behavior of the polar Euler angle $\beta$.\label{fig:td_beta}
		}
	\end{center}
\end{figure}

The integration of the spin-precession dynamics returns arrays expressed in the $L_0:=\hat{L}_\mathrm{N}(f_{\mathrm{ref}})$ frame, whose $z-$axis is aligned with the binary's orbital angular momentum at the reference frequency. However, phenomenological models work with rotations from the co-precessing to the J-frame (J being the binary's \textit{total} angular momentum), where the evolution of the Euler angles look considerably simpler (see \cref{fig:td_beta}). Thus, we first rotate the unit vector aligned with the orbital angular momentum $\hat{L}_\mathrm{N}$ to the J-frame, and subsequently compute $\alpha$ and $\beta$ via
\begin{align}
	\alpha & =\arctan\left(\frac{\hat{L}_{\mathrm{N},y}}{\hat{L}_{\mathrm{N},x}}\right), \\
	\beta  & = \arccos(\hat{L}_{\mathrm{N},z}).
\end{align}

\label{subsec:angles_model}
The third angle, $\gamma$, is computed by numerically integrating the minimal rotation condition~\cite{Boyle:2011gg}:
\begin{equation}
	\dot{\gamma}=-\dot{\alpha}\cos{\beta}.
	\label{eq:min_rotation}
\end{equation}

\subsubsection{Post-inspiral continuation of Euler angles}
The Euler angles computed via numerical integration of the \textmd{SpinTaylor} equations will not typically cover the full frequency range required in data analysis. In the worst case scenario, the integration will immediately fail. In practice, this will happen when the reference frequency requested by the user is too close to merger for the system considered. In this case, \phXPHMST will fall back to a precession-averaged solution. If the reference frequency is sufficiently lower than the merger frequency, the integration will succeed, but the solution will cover only the inspiral cycles of the full signal.

Using information from the high-frequency part of the PN solution, \phXPHMST computes a smooth analytical continuation of the $\alpha$ and $\beta$ angles, from which the third Euler angle, $\gamma$, is then obtained via Eq.\,\eqref{eq:min_rotation}. This continuation is not calibrated to NR in any way (unlike the \textmd{PNR} model presented in \cite{Hamilton:2021pkf}): its main purpose is to ensure that the angles evolve smoothly over the whole frequency range, while also providing the basic functionalities for future integrations with NR-tuned merger-ringdown angles. The angle $\alpha$ is continued via the following function:
\begin{equation}
	\alpha_{\rm{MRD}}(f)=a_0+\frac{a_1}{f^2}+\frac{a_2}{f^4}\,,
	\label{eq:alpha_mrd}
\end{equation}
whereas the angle $\beta$ is extended through using
\begin{equation}
	\beta_{\rm{MRD}}(f)=e^{-\kappa f}\left(\frac{b_1}{f}+\frac{b_2}{f^2}+\frac{b_3}{f^3}\right)+b_0\,,
	\label{eq:beta_mrd}
\end{equation}
with $\kappa=\tau_{21}^{-1} - \tau_{22}^{-1}$, where $\tau_{22}$ and $\tau_{21}$ are the damping times of the fundamental $(2,2)$ and $(2,1)$ quasinormal modes. In the time domain,
$\kappa$ can be approximately associated with the characteristic post-merger damping rate of $\beta$~\cite{OShaughnessy:2012iol,Estelles:2021gvs} and we keep it here as a softening factor for the exponential function, since $\kappa\ll 1$. The high-frequency limit of Eq.~\eqref{eq:beta_mrd} ensures that the precession cone closes up after the merger~\cite{Hamilton:2021pkf}. The coefficient $b_0$ can only take one of the two values $\{0,\pi\}$, with the limit chosen to be the closest to the last available value of $\beta$ at high frequency.

The analytical continuations are attached to the inspiral angles by requiring continuity of the solution and its first derivative at a fixed frequency $f_{\rm{trans}}=0.97\,f_{\rm{max}}^{\rm{PN}}$, where $f_{\rm{max}}^{\rm{PN}}$ denotes the last frequency covered by the PN inspiral solutions. The remaining free coefficient in Eqs.~\eqref{eq:alpha_mrd} and \eqref{eq:beta_mrd} is fixed by demanding that $\alpha_{\rm{MRD}}$ and $\beta_{\rm{MRD}}$ agree with the respective PN solutions at another frequency point in the inspiral range.

As in previous models, the magnitude of the remnant's total angular momentum, $J_\mathrm{f}$, is approximated via the squared sum of two contributions, one perpendicular and one parallel to the orbital angular momentum:
\begin{equation}
	J_\mathrm{f} = \sqrt{\displaystyle S_\perp^2 + (S_\parallelto^\text{\phantom{2}} + L_\mathrm{f})^2}\,,
\end{equation}
where the orbital angular momentum radiated through the coalescence is approximated using aligned-spin NR fits.

In \phXPHMST, we model the in-plane spin contribution as a vectorial sum of the individual in-plane spins. Note that this is different from the MSA model, which uses a precession-averaged expression for $J_\mathrm{f}$ (see Eq.~(4.20) of~\cite{Pratten:2020ceb}). The model described here can be activated by setting \verb|PhenomXPrecVersion: 320| (to activate the new Euler angles), and \verb|PhenomXPFinalSpinMod: 2| (to adopt the above final spin description).

\subsection{Comparison of SpinTaylor and NR Euler angles}
\label{subsec:angles_nr_comparison}

In this subsection, we directly compare the Euler angles extracted from selected NR simulations to those estimated by \phXPHMST. For each NR simulation, we identify the co-precessing frame's axes with the eigenvectors of a matrix expressing the orientation-averaged radiation of the system. This matrix is computed using the $\ell=2$ subset of the Newman-Penrose scalar $\psi_4$~\cite{OShaughnessy:2011pmr}. The $z$-axis of the co-precessing frame, in particular, is taken to coincide with the dominant principal axis of this matrix. In the twisting-up performed by \phXPHMST, the $z$-axis of the co-precessing frame tracks instead the Newtonian orbital angular momentum, $\hat{L}_\mathrm{N}$.

\begin{figure*}%
	\centering
	\subfloat[]{\includegraphics[width=0.60\columnwidth]{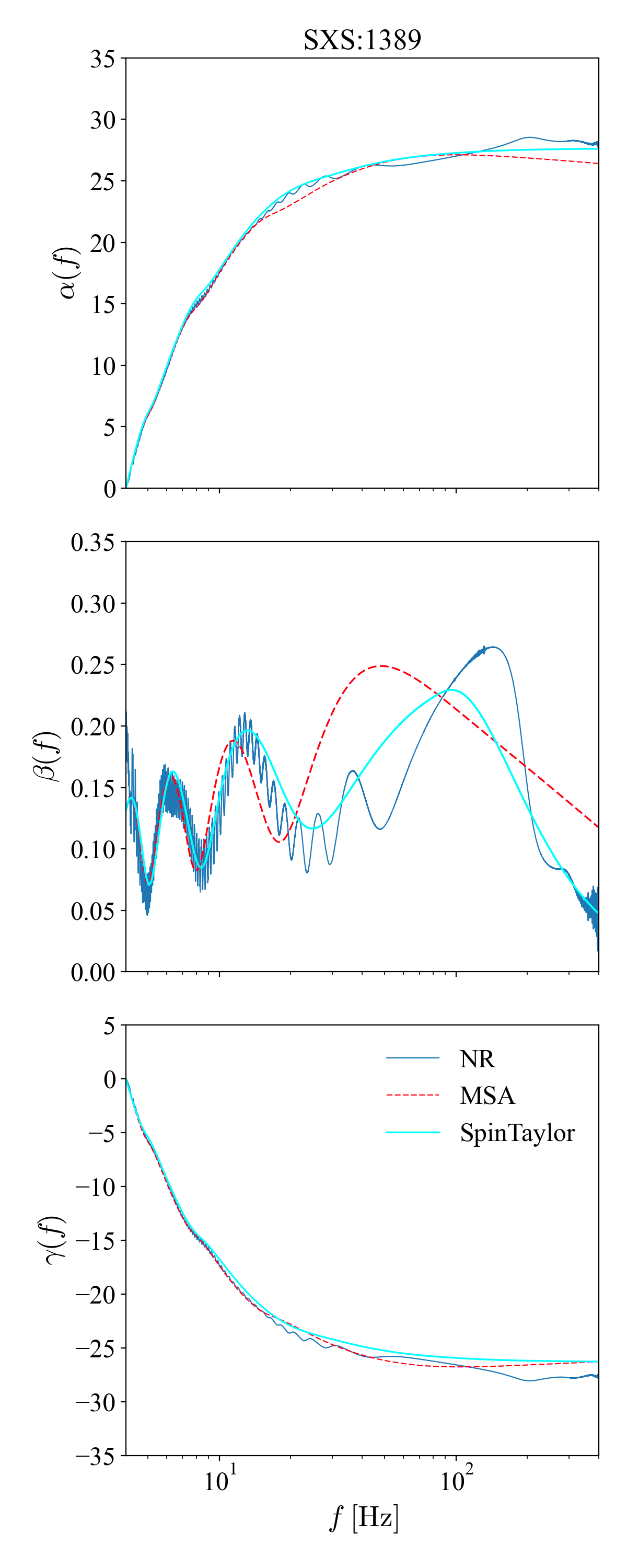}\label{subfiga}}\hfill%
	\subfloat[]{\includegraphics[width=0.60\columnwidth]{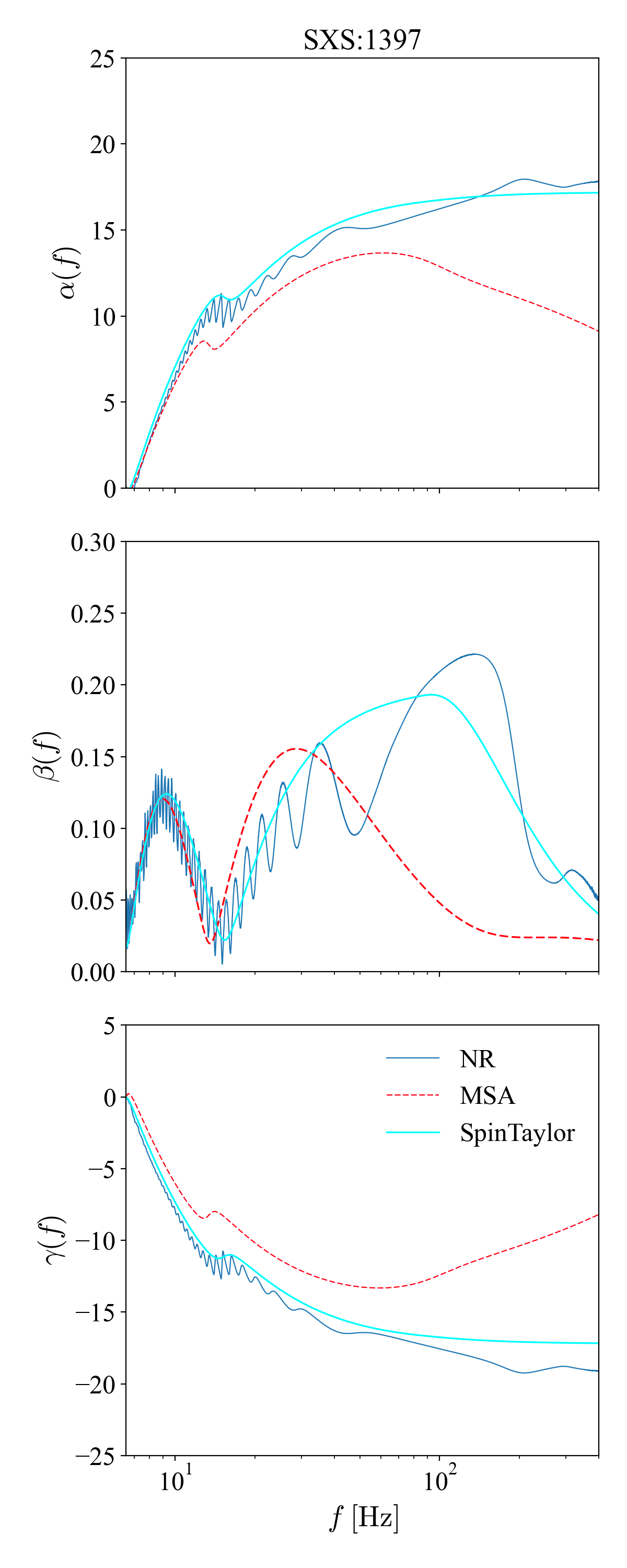}\label{subfigb}}\hfill%
	\subfloat[]{\includegraphics[width=0.60\columnwidth]{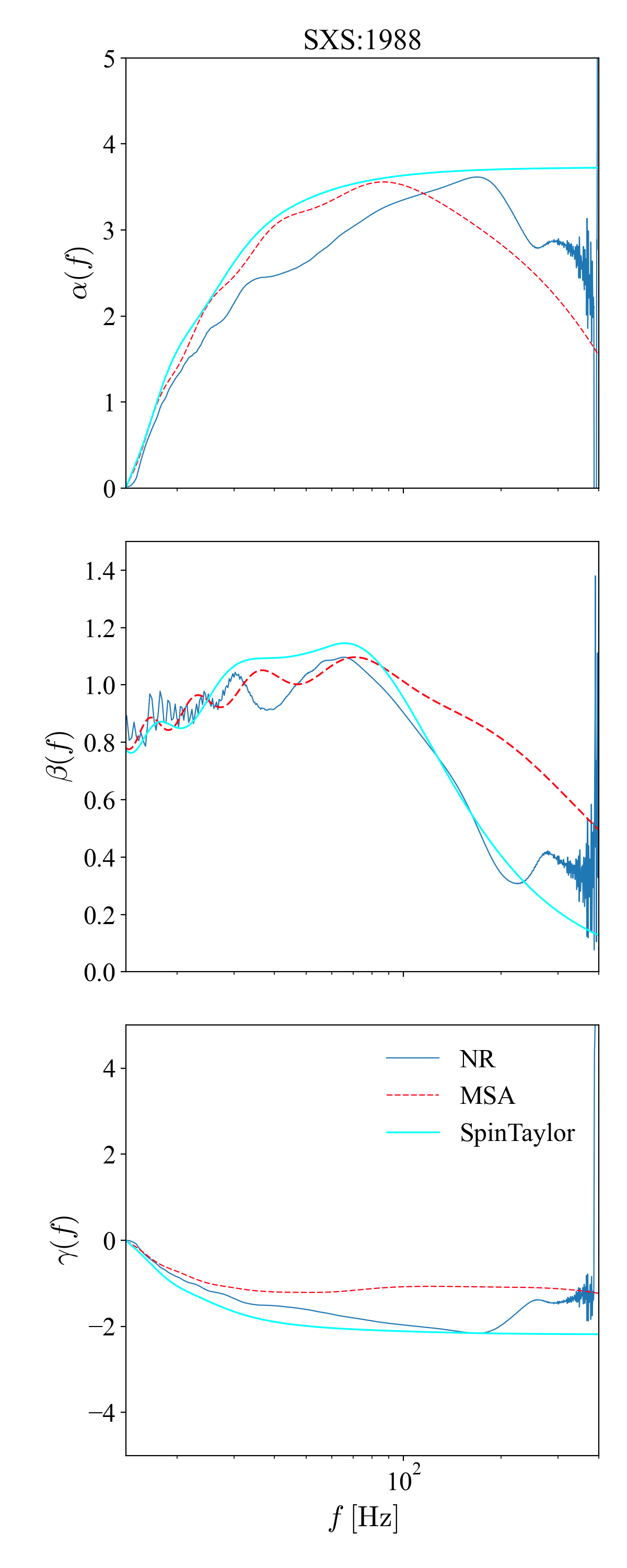}\label{subfigc}}%
	\vspace{-4ex}\caption{
		Each column shows the frequency-domain Euler angles to transform the gravitational-wave signal from the co-precessing to the J-frame, for the NR simulation indicated in the title (see main text for a brief description of the corresponding binary's parameters). The angles extracted from the NR data are shown in blue, whereas those computed with the MSA and SpinTaylor approximations (with the default settings specified in \phX) are shown in red and cyan respectively. In the cases shown here, the SpinTaylor approximation appears to better capture the modulations visible in the NR angles, as well the late-inspiral behaviour of $\beta$. For plotting purposes, we normalize the offsets of the $\alpha$ and $\gamma$ angles so that they take an initial value of zero.}
	\label{fig:euler_angles_fd}
\end{figure*}

In the set of plots presented in \cref{fig:euler_angles_fd}, we compare the full IMR angles estimated by PhenomX models with the ones extracted from a selection of NR waveforms, with different values of the effective precession spin $\chi_p$. To do so, we first rotate the $\ell=2$ set of NR strain modes to the $J$-frame; we then Fourier-transform the modes and apply the eigenvalue method mentioned earlier. Proceeding from left to right, the first simulation, SXS:1389, corresponds to a binary with mass ratio $Q\approx1.6$, effective aligned spin $\chi_\mathrm{eff}\approx-0.06$ and $\chi_p\approx0.28$. The second simulation, SXS:1397, corresponds to a $Q\approx1.7$ system with $\chi_\mathrm{eff}\approx0.25$ and $\chi_p\approx0.72$. Finally, SXS:1988 corresponds to a $Q=4$ system with $\chi_\mathrm{eff}\approx-0.44$ and $\chi_p\approx0.57$. We have fixed the total mass of the systems to be $80\Msun$ and plotted the angles from the minimum frequency available for each NR simulation. The SpinTaylor approximation appears to better capture the (orbit-averaged) modulations visible in the NR angles than the MSA model, as well the late-inspiral behaviour of $\beta$. We can also see that the merger-ringdown continuations of the angles produce sensible results; the end of the inspiral regime roughly coincides with the onset of the $\beta$ decay.

\begin{figure*}[htbp]
	\begin{center}
		\includegraphics[width=0.7\textwidth]{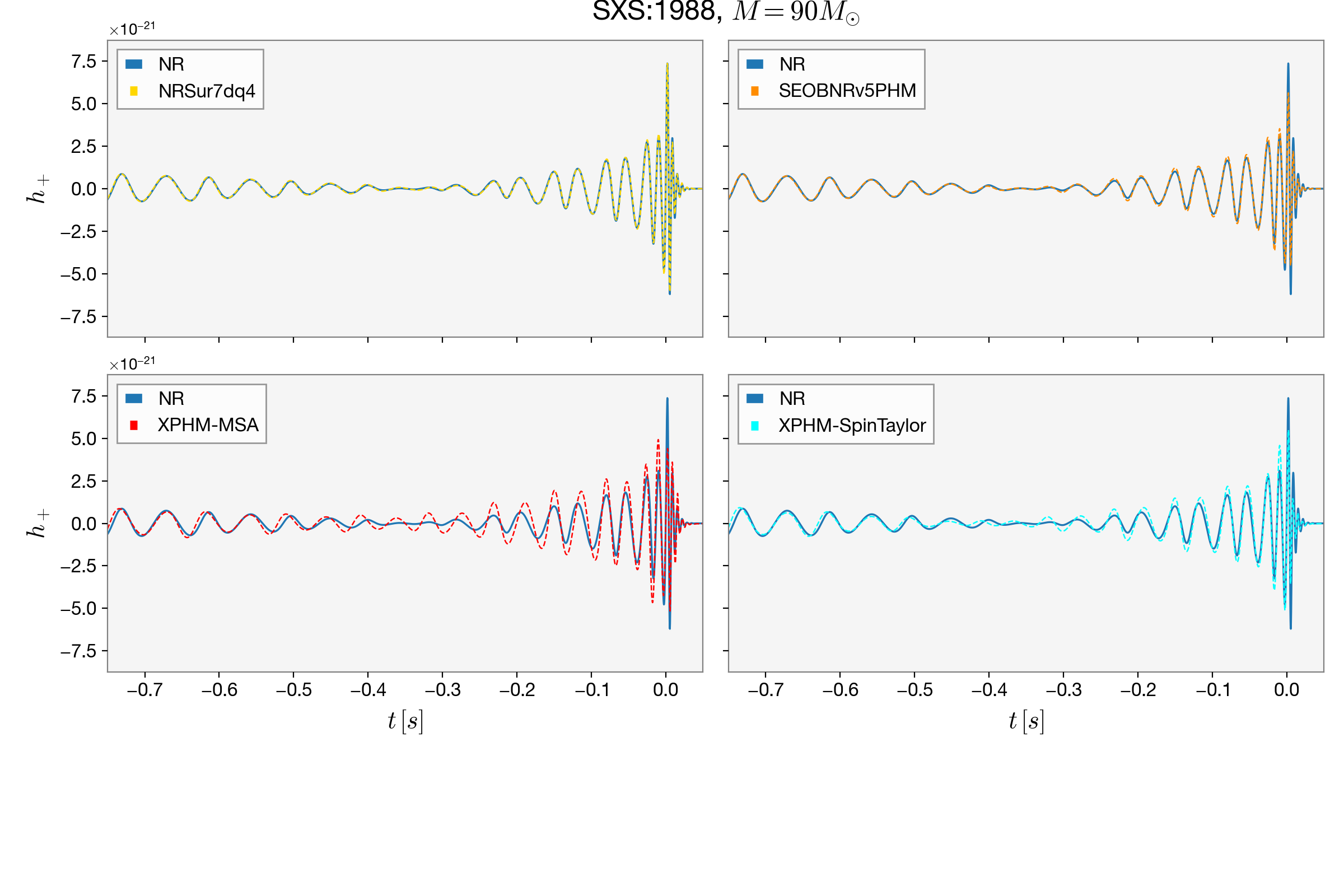}
		\caption{The comparison of the plus polarization $h_{+}$ for a BBH matching the parameters of SXS:1988, with a total mass of $90\Msun$. In the top (bottom) panels we show the overlap of the NR waveform with several time(frequency)-domain waveform models available through the \codename{LALSimulation} or \codename{gwsignal} interfaces.  \label{fig:time_domain_SXS1988}
		}
	\end{center}
\end{figure*}

Fig.~\ref{fig:time_domain_SXS1988} shows the time-domain plus polarization returned by several waveform models for one of the simulations considered here, namely SXS:1988. We overlay the NR data (shown in dark blue) with the predictions of several time-domain (top panels) and frequency-domain models (bottom panels), and pick a binary with inclination $\iota=\pi/2$ rad\footnote{$\iota$ commonly denotes the angle between the orbital angular momentum and the line of sight.} at 15 Hz and total mass $90\Msun$. The time-domain comparison confirms some of the qualitative conclusions drawn from the inspection of the Euler angles (Fig.~\ref{fig:euler_angles_fd}): during the inspiral, the new SpinTaylor version of \phXPHM is closer to the NR data than the MSA version.

\section{Comparison to Numerical Relativity waveforms}\label{sec:nr_comparison}

In this section, we present a comparative analysis of the mismatches between numerical relativity waveforms and the MSA and SpinTaylor versions of \phXPHM compared to those generated by \phXOfourA.

The matches for this study have been computed against the same set of 99 NR precessing SXS waveforms 
used in the original \phXPHM paper between 20 Hz and 2048 Hz and over five total masses $M \in \cbr[0]{50,\ 100,\ 150,\ 200,\ 250}\Msun$, three inclinations $\iota \in \cbr[0]{0,\ \pi/3,\ \pi/2}$, and five values for both the polarization angle of the source and signal phase at the reference frequency.

To prevent correlations between bad matches and low signal-to-noise ratios (SNRs), we compute the SNR-weighted match $\mathcal{\bar{F}}$~\cite{Harry:2016ijz} by averaging over the polarization angle and reference phase as in the following expression
\begin{equation}\label{eq:SNR_match}
	\mathcal{\bar{F}} =
	\sbr[4]{\frac{\sum_{i}\mathcal{F}_i^3\norm[1]{h_{i}^\mathrm{NR}}^3}{\sum_{i}\norm[1]{h_{i}^\mathrm{NR}}^3}}^{1/3},
\end{equation}
where the $i$ index runs over the 25 considered combinations of polarization angle and reference phase; and compute the mismatch $1 - \mathcal{\bar{F}}$ using this quantity. As is customary, the match $\mathcal{F}_i$ between a source gravitational waveform $h_\mathrm{s}$ and a template $h_\mathrm{t}$ is defined as:
\begin{equation}
	\mathcal{F}_i=\max_{t_\mathrm{c},\phi_\mathrm{c},\psi}\frac{\braket{h_\mathrm{s}|h_\mathrm{t}}}{\sqrt{\braket{h_\mathrm{s}|h_\mathrm{s}}\braket{h_\mathrm{t}|h_\mathrm{t}}}}\,,
\end{equation}
where $\braket{\,\cdot\,|\,\cdot\,}$ denotes the usual noise-weighted inner product, $\norm[0]{h}=\sqrt{\braket{h|h}}$ and the match is maximised over the polarization angle $\psi$, the coalescence time $t_\mathrm{c}$ and the coalescence phase $\phi_\mathrm{c}$. In our case, the detector's noise properties are approximated by the design sensitivity Zero-Detuned-High-Power Power Spectral Density~\cite{adligopsd}.

\begin{figure*}
	\centering
	\includegraphics[width=\textwidth]{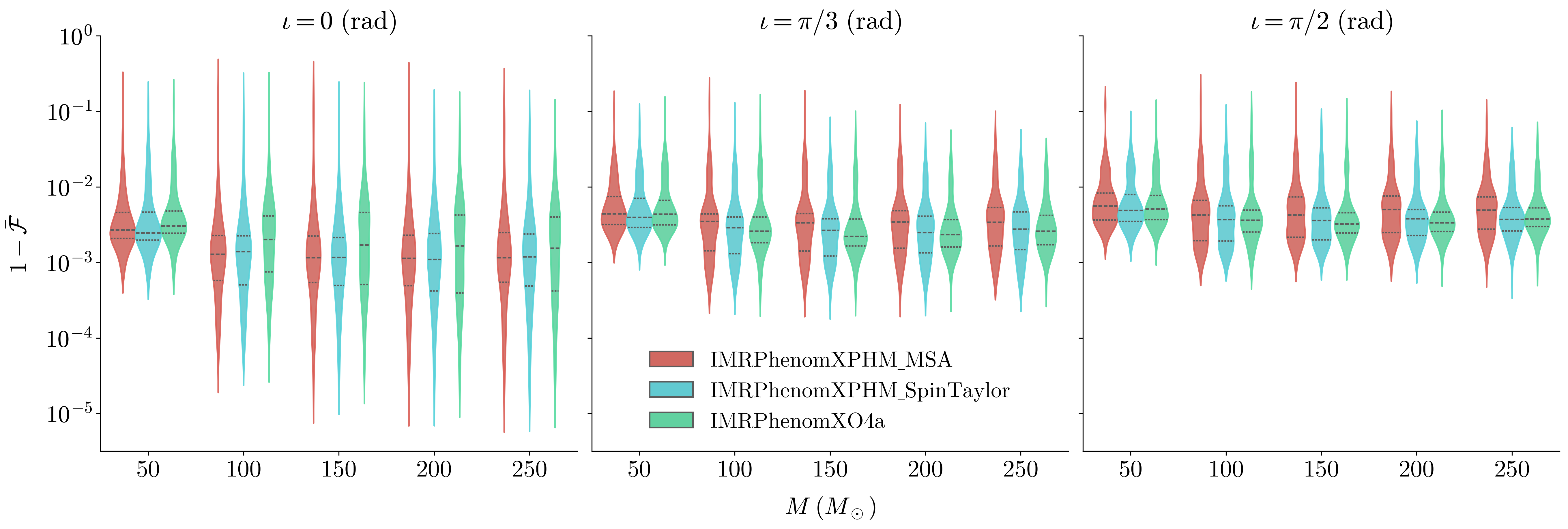}
	\caption{Violin plots showing the SNR-weighted mismatch distributions between 99 precessing SXS waveforms and the MSA and SpinTaylor versions of \phXPHM, compared to those generated by \phXOfourA. The plots are divided by total mass and inclination. The horizontal lines represent the median and quartiles of their respective distribution.}
	\label{fig:mismatch_violins}
\end{figure*}

The resulting mismatch distributions are summarized in \cref{fig:mismatch_violins}, where it can be seen how both versions of the Euler angles are very similar across all tested masses for face-on sources, but for non-zero inclinations, \phXPHMST produces lower median mismatches and narrower spreads than MSA, especially for higher total masses. \phXOfourA displays slightly worse mismatches than both versions of \phXPHM for face-on sources, while its performance is comparable to \phXPHMST for non-zero inclinations, with slightly smaller spreads.

Overall, computing the mean mismatch and its standard error over all cases, we find that \phXPHMST has both the smallest mean SNR-weighted mismatch and standard deviation with a value of $(5.1\,\pm\,0.2)\times 10^{-3}$, followed by \phXOfourA with $(5.45\,\pm\,0.25)\times 10^{-3}$, and the MSA angles with $(6.3\,\pm\,0.4)\times 10^{-3}$.

When looking at the worst matches of each model, we find that these all happen for strongly precessing (high in-plane spins) high mass-ratio binaries (SXS:0057, SXS:0058 and SXS:0062 for some inclinations and SXS:0165 for all inclinations).

However, although all models perform worse for these configurations, we find that \phXOfourA and \phXPHMST perform noticeably better than the MSA angles, as can be seen in \cref{fig:mismatch_violins} by looking at the upper tails of the distributions. This observation can be quantified further by computing the number of outliers for a given match threshold, which we set here to $96.5\%$. We find that \phXOfourA and \phXPHMST have the same number of cases with matches below this fiducial threshold, $21$ out of $1485$, while \phXPHMMSA has $29$.
Therefore, both \phXOfourA and \phXPHMST have the benefit of reducing the number of configurations that are more likely to lead to significant biases in parameter estimation.

\section{Comparison to other models}
\label{sec:model_comparison}

In this subsection, we compare \phXPHMST, along with other state of the art gravitational waveform models, to a set of 5000 \textmd{NRSur7dq4} waveforms, whose parameters have been randomly sampled following a uniform prior in $Q$, with $Q\in[1,6]$ and total mass $M$ with $M\in[80,200]\Msun$ and spins isotropically distributed, with spin magnitudes $\chi_{1,2}\in[0,0.9]$. Hence, the sample includes configurations both in the training and extrapolation range of \textmd{NRSur7dq4}. The inclination of the source is also randomly chosen, following a uniform prior in $\cos\iota$, as well as the signal's polarization. We include in this comparison the frequency-domain phenomenological models discussed in Sec.~\ref{sec:nr_comparison}, as well as the native time-domain models \modelname{SEOBNRv5PHM} and \phTPHM.  We quantify the level of agreement between models via the match $\mathcal{F}$ (no SNR-weighting taken), using the same noise curve as in our previous comparison and setting the minimum and maximum frequency for the match to 25\,Hz and 2048\,Hz, respectively. The lower cutoff is chosen to minimize the failures of \textmd{NRSur7dq4}, whose starting frequency is restricted by the length of the NR waveforms employed in its construction.

\begin{figure*}
	\centering
	\includegraphics[width=0.8\textwidth]{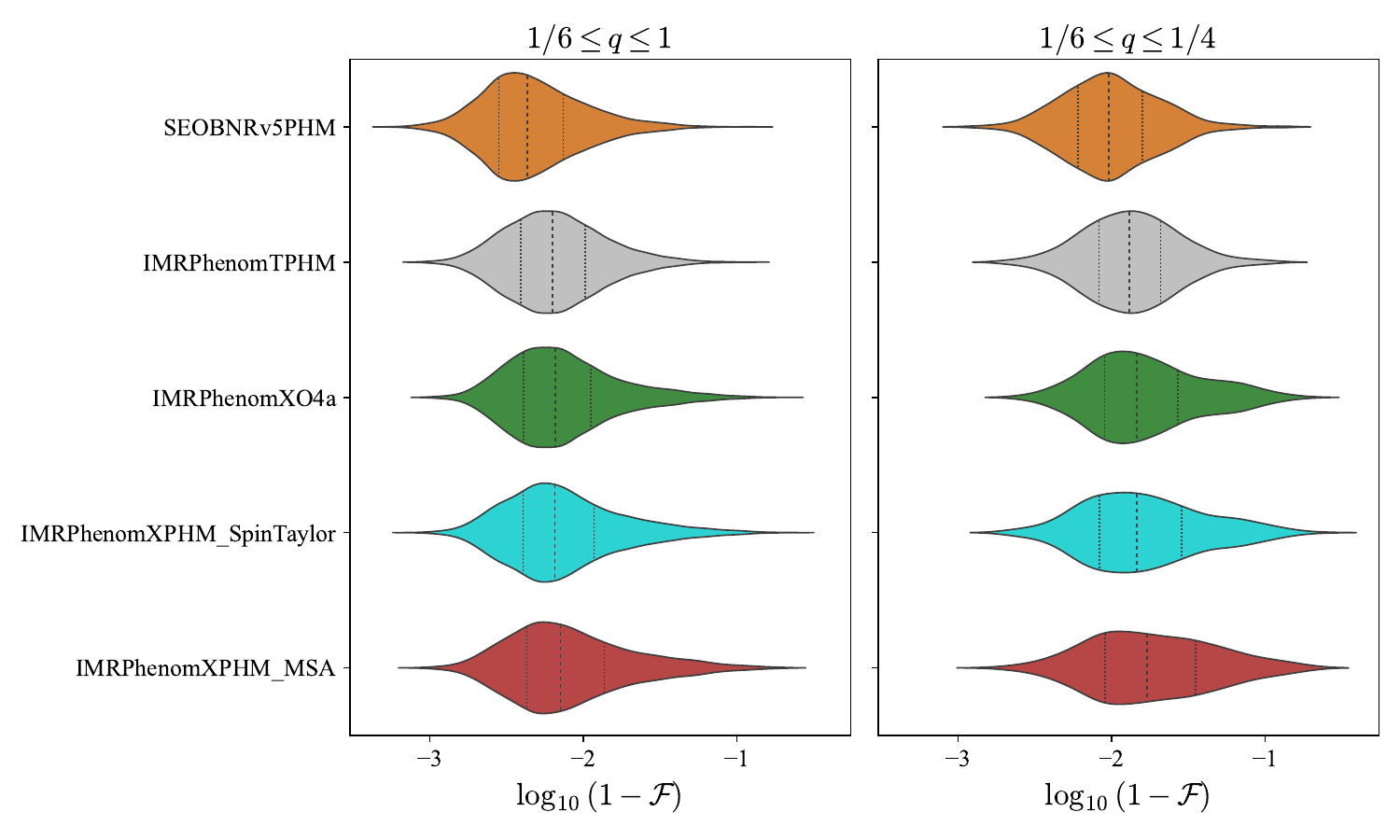}\hspace{3.35cm}
	\caption{Violin plots for the mismatch distributions obtained comparing each model to the precessing NR surrogate \textmd{NRSur7dq4}, over a random sample of 5000 precessing configurations (see main text for further details). The results have been grouped according to the mass ratio of the configurations, as specified in the panels' titles. Dashed (solid) vertical lines mark the median (first and third quartiles) of the plotted distributions.}
	\label{fig:mismatch_surro}
\end{figure*}

The results of the test are presented in Fig.~\ref{fig:mismatch_surro}. In line with previous analyses~\cite{Ramos-Buades:2023ehm,Thompson:2023ase, MacUilliam:2024oif}, \modelname{SEOBNRv5PHM} yields the lowest mismatches against \modelname{NRSur7dq4}. Among phenomenological models, \phTPHM deliver the best matches as well as a reduced tail of outliers; the original \phXPHMMSA implementation has the worst performance. When considering the full sample, the median of the match distributions differ by very small quantities, of the order of $10^{-3}$. However, more significant differences emerge when restricting to very unequal masses, as shown in the right panel of Fig.~\ref{fig:mismatch_surro}. Similarly to what we have observed in the previous comparison against NR simulations, the tails of the distributions are very sensitive to modelling improvements: setting again a fiducial match threshold $\mathcal{F}_*=96.5\%$, \modelname{SEOBNRv5PHM} has the lowest fraction of outliers (defined as the percentage of configurations with $\mathcal{F}<\mathcal{F}_*$), amounting to 1.2\%, followed by \phTPHM with 1.9\%. \phXPHMMSA total the highest percentage of outliers, 8.3\%; both \phXPHMST and \phXOfourA vastly improve on this result, with 5.9\% and 5.2\%, respectively.

It is important to stress that \modelname{SEOBNRv5PHM} features substantial updates to the aligned-spin baseline that go significantly beyond the amplitude recalibration discussed in Subsec.~\ref{subsec:amplitude_model}. Current \modelname{IMRPhenomX/T} models would similarly benefit from a recalibration of the phasing of their aligned-spin sector, which we leave for future work.

\section{Benchmarks}
\label{sec:benchmarks}

\begin{figure}[htp]
	\begin{center}
		\includegraphics[width=0.9\columnwidth]{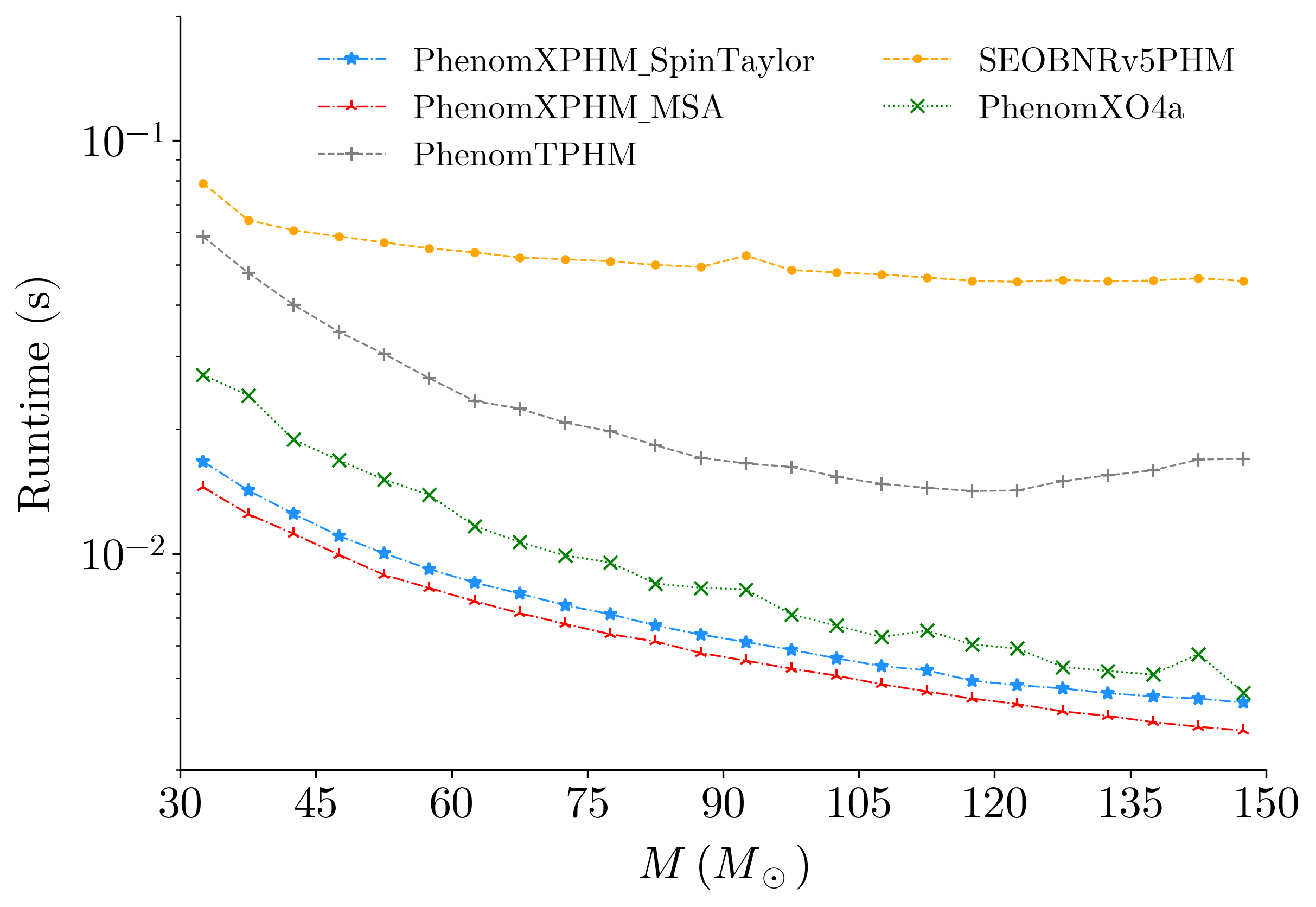}\hspace{0.075\columnwidth}
		\caption{The evaluation times of \phXPHM for the new (SpinTaylor) and previous (MSA) versions of precession, compared to those of \phXOfourA, \phTPHM and \modelname{SEOBNRv5PHM} over a sample of 5000 random precessing black-hole binary configurations, run on a 12-Core Apple M3 Pro.}
		\label{fig:timing_bbh_precessing}
	\end{center}
\end{figure}

In this subsection we compare the evaluation times of \phXPHMST to those of other waveform models including precession and higher harmonics. In particular, we focus on phenomenological models, including its predecessor \phXPHMMSA, and the recent models \phXOfourA and \phTPHM, but we also include the effective-one-body model \modelname{SEOBNRv5PHM} for reference.
Our test proceeds as follows: we generate 5000 random precessing black-hole configurations with mass ratio $Q\in[1,10]$, total mass $M_{T}\in[30,150]\Msun$ and spins isotropically distributed, and call each model activating its default mode content, evaluating\footnote{The evaluation is performed through the \textmd{SimInspiralFD} interface of \codename{LALSimulation}, with the exception of \modelname{SEOBNRv5PHM}, which is implemented in Python and must be called through the new \textmd{GenerateFDWaveform} interface within the \codename{gwsignal} module of \codename{LALSimulation}.} the waveforms over the frequency range $20-2048\ \mathrm{Hz}$, with a frequency step of $\Delta f=0.125\ \mathrm{Hz}$. We then average the individual evaluation times over mass bins of $5\Msun$ in total mass. \Cref{fig:timing_bbh_precessing} shows the results of our test, where for each model we have plotted the mean evaluation time, as a function of total mass. All frequency-domain phenomenological models exhibit similar performance, with the original \phXPHMMSA being the fastest, followed by \phXPHMST and \phXOfourA with slow-up factors of approximately 1.2 and 1.3 respectively.
\phXOfourA has a slightly wider variability in its evaluation times, with the model's evaluation time growing more significantly for lighter systems than that of \phXPHMST. As expected, \modelname{SEOBNRv5PHM} and \phTPHM are the slowest models in frequency domain, \phTPHM being roughly 3 times more expensive than \phXPHM (including the cost of Fourier-transforming the original time-domain polarizations), and, at the same time, \modelname{SEOBNRv5PHM} being roughly 3 times slower than \phTPHM.

The speed of the previous \phXPHMMSA implementation stems from its precession-averaged treatment of the spin dynamics, which does not require the numerical integration of ordinary differential equations. In \phXPHMST, the cost of this operation is amortized in the low-frequency part of the inspiral by performing the integration of the PN spin-precession equations over a coarse time grid and interpolating the resulting Euler angles. The coarseness of the low-frequency grid can be changed to achieve a further speed up of the waveform evaluation; this feature could be useful when analyzing long signals. 

\section{Parameter estimation}
\label{sec:pe}

In this section, we test the performance of the new model by employing it in the analysis of both real (Subsec.~\ref{subsec:real_gw}) and simulated gravitational-wave signals (Subsec.~\ref{subsec:simulated_gw}).

\subsection{Study of real gravitational-wave events}
\label{subsec:real_gw}

Here, we present a re-analysis of three GW events from the GWTC-3 catalog~\cite{KAGRA:2021vkt}: GW191109 (shorthand for GW191109\_010717), GW200129 (shorthand for GW200129\_065458) and GW200225 (shorthand for GW200225\_060421). In all cases, we analyse data frames publicly available thorugh the GWOSC portal~\cite{LIGOScientific:2019lzm}, and employ the software \codename{bilby}~\cite{Ashton:2018jfp} to perform Bayesian parameter estimation, using nested sampling. With the exception of sampler settings, which reflect updated features in \codename{bilby}\footnote{Our analyses use the \textmd{acceptance-walk} stepping method, with \texttt{nlive = 1000} and \texttt{naccept = 60}.}, the remaining analysis settings, including the calibration envelopes and noise curves, match those included in the GWTC-3 data release~\cite{gwtc3-zenodo}.

\begin{figure*}
	\subfloat[]{\includegraphics[width=0.66666\columnwidth]{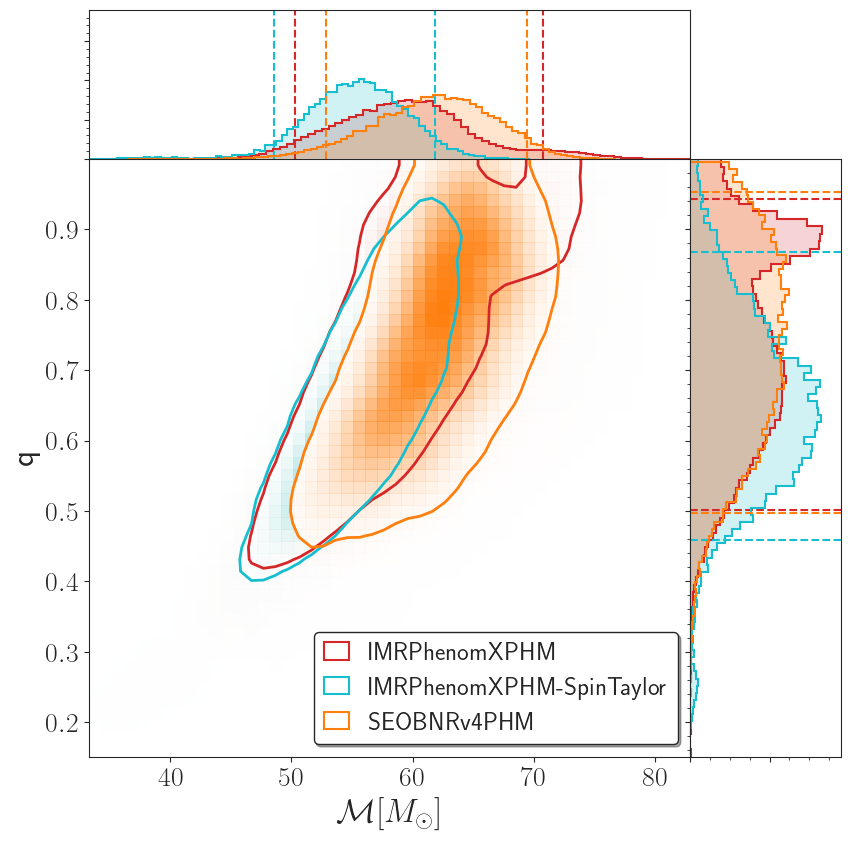}}\hfill
	\subfloat[]{\includegraphics[width=0.66666\columnwidth]{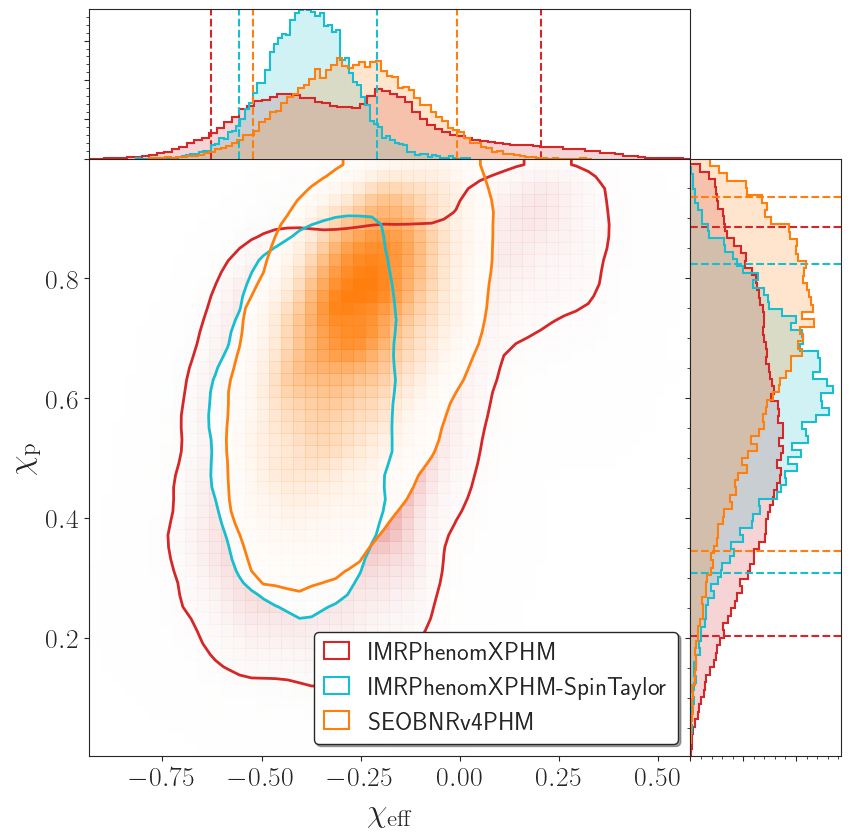}}\hfill
	\subfloat[]{\includegraphics[width=0.66666\columnwidth]{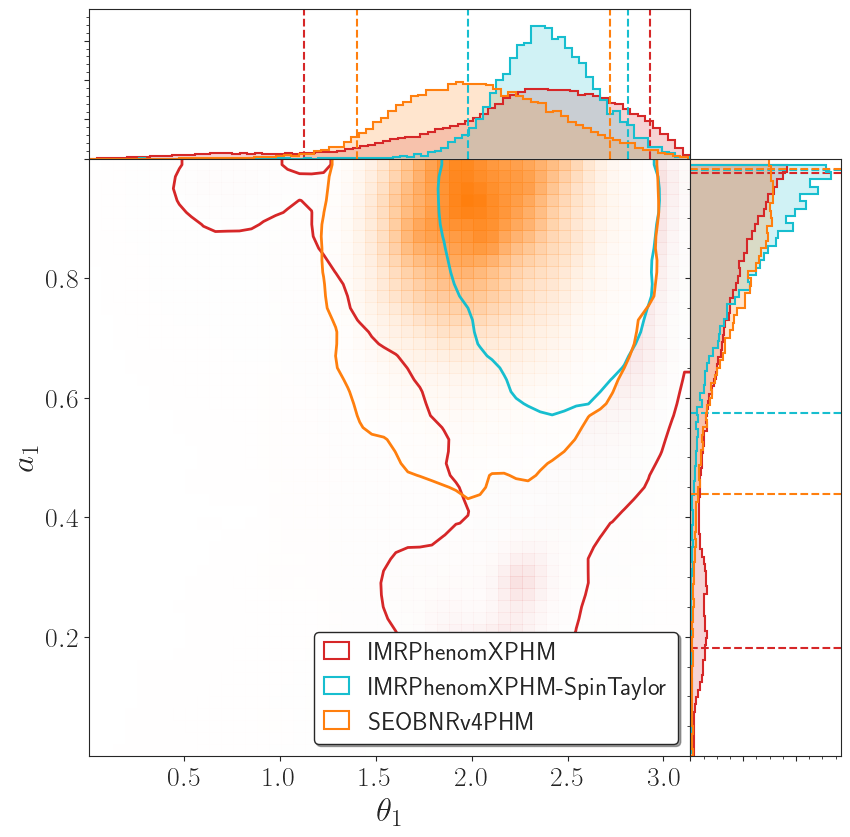}}\hfill
	\caption{Joint posterior distributions for some of the mass and spin parameters for GW191109. The left panel displays chirp mass and mass ratio. The middle panel shows the effective aligned and precessing spin parameters $\chi_{\mathrm{eff}}$ and $\chi_{\mathrm{p}}$, while the right panel shows the primary's spin magnitude and tilt. Posterior samples from the GWTC-3 data release~\cite{gwtc3-zenodo} are shown in red and orange (for \phXPHM and SEOBNRv4PHM, respectively), whereas results for the new \phXPHMST model are shown in cyan. Dashed (solid) lines indicate the 90\% credible intervals for the 1D (2D) posterior distributions.}
	\label{fig:gw191109}
\end{figure*}

GW191109 has attracted some attention due to its support for $\chi_{\mathrm{eff}}<0$, indicating that one or both components have spins anti-aligned with the orbital angular momentum. Negative values of $\chi_{\mathrm{eff}}$ might be the tell-tale sign of a binary's dynamical origin~\cite{Zhang:2023fpp}. Moreover, observations of BBHs with sufficiently negative effective spins could help to constrain the contribution of hierarchical mergers~\cite{Fishbach:2022lzq}. Our analysis, presented in Fig.~\ref{fig:gw191109}, shows that the new model \phXPHMST clearly constrains the primary's spin tilt $\theta_1$ to be larger than $\pi/2$, with a median value of $\sim 2.4$ rad.\footnote{Ref.~\cite{Udall:2024ovp} has questioned whether the support for $\chi_{\mathrm{eff}}<0$ might be due to some imperfect modelling of a glitch affecting the LIGO Livingston detector shortly before merger. We do not address this issue here, and rely on the same deglitched frames employed for the GWTC-3 catalogue.}
Furthermore, the posterior for $\chi_{\mathrm{eff}}$ for  \phXPHMST excludes $\chi_{\mathrm{eff}}>0$ with at least 99.98\% credibility, delivering a much more stringent constraint than \phXPHM and \textmd{SEOBNRv4PHM}. These results are in good agreement with a recent reanalysis of this event performed with the updated model \modelname{SEOBNRv5PHM} (see Fig.~14 of \cite{Ramos-Buades:2023ehm}).

\begin{figure*}
	\subfloat[]{\includegraphics[width=0.8\columnwidth]{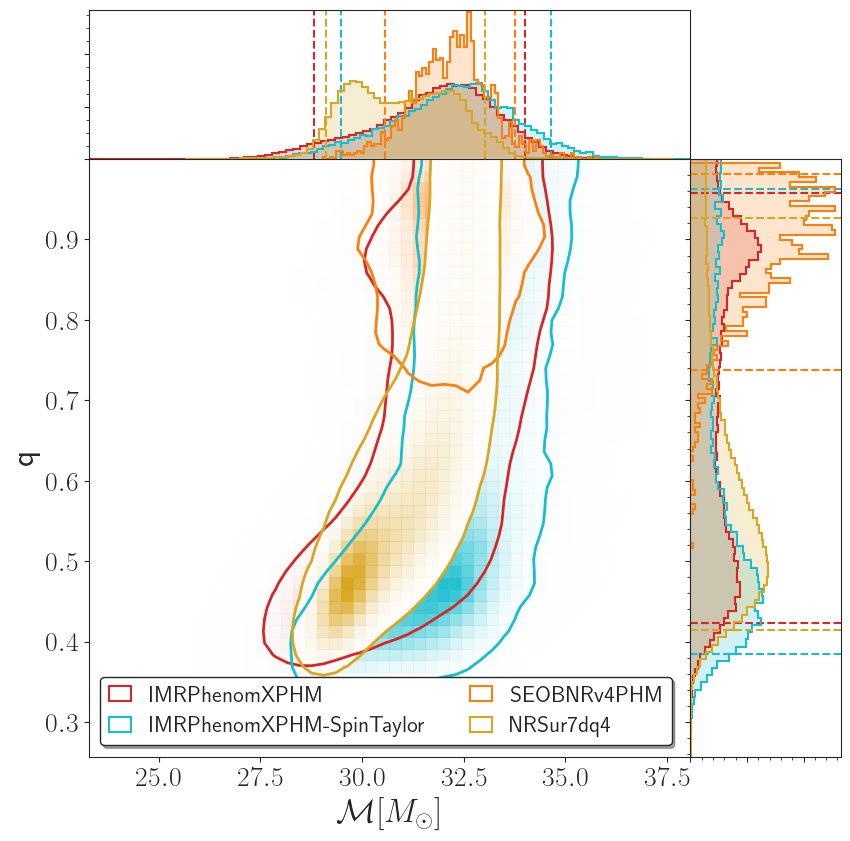}}\hspace{1cm}
	\subfloat[]{\includegraphics[width=0.8\columnwidth]{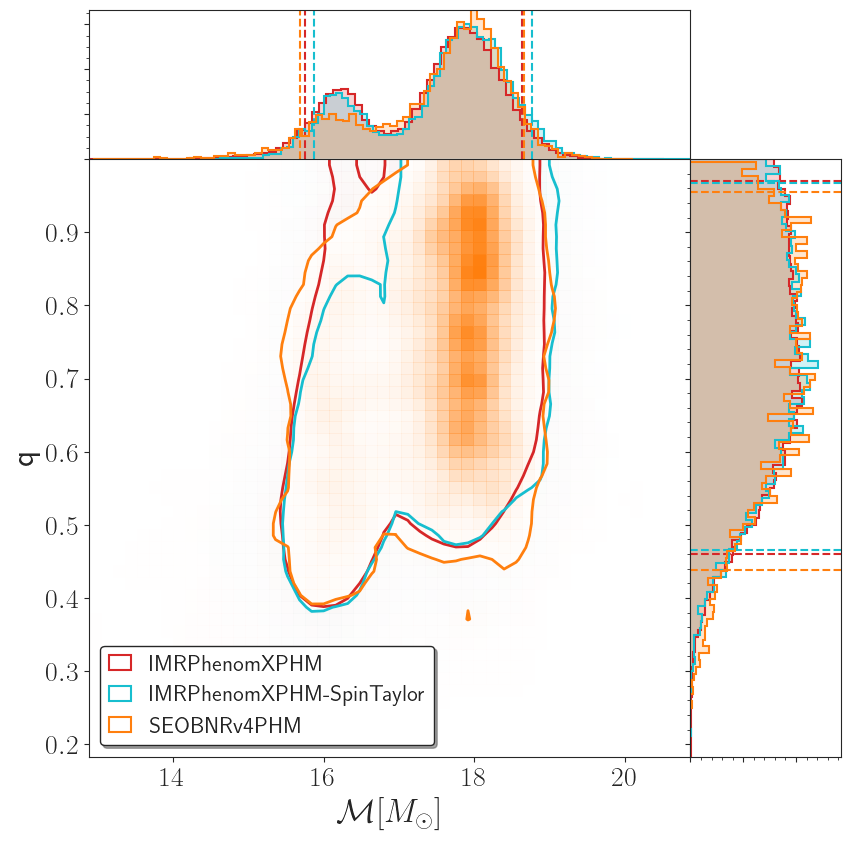}}\\
	\subfloat[]{\includegraphics[width=0.8\columnwidth]{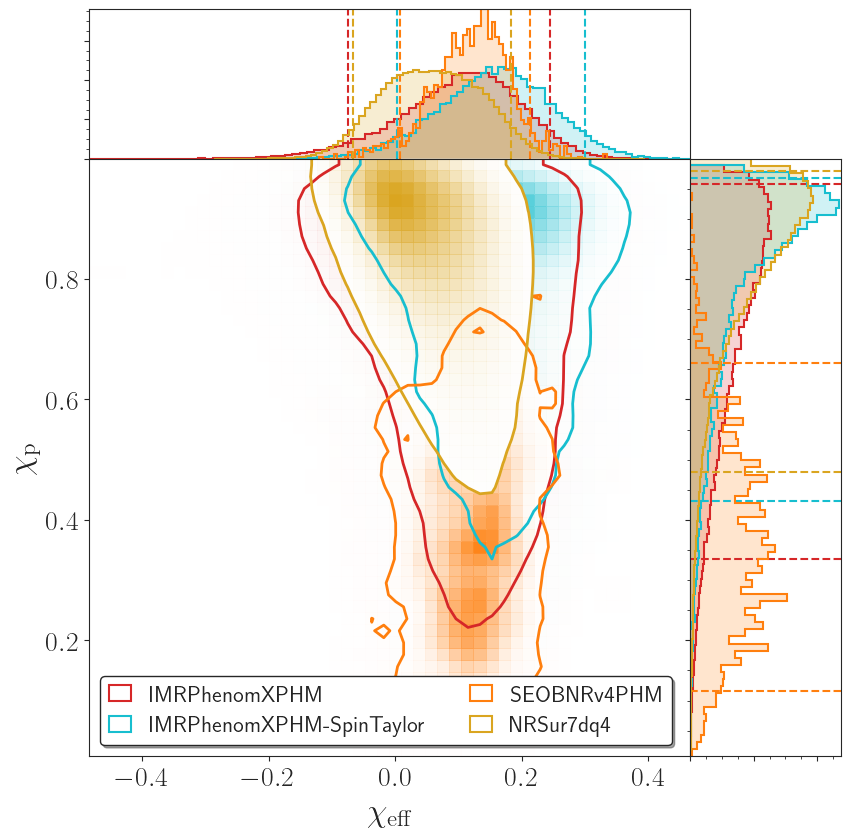}}\hspace{1cm}
	\subfloat[]{\includegraphics[width=0.8\columnwidth]{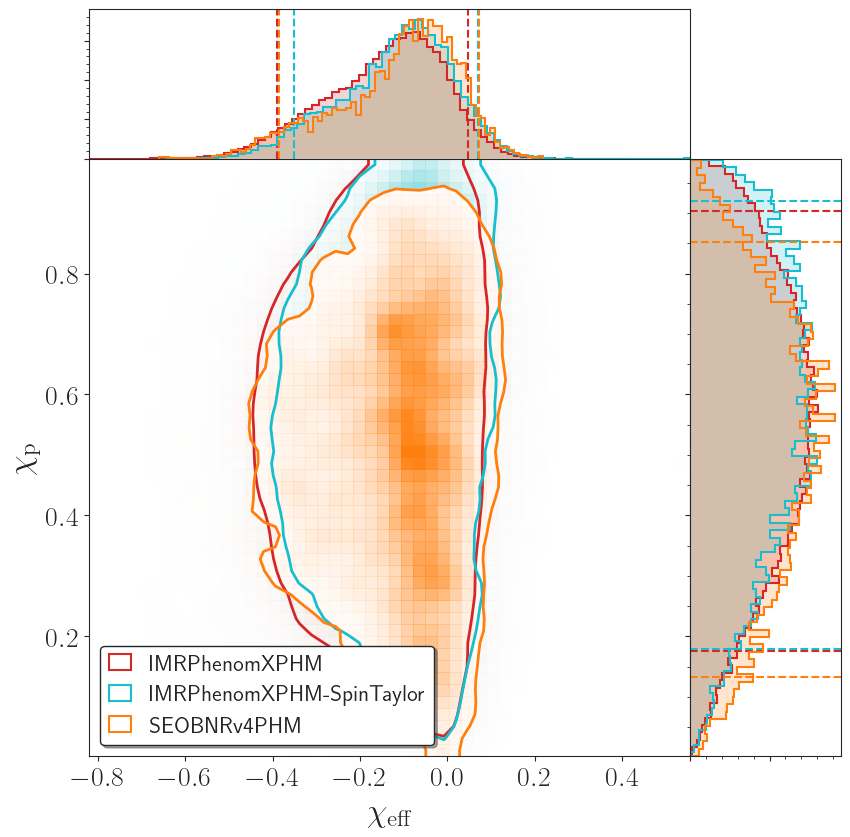}}
	\caption{Joint posterior distributions for some of the spin parameters for the gravitational-wave events GW200129 (left column) and GW200225 (right column). The upper panels show posterior distributions for the chirp mass and mass ratio, while the lower panels show the effective aligned spin and precessing spin parameters $\chi_{\mathrm{eff}}$ and $\chi_{\mathrm{p}}$. Posterior samples from the GWTC-3 data release~\cite{gwtc3-zenodo} are shown in red and orange (for \phXPHM and SEOBNRv4PHM, respectively), whereas results for the new \phXPHMST model are shown in cyan. For GW200129, we also show results obtained with the surrogate model NRSur7dq4, taken from~\cite{hannam_mark_2022_6672460} (in gold). Dashed (solid) lines indicate the 90\% credible intervals for the 1D (2D) posterior distributions.}
	\label{fig:gw200225_gw200129}
\end{figure*}

GW200129 was hailed as the first strongly precessing BBH detected so far~\cite{Hannam:2021pit}. Evidence of precession has been later questioned and attributed to glitch contamination by some authors~\cite{Payne:2022spz}, though other analyses showed that support for precession still stands, regardless of the deglitching technique applied to the strain data~\cite{Macas:2023wiw}. Here, we compare the results obtained with \phXPHMST to the GWTC-3 results as well as to the data release accompanying ~\cite{Hannam:2021pit}, where the event was analyzed with the \textmd{NRSur7dq4} model. The results, shown in Fig.~\ref{fig:gw200225_gw200129}, are in broad agreement with the previous \phXPHM implementation, though we observe the new model shows a slightly increased support for the same $q\sim0.45$ mode favored by the surrogate model. The support for extreme spins is unchanged.

Finally, we reexamine GW200225: similarly to GW191109, this event is interesting from an astrophysical point of view because of its support for negative $\chi_{\mathrm{eff}}$. In this case, we do not find significant differences between different waveform models, which return very similar posterior distributions for both mass and spin parameters, as can be gathered from Fig.~\ref{fig:gw200225_gw200129}.

\subsection{The study of a simulated signal}
\label{subsec:simulated_gw}

In this subsection, we present the result of an injection and recovery study to investigate further the performance of \phXPHMST when the source parameters are known exactly. To this end, we inject a synthetic signal, the SXS simulation \textsc{SXS:1916}, into the LIGO Hanford-Livingston-Virgo network, in zero noise. For the recovery, we employed publicly available O4 noise curves for all detectors~\cite{noise}. The simulated source has right ascension and declination $\text{ra}=5.54$ rad and $\text{dec}=0.09$ rad, coalescence and polarization angles $\phi_\mathrm{c}=\psi=0$ rad, and it is placed at a luminosity distance $D_\mathrm{L}=1$ Gpc, with the trigger's geocentric time set to 1249852257.0 s. The inclination of the total angular momentum with respect to the line of sight was taken to be $\theta_{\mathrm{JN}}\sim 0.5$ rad. \textsc{SXS:1916} corresponds to a black hole binary with mass ratio $Q=4$, with dimensionless spins at the reference frequency (20 Hz) $\vec{\chi}_{1}\approx[0.243,\,0.762,\,0.010]$ and $\vec{\chi}_{2}\approx[0.530,\,-0.599,\,-0.033]$. These values translate into an effective aligned-spin parameter $\chi_{\mathrm{eff}}\approx 1.52\times 10^{-3}$ and precession spin $\chi_\mathrm{p}\approx 0.800$. We choose the total mass of the system to be $M\approx 58.3 \Msun$.

\begin{figure*}
	\centering
	\subfloat[]{\includegraphics[width=0.8\columnwidth]{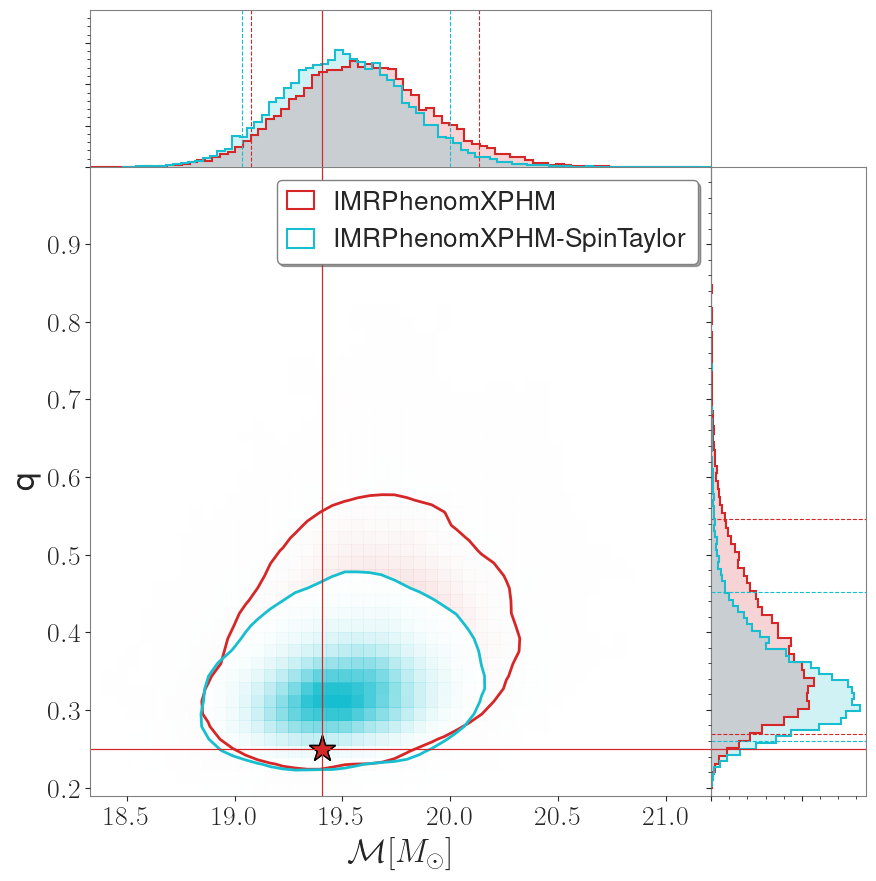}}\hspace{1cm}
	\subfloat[]{\includegraphics[width=0.8\columnwidth]{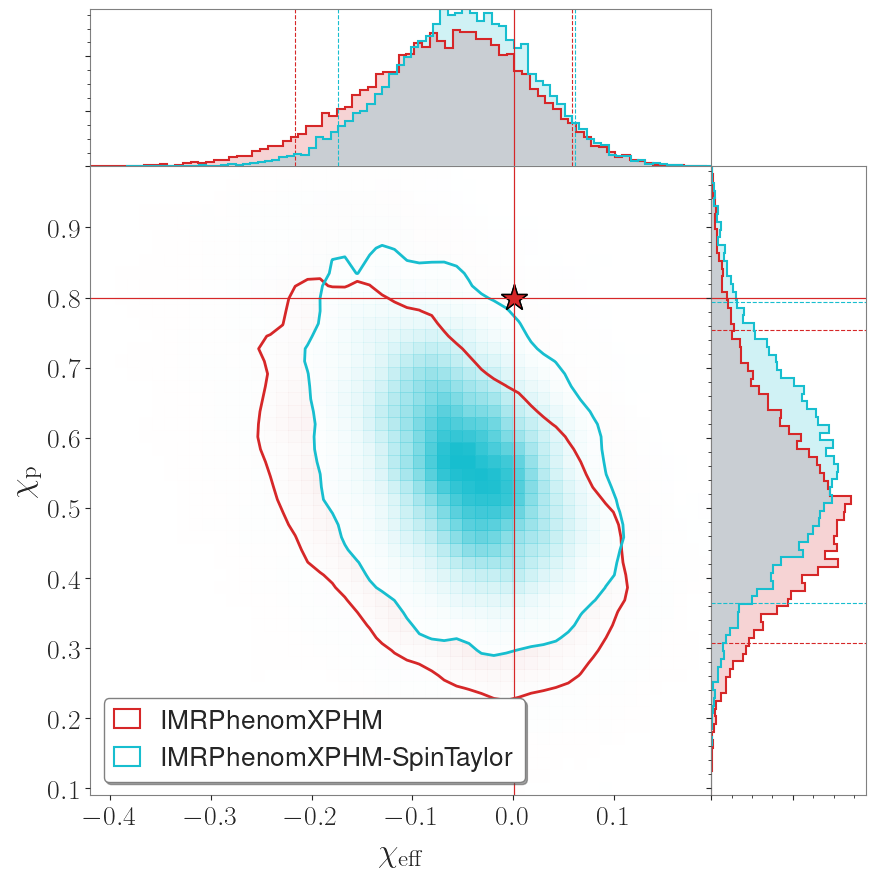}}\\
	\subfloat[]{\includegraphics[width=0.8\columnwidth]{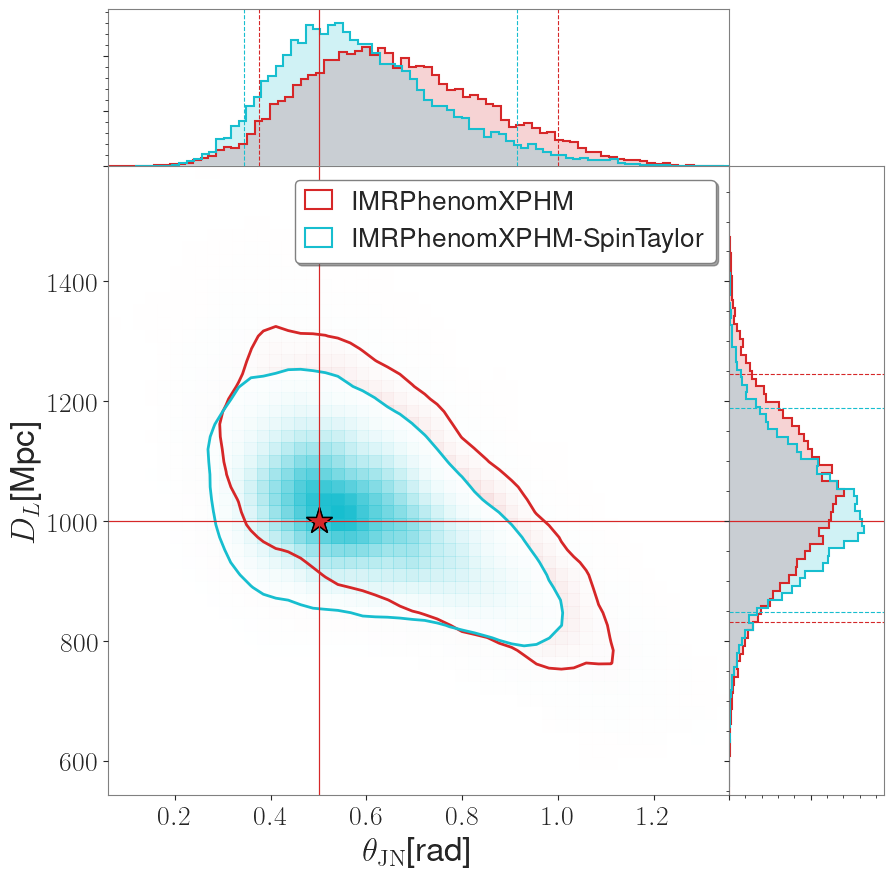}}\hspace{1cm}
	\subfloat[]{\includegraphics[width=0.8\columnwidth]{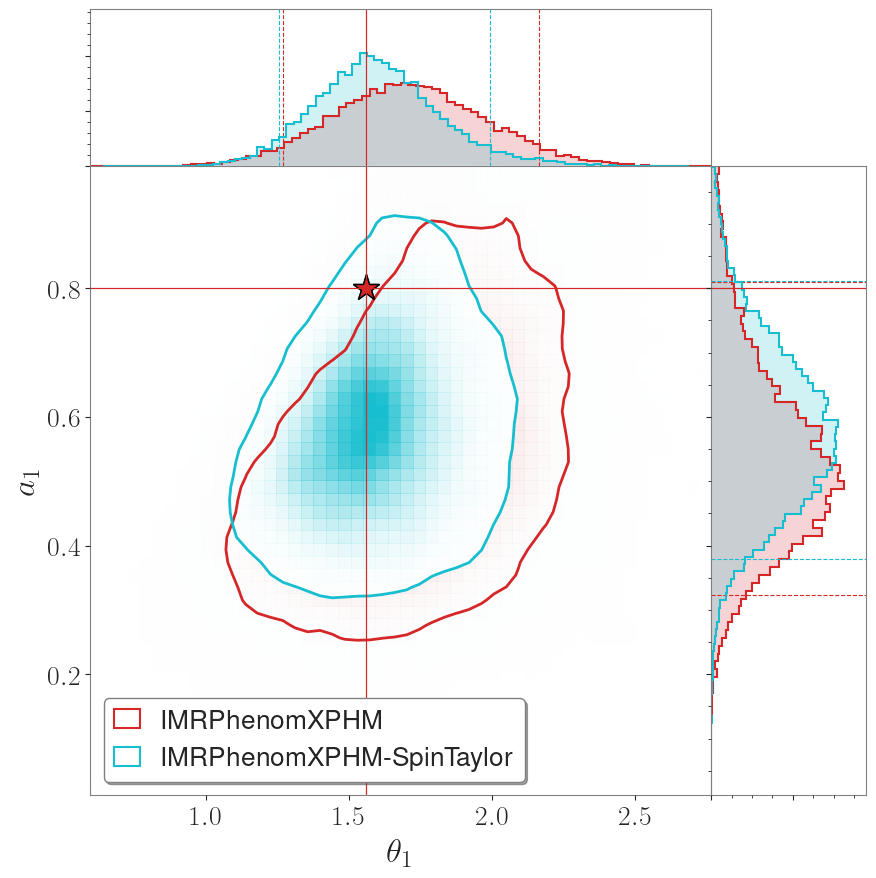}}
	\caption{Joint posterior distributions for some of the mass and spin parameters for the synthetic signal described in Subsec.~\ref{subsec:simulated_gw}. In the top row, the left panel shows posterior distributions for the chirp mass and mass ratio, while the right panel shows the effective aligned and precessing spin parameters $\chi_{\mathrm{eff}}$ and $\chi_{\mathrm{p}}$; in the bottom row, the left panel shows the luminosity distance and the inclination angle $\theta_{\mathrm{JN}}$, whereas the right panel shows the primary's spin magnitude $a_1$ and tilt $\theta_1$. The standard version of \phXPHM is shown in red, whereas the \phXPHMST results are shown in cyan, following the same color scheme previously adopted. Dashed (solid) lines indicate the 90\% credible intervals for the 1D (2D) posterior distributions.}
	\label{fig:sxs1916}
\end{figure*}

\begin{figure}
	\includegraphics[width=\columnwidth]{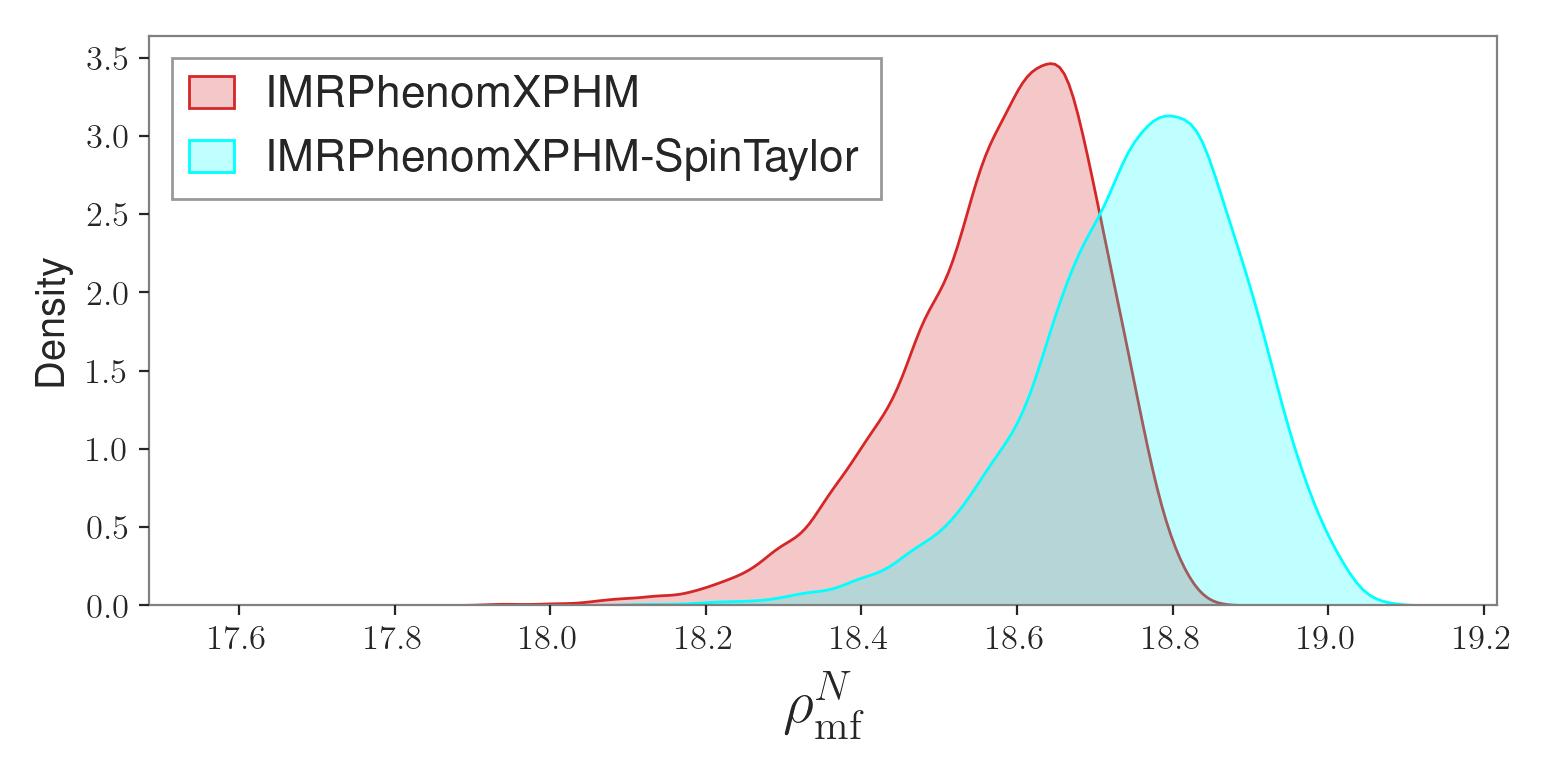}
	\caption{Matched filter SNR recovered by \phXPHM in its default version using MSA angles (red) and \phXPHMST (cyan) for the synthetic signal based on \textsc{SXS:1916}. See main text for further details.}
	\label{fig:sxs1916_snr}
\end{figure}

We present some salient results of this injection study in Fig.~\ref{fig:sxs1916}. \phXPHMST correctly recovers almost all of the injected parameters, with the exception of $\chi_\mathrm{p}$ and $q$, which are just slightly underestimated and overestimated, respectively. However, we also note that the injected mass parameters do lie inside the 90\% bounds of their joint posteriors.  We can also see that the posterior distribution for $\chi_\mathrm{p}$ is slightly shifted towards higher values (i.e., closer to the injected value) with respect to the standard \phXPHM(MSA) result. One can also appreciate that \phXPHMST always returns narrower posteriors, delivering overall better constraints on the source parameters. This holds true, in particular, for the primary's spin tilt and effective spin parameter $\chi_\mathrm{eff}$. The superior performance of \phXPHMST is also evident from the network matched filter SNR recovered by the two models, which is shown in Fig.~\ref{fig:sxs1916_snr}, as well as from the $\log_{10}$ Bayes factor for \phXPHMST (ST) over \phXPHMMSA, $\mathcal{B}_\mathrm{ST/MSA}\sim0.99$, signaling a substantial preference for new model.

\section{Conclusions}\label{sec:conclusions}
In this paper, we presented \phXPHMST, a new flavor of the waveform model \phXPHM. The model features two main improvements: the first is a recalibration of the amplitudes of the aligned-spin higher harmonics, which is at the same time simpler and more robust (Subsec.~\ref{subsec:amplitude_model}). The new recalibration drastically reduces the risk of encountering problematic behavior of the waveform due to poor extrapolation over parameter space, but has a negligible impact on parameter estimation.

The second and main improvement consists of a new model of the inspiral Euler angles based on numerical solutions to the PN spin-precession equations (Subsec.~\ref{subsec:angles_model}). In the merger-ringdown region, where the PN approximation breaks down, the angles are smoothly continued by means of simple analytical expressions,  ensuring the resulting inspiral-merger model is well behaved throughout the coalescence.

Despite the simplicity of this approach, which does not involve any explicit calibration of the precessing sector to NR waveforms, the new model can better reproduce some features of the Euler angles extracted from full NR simulations (Subsec.~\ref{subsec:angles_nr_comparison}). This is an important result, since these angles are at the core of the twisting-up approximation employed by many precessing waveform models. The visual agreement observed in the comparison of the Euler angles translates into an increased faithfulness of \phXPHMST with respect to both NR (Sec.~\ref{sec:nr_comparison}) and NRSur7dq4 waveforms (Sec.~\ref{sec:model_comparison}).

In Sec.~\ref{sec:benchmarks}, we showed that the improved faithfulness of \phXPHMST comes with a modest computational overhead for the typical total masses of binary black holes observed by current detectors.

Finally, in Sec.~\ref{sec:pe}, we have shown that the new model returns well-behaved posteriors when re-analysing both real gravitational-wave events and a synthetic signal based on a precessing NR waveform. Remarkably, \phXPHMST often delivers narrower, more constraining posterior distributions than its predecessor. This property might be very useful to achieve a better characterisation of the the population of compact binaries.

There are several possible extensions of the work presented in this paper. We are already working on an improved version of this model incorporating the features introduced in \phXOfourA, including mode asymmetries~\cite{Ghosh:2023mhc}, calibrated merger-ringdown angles~\cite{Hamilton:2021pkf} and effective ringdown frequencies accounting for precession effects~\cite{Hamilton:2023znn}. As we have previously remarked, current phenomenological models could also profit from a more thorough recalibration of the aligned-spin sector. Another possible improvement is the extension of the mode content of \modelname{IMRPhenom} models: this would be beneficial both for current and future detectors such as LISA. Finally, the twisting-up method described here could be extended to include eccentricity in templates for binary black holes~\cite{Liu:2023ldr,Gamba:2024cvy}. Continuous updates to \modelname{IMRPhenomX/T} models will be key to ensuring that they can meet the accuracy requirements posed by upcoming observations.

\section{Acknowledgements}

We thank Nathan Johnson-McDaniel for providing the initial motivation for some of this work. We thank Jonathan Thompson for providing useful comments and carefully proofreading this manuscript.

We thank Maria Haney, N.V. Krishnendu, Shubhanshu Tiwari and Nathan Johnson-McDaniel for being part of the internal LIGO review team for \phXPHMST. We thank Eleanor Hamilton and Geraint Pratten for useful discussions.

This work was supported by the Universitat de les Illes Balears (UIB); the Spanish Agencia Estatal de Investigación grants PID2022-138626NB-I00, RED2022-134204-E, RED2022-134411-T, funded by MICIU/AEI/10.13039/501100011033 and the ERDF/EU; and the Comunitat Autònoma de les Illes Balears through the Servei de Recerca i Desenvolupament and the Conselleria d'Educació i Universitats with funds from the Tourist Stay Tax Law (PDR2020/1i1 - ITS2017-006), from the European Union - NextGenerationEU/PRTR-C17.I1 (SINCO2022/6719) and from the European Union - European Regional Development Fund (ERDF) (SINCO2022/18146).
SA acknowledges the University College Dublin Ad Astra Fund. CG is supported by the Swiss National Science Foundation (SNSF) Ambizione grant PZ00P2\_223711.
The authors thank the Supercomputing and Bioinnovation Center (SCBI) of the University of Malaga and the Red Española de Supercomputación through the computing grant AECT-2024-2-0017. The authors are grateful for computational resources provided by the LIGO Laboratory and supported by National Science Foundation Grants PHY-0757058 and PHY-0823459.

This material is based upon work supported by NSF's LIGO Laboratory which is a major facility fully funded by the National Science Foundation. This research has made use of data obtained from the Gravitational Wave Open Science Center~\cite{LIGOScientific:2019lzm}, a service of LIGO Laboratory, the LIGO Scientific Collaboration, Virgo Collaboration and KAGRA. LIGO Laboratory and Advanced LIGO are funded by the United States NSF as well as the Science and Technology Facilities Council (STFC) of the United Kingdom, the Max-Planck-Society (MPS), and the State of Niedersachsen/Germany for support of the construction of Advanced LIGO and construction and operation of the GEO600 detector. Additional support for Advanced LIGO was provided by the Australian Research Council. Virgo is funded, through the European Gravitational Observatory (EGO), by the French Centre National de Recherche Scientifique (CNRS), the Italian Istituto Nazionale di Fisica Nucleare (INFN) and the Dutch Nikhef, with contributions by institutions from Belgium, Germany, Greece, Hungary, Ireland, Japan, Monaco, Poland, Portugal, Spain. KAGRA is supported by Ministry of Education, Culture, Sports, Science and Technology (MEXT), Japan Society for the Promotion of Science (JSPS) in Japan; National Research Foundation (NRF) and Ministry of Science and ICT (MSIT) in Korea; Academia Sinica (AS) and National Science and Technology Council (NSTC) in Taiwan.
This work made use of the software packages  \textmd{bilby}~\cite{Ashton:2018jfp}, \textmd{dynesty}~\cite{Speagle_2020}, \textmd{LALSuite}~\cite{lalsuite}, \textmd{matplotlib}~\cite{Hunter:2007ouj}, \textmd{numpy}~\cite{Harris:2020xlr}, \textmd{PyCBC}~\cite{PyCBC}, and \text{scipy}~\cite{Virtanen:2019joe}.


\vfill
\let\c\Originalcdefinition %
\let\d\Originalddefinition %
\let\i\Originalidefinition

\bibliography{references}

\providecommand{\noopsort}[1]{}\providecommand{\singleletter}[1]{#1}%
\begin{thebibliography}{92}%
\makeatletter
\providecommand \@ifxundefined [1]{%
 \@ifx{#1\undefined}
}%
\providecommand \@ifnum [1]{%
 \ifnum #1\expandafter \@firstoftwo
 \else \expandafter \@secondoftwo
 \fi
}%
\providecommand \@ifx [1]{%
 \ifx #1\expandafter \@firstoftwo
 \else \expandafter \@secondoftwo
 \fi
}%
\providecommand \natexlab [1]{#1}%
\providecommand \enquote  [1]{``#1''}%
\providecommand \bibnamefont  [1]{#1}%
\providecommand \bibfnamefont [1]{#1}%
\providecommand \citenamefont [1]{#1}%
\providecommand \href@noop [0]{\@secondoftwo}%
\providecommand \href [0]{\begingroup \@sanitize@url \@href}%
\providecommand \@href[1]{\@@startlink{#1}\@@href}%
\providecommand \@@href[1]{\endgroup#1\@@endlink}%
\providecommand \@sanitize@url [0]{\catcode `\\12\catcode `\$12\catcode
  `\&12\catcode `\#12\catcode `\^12\catcode `\_12\catcode `\%12\relax}%
\providecommand \@@startlink[1]{}%
\providecommand \@@endlink[0]{}%
\providecommand \url  [0]{\begingroup\@sanitize@url \@url }%
\providecommand \@url [1]{\endgroup\@href {#1}{\urlprefix }}%
\providecommand \urlprefix  [0]{URL }%
\providecommand \Eprint [0]{\href }%
\providecommand \doibase [0]{http://dx.doi.org/}%
\providecommand \selectlanguage [0]{\@gobble}%
\providecommand \bibinfo  [0]{\@secondoftwo}%
\providecommand \bibfield  [0]{\@secondoftwo}%
\providecommand \translation [1]{[#1]}%
\providecommand \BibitemOpen [0]{}%
\providecommand \bibitemStop [0]{}%
\providecommand \bibitemNoStop [0]{.\EOS\space}%
\providecommand \EOS [0]{\spacefactor3000\relax}%
\providecommand \BibitemShut  [1]{\csname bibitem#1\endcsname}%
\let\auto@bib@innerbib\@empty
\bibitem [{\citenamefont {Aasi}\ \emph {et~al.}(2015)\citenamefont {Aasi} \emph
  {et~al.}}]{LIGOScientific:2014pky}%
  \BibitemOpen
  \bibfield  {author} {\bibinfo {author} {\bibfnamefont {J.}~\bibnamefont
  {Aasi}} \emph {et~al.} (\bibinfo {collaboration} {LIGO Scientific}),\ }\href
  {\doibase 10.1088/0264-9381/32/7/074001} {\bibfield  {journal} {\bibinfo
  {journal} {Class. Quant. Grav.}\ }\textbf {\bibinfo {volume} {32}},\ \bibinfo
  {pages} {074001} (\bibinfo {year} {2015})},\ \Eprint
  {http://arxiv.org/abs/1411.4547} {arXiv:1411.4547 [gr-qc]} \BibitemShut
  {NoStop}%
\bibitem [{\citenamefont {Acernese}\ \emph {et~al.}(2015)\citenamefont
  {Acernese} \emph {et~al.}}]{VIRGO:2014yos}%
  \BibitemOpen
  \bibfield  {author} {\bibinfo {author} {\bibfnamefont {F.}~\bibnamefont
  {Acernese}} \emph {et~al.} (\bibinfo {collaboration} {VIRGO}),\ }\href
  {\doibase 10.1088/0264-9381/32/2/024001} {\bibfield  {journal} {\bibinfo
  {journal} {Class. Quant. Grav.}\ }\textbf {\bibinfo {volume} {32}},\ \bibinfo
  {pages} {024001} (\bibinfo {year} {2015})},\ \Eprint
  {http://arxiv.org/abs/1408.3978} {arXiv:1408.3978 [gr-qc]} \BibitemShut
  {NoStop}%
\bibitem [{\citenamefont {Akutsu}\ \emph {et~al.}(2021)\citenamefont {Akutsu}
  \emph {et~al.}}]{KAGRA:2020tym}%
  \BibitemOpen
  \bibfield  {author} {\bibinfo {author} {\bibfnamefont {T.}~\bibnamefont
  {Akutsu}} \emph {et~al.} (\bibinfo {collaboration} {KAGRA}),\ }\href
  {\doibase 10.1093/ptep/ptaa125} {\bibfield  {journal} {\bibinfo  {journal}
  {PTEP}\ }\textbf {\bibinfo {volume} {2021}},\ \bibinfo {pages} {05A101}
  (\bibinfo {year} {2021})},\ \Eprint {http://arxiv.org/abs/2005.05574}
  {arXiv:2005.05574 [physics.ins-det]} \BibitemShut {NoStop}%
\bibitem [{\citenamefont {Abbott}\ \emph
  {et~al.}(2021{\natexlab{a}})\citenamefont {Abbott} \emph
  {et~al.}}]{LIGOScientific:2021djp}%
  \BibitemOpen
  \bibfield  {author} {\bibinfo {author} {\bibfnamefont {R.}~\bibnamefont
  {Abbott}} \emph {et~al.} (\bibinfo {collaboration} {LIGO Scientific, VIRGO,
  KAGRA}),\ }\href@noop {} {\  (\bibinfo {year} {2021}{\natexlab{a}})},\
  \Eprint {http://arxiv.org/abs/2111.03606} {arXiv:2111.03606 [gr-qc]}
  \BibitemShut {NoStop}%
\bibitem [{gra()}]{gracedb}%
  \BibitemOpen
  \href@noop {} {\enquote {\bibinfo {title} {{GraceDB — Gravitational-Wave
  Candidate Event Database}},}\ }\bibinfo {howpublished}
  {\url{https://gracedb.ligo.org/superevents/public/O4/}}\BibitemShut {NoStop}%
\bibitem [{\citenamefont {Kidder}\ \emph {et~al.}(1993)\citenamefont {Kidder},
  \citenamefont {Will},\ and\ \citenamefont {Wiseman}}]{Kidder:1992fr}%
  \BibitemOpen
  \bibfield  {author} {\bibinfo {author} {\bibfnamefont {L.~E.}\ \bibnamefont
  {Kidder}}, \bibinfo {author} {\bibfnamefont {C.~M.}\ \bibnamefont {Will}}, \
  and\ \bibinfo {author} {\bibfnamefont {A.~G.}\ \bibnamefont {Wiseman}},\
  }\href {\doibase 10.1103/PhysRevD.47.R4183} {\bibfield  {journal} {\bibinfo
  {journal} {Phys. Rev. D}\ }\textbf {\bibinfo {volume} {47}},\ \bibinfo
  {pages} {R4183} (\bibinfo {year} {1993})},\ \Eprint
  {http://arxiv.org/abs/gr-qc/9211025} {arXiv:gr-qc/9211025} \BibitemShut
  {NoStop}%
\bibitem [{\citenamefont {Kidder}(1995)}]{Kidder:1995zr}%
  \BibitemOpen
  \bibfield  {author} {\bibinfo {author} {\bibfnamefont {L.~E.}\ \bibnamefont
  {Kidder}},\ }\href {\doibase 10.1103/PhysRevD.52.821} {\bibfield  {journal}
  {\bibinfo  {journal} {Phys. Rev. D}\ }\textbf {\bibinfo {volume} {52}},\
  \bibinfo {pages} {821} (\bibinfo {year} {1995})},\ \Eprint
  {http://arxiv.org/abs/gr-qc/9506022} {arXiv:gr-qc/9506022} \BibitemShut
  {NoStop}%
\bibitem [{\citenamefont {Cutler}\ and\ \citenamefont
  {Flanagan}(1994)}]{Cutler:1994ys}%
  \BibitemOpen
  \bibfield  {author} {\bibinfo {author} {\bibfnamefont {C.}~\bibnamefont
  {Cutler}}\ and\ \bibinfo {author} {\bibfnamefont {E.~E.}\ \bibnamefont
  {Flanagan}},\ }\href {\doibase 10.1103/PhysRevD.49.2658} {\bibfield
  {journal} {\bibinfo  {journal} {Phys. Rev. D}\ }\textbf {\bibinfo {volume}
  {49}},\ \bibinfo {pages} {2658} (\bibinfo {year} {1994})},\ \Eprint
  {http://arxiv.org/abs/gr-qc/9402014} {arXiv:gr-qc/9402014} \BibitemShut
  {NoStop}%
\bibitem [{\citenamefont {Poisson}\ and\ \citenamefont
  {Will}(1995)}]{Poisson:1995ef}%
  \BibitemOpen
  \bibfield  {author} {\bibinfo {author} {\bibfnamefont {E.}~\bibnamefont
  {Poisson}}\ and\ \bibinfo {author} {\bibfnamefont {C.~M.}\ \bibnamefont
  {Will}},\ }\href {\doibase 10.1103/PhysRevD.52.848} {\bibfield  {journal}
  {\bibinfo  {journal} {Phys. Rev. D}\ }\textbf {\bibinfo {volume} {52}},\
  \bibinfo {pages} {848} (\bibinfo {year} {1995})},\ \Eprint
  {http://arxiv.org/abs/gr-qc/9502040} {arXiv:gr-qc/9502040} \BibitemShut
  {NoStop}%
\bibitem [{\citenamefont {Damour}(2001)}]{Damour:2001tu}%
  \BibitemOpen
  \bibfield  {author} {\bibinfo {author} {\bibfnamefont {T.}~\bibnamefont
  {Damour}},\ }\href {\doibase 10.1103/PhysRevD.64.124013} {\bibfield
  {journal} {\bibinfo  {journal} {Phys. Rev. D}\ }\textbf {\bibinfo {volume}
  {64}},\ \bibinfo {pages} {124013} (\bibinfo {year} {2001})},\ \Eprint
  {http://arxiv.org/abs/gr-qc/0103018} {arXiv:gr-qc/0103018} \BibitemShut
  {NoStop}%
\bibitem [{\citenamefont {Racine}(2008)}]{Racine:2008qv}%
  \BibitemOpen
  \bibfield  {author} {\bibinfo {author} {\bibfnamefont {E.}~\bibnamefont
  {Racine}},\ }\href {\doibase 10.1103/PhysRevD.78.044021} {\bibfield
  {journal} {\bibinfo  {journal} {Phys. Rev.}\ }\textbf {\bibinfo {volume}
  {D78}},\ \bibinfo {pages} {044021} (\bibinfo {year} {2008})},\ \Eprint
  {http://arxiv.org/abs/0803.1820} {arXiv:0803.1820 [gr-qc]} \BibitemShut
  {NoStop}%
\bibitem [{\citenamefont {Ajith}(2011)}]{Ajith:2011ec}%
  \BibitemOpen
  \bibfield  {author} {\bibinfo {author} {\bibfnamefont {P.}~\bibnamefont
  {Ajith}},\ }\href {\doibase 10.1103/PhysRevD.84.084037} {\bibfield  {journal}
  {\bibinfo  {journal} {Phys. Rev. D}\ }\textbf {\bibinfo {volume} {84}},\
  \bibinfo {pages} {084037} (\bibinfo {year} {2011})},\ \Eprint
  {http://arxiv.org/abs/1107.1267} {arXiv:1107.1267 [gr-qc]} \BibitemShut
  {NoStop}%
\bibitem [{\citenamefont {Schmidt}\ \emph {et~al.}(2011)\citenamefont
  {Schmidt}, \citenamefont {Hannam}, \citenamefont {Husa},\ and\ \citenamefont
  {Ajith}}]{Schmidt:2010it}%
  \BibitemOpen
  \bibfield  {author} {\bibinfo {author} {\bibfnamefont {P.}~\bibnamefont
  {Schmidt}}, \bibinfo {author} {\bibfnamefont {M.}~\bibnamefont {Hannam}},
  \bibinfo {author} {\bibfnamefont {S.}~\bibnamefont {Husa}}, \ and\ \bibinfo
  {author} {\bibfnamefont {P.}~\bibnamefont {Ajith}},\ }\href {\doibase
  10.1103/PhysRevD.84.024046} {\bibfield  {journal} {\bibinfo  {journal} {Phys.
  Rev. D}\ }\textbf {\bibinfo {volume} {84}},\ \bibinfo {pages} {024046}
  (\bibinfo {year} {2011})},\ \Eprint {http://arxiv.org/abs/1012.2879}
  {arXiv:1012.2879 [gr-qc]} \BibitemShut {NoStop}%
\bibitem [{\citenamefont {Baird}\ \emph {et~al.}(2013)\citenamefont {Baird},
  \citenamefont {Fairhurst}, \citenamefont {Hannam},\ and\ \citenamefont
  {Murphy}}]{Baird:2012cu}%
  \BibitemOpen
  \bibfield  {author} {\bibinfo {author} {\bibfnamefont {E.}~\bibnamefont
  {Baird}}, \bibinfo {author} {\bibfnamefont {S.}~\bibnamefont {Fairhurst}},
  \bibinfo {author} {\bibfnamefont {M.}~\bibnamefont {Hannam}}, \ and\ \bibinfo
  {author} {\bibfnamefont {P.}~\bibnamefont {Murphy}},\ }\href {\doibase
  10.1103/PhysRevD.87.024035} {\bibfield  {journal} {\bibinfo  {journal} {Phys.
  Rev. D}\ }\textbf {\bibinfo {volume} {87}},\ \bibinfo {pages} {024035}
  (\bibinfo {year} {2013})},\ \Eprint {http://arxiv.org/abs/1211.0546}
  {arXiv:1211.0546 [gr-qc]} \BibitemShut {NoStop}%
\bibitem [{\citenamefont {Hannam}\ \emph {et~al.}(2014)\citenamefont {Hannam},
  \citenamefont {Schmidt}, \citenamefont {Boh\'e}, \citenamefont {Haegel},
  \citenamefont {Husa}, \citenamefont {Ohme}, \citenamefont {Pratten},\ and\
  \citenamefont {P\"urrer}}]{Hannam:2013oca}%
  \BibitemOpen
  \bibfield  {author} {\bibinfo {author} {\bibfnamefont {M.}~\bibnamefont
  {Hannam}}, \bibinfo {author} {\bibfnamefont {P.}~\bibnamefont {Schmidt}},
  \bibinfo {author} {\bibfnamefont {A.}~\bibnamefont {Boh\'e}}, \bibinfo
  {author} {\bibfnamefont {L.}~\bibnamefont {Haegel}}, \bibinfo {author}
  {\bibfnamefont {S.}~\bibnamefont {Husa}}, \bibinfo {author} {\bibfnamefont
  {F.}~\bibnamefont {Ohme}}, \bibinfo {author} {\bibfnamefont {G.}~\bibnamefont
  {Pratten}}, \ and\ \bibinfo {author} {\bibfnamefont {M.}~\bibnamefont
  {P\"urrer}},\ }\href {\doibase 10.1103/PhysRevLett.113.151101} {\bibfield
  {journal} {\bibinfo  {journal} {Phys. Rev. Lett.}\ }\textbf {\bibinfo
  {volume} {113}},\ \bibinfo {pages} {151101} (\bibinfo {year} {2014})},\
  \Eprint {http://arxiv.org/abs/1308.3271} {arXiv:1308.3271 [gr-qc]}
  \BibitemShut {NoStop}%
\bibitem [{\citenamefont {Arun}\ \emph {et~al.}(2009)\citenamefont {Arun},
  \citenamefont {Buonanno}, \citenamefont {Faye},\ and\ \citenamefont
  {Ochsner}}]{Arun:2008kb}%
  \BibitemOpen
  \bibfield  {author} {\bibinfo {author} {\bibfnamefont {K.~G.}\ \bibnamefont
  {Arun}}, \bibinfo {author} {\bibfnamefont {A.}~\bibnamefont {Buonanno}},
  \bibinfo {author} {\bibfnamefont {G.}~\bibnamefont {Faye}}, \ and\ \bibinfo
  {author} {\bibfnamefont {E.}~\bibnamefont {Ochsner}},\ }\href {\doibase
  10.1103/PhysRevD.79.104023} {\bibfield  {journal} {\bibinfo  {journal} {Phys.
  Rev. D}\ }\textbf {\bibinfo {volume} {79}},\ \bibinfo {pages} {104023}
  (\bibinfo {year} {2009})},\ \bibinfo {note} {[Erratum: Phys.Rev.D 84, 049901
  (2011)]},\ \Eprint {http://arxiv.org/abs/0810.5336} {arXiv:0810.5336 [gr-qc]}
  \BibitemShut {NoStop}%
\bibitem [{\citenamefont {Apostolatos}\ \emph {et~al.}(1994)\citenamefont
  {Apostolatos}, \citenamefont {Cutler}, \citenamefont {Sussman},\ and\
  \citenamefont {Thorne}}]{Apostolatos:1994mx}%
  \BibitemOpen
  \bibfield  {author} {\bibinfo {author} {\bibfnamefont {T.~A.}\ \bibnamefont
  {Apostolatos}}, \bibinfo {author} {\bibfnamefont {C.}~\bibnamefont {Cutler}},
  \bibinfo {author} {\bibfnamefont {G.~J.}\ \bibnamefont {Sussman}}, \ and\
  \bibinfo {author} {\bibfnamefont {K.~S.}\ \bibnamefont {Thorne}},\ }\href
  {\doibase 10.1103/PhysRevD.49.6274} {\bibfield  {journal} {\bibinfo
  {journal} {Phys. Rev. D}\ }\textbf {\bibinfo {volume} {49}},\ \bibinfo
  {pages} {6274} (\bibinfo {year} {1994})}\BibitemShut {NoStop}%
\bibitem [{\citenamefont {Abbott}\ \emph {et~al.}(2019)\citenamefont {Abbott}
  \emph {et~al.}}]{LIGOScientific:2018mvr}%
  \BibitemOpen
  \bibfield  {author} {\bibinfo {author} {\bibfnamefont {B.~P.}\ \bibnamefont
  {Abbott}} \emph {et~al.} (\bibinfo {collaboration} {LIGO Scientific,
  Virgo}),\ }\href {\doibase 10.1103/PhysRevX.9.031040} {\bibfield  {journal}
  {\bibinfo  {journal} {Phys. Rev.}\ }\textbf {\bibinfo {volume} {X9}},\
  \bibinfo {pages} {031040} (\bibinfo {year} {2019})},\ \Eprint
  {http://arxiv.org/abs/1811.12907} {arXiv:1811.12907 [astro-ph.HE]}
  \BibitemShut {NoStop}%
\bibitem [{\citenamefont {Abbott}\ \emph
  {et~al.}(2021{\natexlab{b}})\citenamefont {Abbott} \emph
  {et~al.}}]{LIGOScientific:2020kqk}%
  \BibitemOpen
  \bibfield  {author} {\bibinfo {author} {\bibfnamefont {R.}~\bibnamefont
  {Abbott}} \emph {et~al.} (\bibinfo {collaboration} {LIGO Scientific,
  Virgo}),\ }\href {\doibase 10.3847/2041-8213/abe949} {\bibfield  {journal}
  {\bibinfo  {journal} {Astrophys. J. Lett.}\ }\textbf {\bibinfo {volume}
  {913}},\ \bibinfo {pages} {L7} (\bibinfo {year} {2021}{\natexlab{b}})},\
  \Eprint {http://arxiv.org/abs/2010.14533} {arXiv:2010.14533 [astro-ph.HE]}
  \BibitemShut {NoStop}%
\bibitem [{\citenamefont {Abbott}\ \emph {et~al.}(2024)\citenamefont {Abbott}
  \emph {et~al.}}]{LIGOScientific:2021usb}%
  \BibitemOpen
  \bibfield  {author} {\bibinfo {author} {\bibfnamefont {R.}~\bibnamefont
  {Abbott}} \emph {et~al.} (\bibinfo {collaboration} {LIGO Scientific,
  VIRGO}),\ }\href {\doibase 10.1103/PhysRevD.109.022001} {\bibfield  {journal}
  {\bibinfo  {journal} {Phys. Rev. D}\ }\textbf {\bibinfo {volume} {109}},\
  \bibinfo {pages} {022001} (\bibinfo {year} {2024})},\ \Eprint
  {http://arxiv.org/abs/2108.01045} {arXiv:2108.01045 [gr-qc]} \BibitemShut
  {NoStop}%
\bibitem [{\citenamefont {Abbott}\ \emph
  {et~al.}(2023{\natexlab{a}})\citenamefont {Abbott} \emph
  {et~al.}}]{KAGRA:2021vkt}%
  \BibitemOpen
  \bibfield  {author} {\bibinfo {author} {\bibfnamefont {R.}~\bibnamefont
  {Abbott}} \emph {et~al.} (\bibinfo {collaboration} {KAGRA, VIRGO, LIGO
  Scientific}),\ }\href {\doibase 10.1103/PhysRevX.13.041039} {\bibfield
  {journal} {\bibinfo  {journal} {Phys. Rev. X}\ }\textbf {\bibinfo {volume}
  {13}},\ \bibinfo {pages} {041039} (\bibinfo {year} {2023}{\natexlab{a}})},\
  \Eprint {http://arxiv.org/abs/2111.03606} {arXiv:2111.03606 [gr-qc]}
  \BibitemShut {NoStop}%
\bibitem [{\citenamefont {Abbott}\ \emph
  {et~al.}(2023{\natexlab{b}})\citenamefont {Abbott} \emph
  {et~al.}}]{KAGRA:2021duu}%
  \BibitemOpen
  \bibfield  {author} {\bibinfo {author} {\bibfnamefont {R.}~\bibnamefont
  {Abbott}} \emph {et~al.} (\bibinfo {collaboration} {KAGRA, VIRGO, LIGO
  Scientific}),\ }\href {\doibase 10.1103/PhysRevX.13.011048} {\bibfield
  {journal} {\bibinfo  {journal} {Phys. Rev. X}\ }\textbf {\bibinfo {volume}
  {13}},\ \bibinfo {pages} {011048} (\bibinfo {year} {2023}{\natexlab{b}})},\
  \Eprint {http://arxiv.org/abs/2111.03634} {arXiv:2111.03634 [astro-ph.HE]}
  \BibitemShut {NoStop}%
\bibitem [{\citenamefont {Hannam}\ \emph
  {et~al.}(2022{\natexlab{a}})\citenamefont {Hannam} \emph
  {et~al.}}]{Hannam:2021pit}%
  \BibitemOpen
  \bibfield  {author} {\bibinfo {author} {\bibfnamefont {M.}~\bibnamefont
  {Hannam}} \emph {et~al.},\ }\href {\doibase 10.1038/s41586-022-05212-z}
  {\bibfield  {journal} {\bibinfo  {journal} {Nature}\ }\textbf {\bibinfo
  {volume} {610}},\ \bibinfo {pages} {652} (\bibinfo {year}
  {2022}{\natexlab{a}})},\ \Eprint {http://arxiv.org/abs/2112.11300}
  {arXiv:2112.11300 [gr-qc]} \BibitemShut {NoStop}%
\bibitem [{\citenamefont {Payne}\ \emph {et~al.}(2022)\citenamefont {Payne},
  \citenamefont {Hourihane}, \citenamefont {Golomb}, \citenamefont {Udall},
  \citenamefont {Udall}, \citenamefont {Davis},\ and\ \citenamefont
  {Chatziioannou}}]{Payne:2022spz}%
  \BibitemOpen
  \bibfield  {author} {\bibinfo {author} {\bibfnamefont {E.}~\bibnamefont
  {Payne}}, \bibinfo {author} {\bibfnamefont {S.}~\bibnamefont {Hourihane}},
  \bibinfo {author} {\bibfnamefont {J.}~\bibnamefont {Golomb}}, \bibinfo
  {author} {\bibfnamefont {R.}~\bibnamefont {Udall}}, \bibinfo {author}
  {\bibfnamefont {R.}~\bibnamefont {Udall}}, \bibinfo {author} {\bibfnamefont
  {D.}~\bibnamefont {Davis}}, \ and\ \bibinfo {author} {\bibfnamefont
  {K.}~\bibnamefont {Chatziioannou}},\ }\href {\doibase
  10.1103/PhysRevD.106.104017} {\bibfield  {journal} {\bibinfo  {journal}
  {Phys. Rev. D}\ }\textbf {\bibinfo {volume} {106}},\ \bibinfo {pages}
  {104017} (\bibinfo {year} {2022})},\ \Eprint
  {http://arxiv.org/abs/2206.11932} {arXiv:2206.11932 [gr-qc]} \BibitemShut
  {NoStop}%
\bibitem [{\citenamefont {Macas}\ \emph {et~al.}(2024)\citenamefont {Macas},
  \citenamefont {Lundgren},\ and\ \citenamefont {Ashton}}]{Macas:2023wiw}%
  \BibitemOpen
  \bibfield  {author} {\bibinfo {author} {\bibfnamefont {R.}~\bibnamefont
  {Macas}}, \bibinfo {author} {\bibfnamefont {A.}~\bibnamefont {Lundgren}}, \
  and\ \bibinfo {author} {\bibfnamefont {G.}~\bibnamefont {Ashton}},\ }\href
  {\doibase 10.1103/PhysRevD.109.062006} {\bibfield  {journal} {\bibinfo
  {journal} {Phys. Rev. D}\ }\textbf {\bibinfo {volume} {109}},\ \bibinfo
  {pages} {062006} (\bibinfo {year} {2024})},\ \Eprint
  {http://arxiv.org/abs/2311.09921} {arXiv:2311.09921 [gr-qc]} \BibitemShut
  {NoStop}%
\bibitem [{\citenamefont {Mandel}\ and\ \citenamefont
  {Farmer}(2022)}]{Mandel:2018hfr}%
  \BibitemOpen
  \bibfield  {author} {\bibinfo {author} {\bibfnamefont {I.}~\bibnamefont
  {Mandel}}\ and\ \bibinfo {author} {\bibfnamefont {A.}~\bibnamefont
  {Farmer}},\ }\href {\doibase 10.1016/j.physrep.2022.01.003} {\bibfield
  {journal} {\bibinfo  {journal} {Phys. Rept.}\ }\textbf {\bibinfo {volume}
  {955}},\ \bibinfo {pages} {1} (\bibinfo {year} {2022})},\ \Eprint
  {http://arxiv.org/abs/1806.05820} {arXiv:1806.05820 [astro-ph.HE]}
  \BibitemShut {NoStop}%
\bibitem [{\citenamefont {Mapelli}(2021)}]{Mapelli:2021taw}%
  \BibitemOpen
  \bibfield  {author} {\bibinfo {author} {\bibfnamefont {M.}~\bibnamefont
  {Mapelli}},\ }\enquote {\bibinfo {title} {{Formation Channels of Single and
  Binary Stellar-Mass Black Holes}},}\ \ (\bibinfo {year} {2021})\ \Eprint
  {http://arxiv.org/abs/2106.00699} {arXiv:2106.00699 [astro-ph.HE]}
  \BibitemShut {NoStop}%
\bibitem [{\citenamefont {Gerosa}\ \emph {et~al.}(2013)\citenamefont {Gerosa},
  \citenamefont {Kesden}, \citenamefont {Berti}, \citenamefont
  {O'Shaughnessy},\ and\ \citenamefont {Sperhake}}]{Gerosa:2013laa}%
  \BibitemOpen
  \bibfield  {author} {\bibinfo {author} {\bibfnamefont {D.}~\bibnamefont
  {Gerosa}}, \bibinfo {author} {\bibfnamefont {M.}~\bibnamefont {Kesden}},
  \bibinfo {author} {\bibfnamefont {E.}~\bibnamefont {Berti}}, \bibinfo
  {author} {\bibfnamefont {R.}~\bibnamefont {O'Shaughnessy}}, \ and\ \bibinfo
  {author} {\bibfnamefont {U.}~\bibnamefont {Sperhake}},\ }\href {\doibase
  10.1103/PhysRevD.87.104028} {\bibfield  {journal} {\bibinfo  {journal} {Phys.
  Rev. D}\ }\textbf {\bibinfo {volume} {87}},\ \bibinfo {pages} {104028}
  (\bibinfo {year} {2013})},\ \Eprint {http://arxiv.org/abs/1302.4442}
  {arXiv:1302.4442 [gr-qc]} \BibitemShut {NoStop}%
\bibitem [{\citenamefont {Vitale}\ \emph {et~al.}(2017)\citenamefont {Vitale},
  \citenamefont {Lynch}, \citenamefont {Sturani},\ and\ \citenamefont
  {Graff}}]{Vitale:2015tea}%
  \BibitemOpen
  \bibfield  {author} {\bibinfo {author} {\bibfnamefont {S.}~\bibnamefont
  {Vitale}}, \bibinfo {author} {\bibfnamefont {R.}~\bibnamefont {Lynch}},
  \bibinfo {author} {\bibfnamefont {R.}~\bibnamefont {Sturani}}, \ and\
  \bibinfo {author} {\bibfnamefont {P.}~\bibnamefont {Graff}},\ }\href
  {\doibase 10.1088/1361-6382/aa552e} {\bibfield  {journal} {\bibinfo
  {journal} {Class. Quant. Grav.}\ }\textbf {\bibinfo {volume} {34}},\ \bibinfo
  {pages} {03LT01} (\bibinfo {year} {2017})},\ \Eprint
  {http://arxiv.org/abs/1503.04307} {arXiv:1503.04307 [gr-qc]} \BibitemShut
  {NoStop}%
\bibitem [{\citenamefont {Rodriguez}\ \emph {et~al.}(2016)\citenamefont
  {Rodriguez}, \citenamefont {Zevin}, \citenamefont {Pankow}, \citenamefont
  {Kalogera},\ and\ \citenamefont {Rasio}}]{Rodriguez:2016vmx}%
  \BibitemOpen
  \bibfield  {author} {\bibinfo {author} {\bibfnamefont {C.~L.}\ \bibnamefont
  {Rodriguez}}, \bibinfo {author} {\bibfnamefont {M.}~\bibnamefont {Zevin}},
  \bibinfo {author} {\bibfnamefont {C.}~\bibnamefont {Pankow}}, \bibinfo
  {author} {\bibfnamefont {V.}~\bibnamefont {Kalogera}}, \ and\ \bibinfo
  {author} {\bibfnamefont {F.~A.}\ \bibnamefont {Rasio}},\ }\href {\doibase
  10.3847/2041-8205/832/1/L2} {\bibfield  {journal} {\bibinfo  {journal}
  {Astrophys. J. Lett.}\ }\textbf {\bibinfo {volume} {832}},\ \bibinfo {pages}
  {L2} (\bibinfo {year} {2016})},\ \Eprint {http://arxiv.org/abs/1609.05916}
  {arXiv:1609.05916 [astro-ph.HE]} \BibitemShut {NoStop}%
\bibitem [{\citenamefont {Stevenson}\ \emph {et~al.}(2017)\citenamefont
  {Stevenson}, \citenamefont {Berry},\ and\ \citenamefont
  {Mandel}}]{Stevenson:2017dlk}%
  \BibitemOpen
  \bibfield  {author} {\bibinfo {author} {\bibfnamefont {S.}~\bibnamefont
  {Stevenson}}, \bibinfo {author} {\bibfnamefont {C.~P.~L.}\ \bibnamefont
  {Berry}}, \ and\ \bibinfo {author} {\bibfnamefont {I.}~\bibnamefont
  {Mandel}},\ }\href {\doibase 10.1093/mnras/stx1764} {\bibfield  {journal}
  {\bibinfo  {journal} {Mon. Not. Roy. Astron. Soc.}\ }\textbf {\bibinfo
  {volume} {471}},\ \bibinfo {pages} {2801} (\bibinfo {year} {2017})},\ \Eprint
  {http://arxiv.org/abs/1703.06873} {arXiv:1703.06873 [astro-ph.HE]}
  \BibitemShut {NoStop}%
\bibitem [{\citenamefont {Hoy}\ \emph {et~al.}(2024)\citenamefont {Hoy},
  \citenamefont {Fairhurst},\ and\ \citenamefont {Mandel}}]{Hoy:2024qpy}%
  \BibitemOpen
  \bibfield  {author} {\bibinfo {author} {\bibfnamefont {C.}~\bibnamefont
  {Hoy}}, \bibinfo {author} {\bibfnamefont {S.}~\bibnamefont {Fairhurst}}, \
  and\ \bibinfo {author} {\bibfnamefont {I.}~\bibnamefont {Mandel}},\
  }\href@noop {} {\  (\bibinfo {year} {2024})},\ \Eprint
  {http://arxiv.org/abs/2408.03410} {arXiv:2408.03410 [gr-qc]} \BibitemShut
  {NoStop}%
\bibitem [{\citenamefont {Varma}\ \emph {et~al.}(2019)\citenamefont {Varma},
  \citenamefont {Field}, \citenamefont {Scheel}, \citenamefont {Blackman},
  \citenamefont {Gerosa}, \citenamefont {Stein}, \citenamefont {Kidder},\ and\
  \citenamefont {Pfeiffer}}]{Varma:2019csw}%
  \BibitemOpen
  \bibfield  {author} {\bibinfo {author} {\bibfnamefont {V.}~\bibnamefont
  {Varma}}, \bibinfo {author} {\bibfnamefont {S.~E.}\ \bibnamefont {Field}},
  \bibinfo {author} {\bibfnamefont {M.~A.}\ \bibnamefont {Scheel}}, \bibinfo
  {author} {\bibfnamefont {J.}~\bibnamefont {Blackman}}, \bibinfo {author}
  {\bibfnamefont {D.}~\bibnamefont {Gerosa}}, \bibinfo {author} {\bibfnamefont
  {L.~C.}\ \bibnamefont {Stein}}, \bibinfo {author} {\bibfnamefont {L.~E.}\
  \bibnamefont {Kidder}}, \ and\ \bibinfo {author} {\bibfnamefont {H.~P.}\
  \bibnamefont {Pfeiffer}},\ }\href {\doibase 10.1103/PhysRevResearch.1.033015}
  {\bibfield  {journal} {\bibinfo  {journal} {Phys. Rev. Research.}\ }\textbf
  {\bibinfo {volume} {1}},\ \bibinfo {pages} {033015} (\bibinfo {year}
  {2019})},\ \Eprint {http://arxiv.org/abs/1905.09300} {arXiv:1905.09300
  [gr-qc]} \BibitemShut {NoStop}%
\bibitem [{\citenamefont {Islam}\ \emph {et~al.}(2023)\citenamefont {Islam},
  \citenamefont {Vajpeyi}, \citenamefont {Shaik}, \citenamefont {Haster},
  \citenamefont {Varma}, \citenamefont {Field}, \citenamefont {Lange},
  \citenamefont {O'Shaughnessy},\ and\ \citenamefont {Smith}}]{Islam:2023zzj}%
  \BibitemOpen
  \bibfield  {author} {\bibinfo {author} {\bibfnamefont {T.}~\bibnamefont
  {Islam}}, \bibinfo {author} {\bibfnamefont {A.}~\bibnamefont {Vajpeyi}},
  \bibinfo {author} {\bibfnamefont {F.~H.}\ \bibnamefont {Shaik}}, \bibinfo
  {author} {\bibfnamefont {C.-J.}\ \bibnamefont {Haster}}, \bibinfo {author}
  {\bibfnamefont {V.}~\bibnamefont {Varma}}, \bibinfo {author} {\bibfnamefont
  {S.~E.}\ \bibnamefont {Field}}, \bibinfo {author} {\bibfnamefont
  {J.}~\bibnamefont {Lange}}, \bibinfo {author} {\bibfnamefont
  {R.}~\bibnamefont {O'Shaughnessy}}, \ and\ \bibinfo {author} {\bibfnamefont
  {R.}~\bibnamefont {Smith}},\ }\href@noop {} {\  (\bibinfo {year} {2023})},\
  \Eprint {http://arxiv.org/abs/2309.14473} {arXiv:2309.14473 [gr-qc]}
  \BibitemShut {NoStop}%
\bibitem [{\citenamefont {Buonanno}\ and\ \citenamefont
  {Damour}(1999)}]{Buonanno:1998gg}%
  \BibitemOpen
  \bibfield  {author} {\bibinfo {author} {\bibfnamefont {A.}~\bibnamefont
  {Buonanno}}\ and\ \bibinfo {author} {\bibfnamefont {T.}~\bibnamefont
  {Damour}},\ }\href {\doibase 10.1103/PhysRevD.59.084006} {\bibfield
  {journal} {\bibinfo  {journal} {Phys. Rev. D}\ }\textbf {\bibinfo {volume}
  {59}},\ \bibinfo {pages} {084006} (\bibinfo {year} {1999})},\ \Eprint
  {http://arxiv.org/abs/gr-qc/9811091} {arXiv:gr-qc/9811091} \BibitemShut
  {NoStop}%
\bibitem [{\citenamefont {Buonanno}\ and\ \citenamefont
  {Damour}(2000)}]{Buonanno:2000ef}%
  \BibitemOpen
  \bibfield  {author} {\bibinfo {author} {\bibfnamefont {A.}~\bibnamefont
  {Buonanno}}\ and\ \bibinfo {author} {\bibfnamefont {T.}~\bibnamefont
  {Damour}},\ }\href {\doibase 10.1103/PhysRevD.62.064015} {\bibfield
  {journal} {\bibinfo  {journal} {Phys. Rev. D}\ }\textbf {\bibinfo {volume}
  {62}},\ \bibinfo {pages} {064015} (\bibinfo {year} {2000})},\ \Eprint
  {http://arxiv.org/abs/gr-qc/0001013} {arXiv:gr-qc/0001013} \BibitemShut
  {NoStop}%
\bibitem [{\citenamefont {Damour}\ \emph {et~al.}(2000)\citenamefont {Damour},
  \citenamefont {Jaranowski},\ and\ \citenamefont {Schaefer}}]{Damour:2000we}%
  \BibitemOpen
  \bibfield  {author} {\bibinfo {author} {\bibfnamefont {T.}~\bibnamefont
  {Damour}}, \bibinfo {author} {\bibfnamefont {P.}~\bibnamefont {Jaranowski}},
  \ and\ \bibinfo {author} {\bibfnamefont {G.}~\bibnamefont {Schaefer}},\
  }\href {\doibase 10.1103/PhysRevD.62.084011} {\bibfield  {journal} {\bibinfo
  {journal} {Phys. Rev. D}\ }\textbf {\bibinfo {volume} {62}},\ \bibinfo
  {pages} {084011} (\bibinfo {year} {2000})},\ \Eprint
  {http://arxiv.org/abs/gr-qc/0005034} {arXiv:gr-qc/0005034} \BibitemShut
  {NoStop}%
\bibitem [{\citenamefont {Ramos-Buades}\ \emph {et~al.}(2023)\citenamefont
  {Ramos-Buades}, \citenamefont {Buonanno}, \citenamefont {Estell\'es},
  \citenamefont {Khalil}, \citenamefont {Mihaylov}, \citenamefont {Ossokine},
  \citenamefont {Pompili},\ and\ \citenamefont
  {Shiferaw}}]{Ramos-Buades:2023ehm}%
  \BibitemOpen
  \bibfield  {author} {\bibinfo {author} {\bibfnamefont {A.}~\bibnamefont
  {Ramos-Buades}}, \bibinfo {author} {\bibfnamefont {A.}~\bibnamefont
  {Buonanno}}, \bibinfo {author} {\bibfnamefont {H.}~\bibnamefont
  {Estell\'es}}, \bibinfo {author} {\bibfnamefont {M.}~\bibnamefont {Khalil}},
  \bibinfo {author} {\bibfnamefont {D.~P.}\ \bibnamefont {Mihaylov}}, \bibinfo
  {author} {\bibfnamefont {S.}~\bibnamefont {Ossokine}}, \bibinfo {author}
  {\bibfnamefont {L.}~\bibnamefont {Pompili}}, \ and\ \bibinfo {author}
  {\bibfnamefont {M.}~\bibnamefont {Shiferaw}},\ }\href {\doibase
  10.1103/PhysRevD.108.124037} {\bibfield  {journal} {\bibinfo  {journal}
  {Phys. Rev. D}\ }\textbf {\bibinfo {volume} {108}},\ \bibinfo {pages}
  {124037} (\bibinfo {year} {2023})},\ \Eprint
  {http://arxiv.org/abs/2303.18046} {arXiv:2303.18046 [gr-qc]} \BibitemShut
  {NoStop}%
\bibitem [{\citenamefont {Khalil}\ \emph {et~al.}(2023)\citenamefont {Khalil},
  \citenamefont {Buonanno}, \citenamefont {Estelles}, \citenamefont {Mihaylov},
  \citenamefont {Ossokine}, \citenamefont {Pompili},\ and\ \citenamefont
  {Ramos-Buades}}]{Khalil:2023kep}%
  \BibitemOpen
  \bibfield  {author} {\bibinfo {author} {\bibfnamefont {M.}~\bibnamefont
  {Khalil}}, \bibinfo {author} {\bibfnamefont {A.}~\bibnamefont {Buonanno}},
  \bibinfo {author} {\bibfnamefont {H.}~\bibnamefont {Estelles}}, \bibinfo
  {author} {\bibfnamefont {D.~P.}\ \bibnamefont {Mihaylov}}, \bibinfo {author}
  {\bibfnamefont {S.}~\bibnamefont {Ossokine}}, \bibinfo {author}
  {\bibfnamefont {L.}~\bibnamefont {Pompili}}, \ and\ \bibinfo {author}
  {\bibfnamefont {A.}~\bibnamefont {Ramos-Buades}},\ }\href {\doibase
  10.1103/PhysRevD.108.124036} {\bibfield  {journal} {\bibinfo  {journal}
  {Phys. Rev. D}\ }\textbf {\bibinfo {volume} {108}},\ \bibinfo {pages}
  {124036} (\bibinfo {year} {2023})},\ \Eprint
  {http://arxiv.org/abs/2303.18143} {arXiv:2303.18143 [gr-qc]} \BibitemShut
  {NoStop}%
\bibitem [{\citenamefont {Mihaylov}\ \emph {et~al.}(2023)\citenamefont
  {Mihaylov}, \citenamefont {Ossokine}, \citenamefont {Buonanno}, \citenamefont
  {Estelles}, \citenamefont {Pompili}, \citenamefont {P\"urrer},\ and\
  \citenamefont {Ramos-Buades}}]{Mihaylov:2023bkc}%
  \BibitemOpen
  \bibfield  {author} {\bibinfo {author} {\bibfnamefont {D.~P.}\ \bibnamefont
  {Mihaylov}}, \bibinfo {author} {\bibfnamefont {S.}~\bibnamefont {Ossokine}},
  \bibinfo {author} {\bibfnamefont {A.}~\bibnamefont {Buonanno}}, \bibinfo
  {author} {\bibfnamefont {H.}~\bibnamefont {Estelles}}, \bibinfo {author}
  {\bibfnamefont {L.}~\bibnamefont {Pompili}}, \bibinfo {author} {\bibfnamefont
  {M.}~\bibnamefont {P\"urrer}}, \ and\ \bibinfo {author} {\bibfnamefont
  {A.}~\bibnamefont {Ramos-Buades}},\ }\href@noop {} {\  (\bibinfo {year}
  {2023})},\ \Eprint {http://arxiv.org/abs/2303.18203} {arXiv:2303.18203
  [gr-qc]} \BibitemShut {NoStop}%
\bibitem [{\citenamefont {Pompili}\ \emph {et~al.}(2023)\citenamefont {Pompili}
  \emph {et~al.}}]{Pompili:2023tna}%
  \BibitemOpen
  \bibfield  {author} {\bibinfo {author} {\bibfnamefont {L.}~\bibnamefont
  {Pompili}} \emph {et~al.},\ }\href {\doibase 10.1103/PhysRevD.108.124035}
  {\bibfield  {journal} {\bibinfo  {journal} {Phys. Rev. D}\ }\textbf {\bibinfo
  {volume} {108}},\ \bibinfo {pages} {124035} (\bibinfo {year} {2023})},\
  \Eprint {http://arxiv.org/abs/2303.18039} {arXiv:2303.18039 [gr-qc]}
  \BibitemShut {NoStop}%
\bibitem [{\citenamefont {Nagar}\ \emph {et~al.}(2018)\citenamefont {Nagar}
  \emph {et~al.}}]{Nagar:2018zoe}%
  \BibitemOpen
  \bibfield  {author} {\bibinfo {author} {\bibfnamefont {A.}~\bibnamefont
  {Nagar}} \emph {et~al.},\ }\href {\doibase 10.1103/PhysRevD.98.104052}
  {\bibfield  {journal} {\bibinfo  {journal} {Phys. Rev. D}\ }\textbf {\bibinfo
  {volume} {98}},\ \bibinfo {pages} {104052} (\bibinfo {year} {2018})},\
  \Eprint {http://arxiv.org/abs/1806.01772} {arXiv:1806.01772 [gr-qc]}
  \BibitemShut {NoStop}%
\bibitem [{\citenamefont {Akcay}\ \emph {et~al.}(2021)\citenamefont {Akcay},
  \citenamefont {Gamba},\ and\ \citenamefont {Bernuzzi}}]{Akcay:2020qrj}%
  \BibitemOpen
  \bibfield  {author} {\bibinfo {author} {\bibfnamefont {S.}~\bibnamefont
  {Akcay}}, \bibinfo {author} {\bibfnamefont {R.}~\bibnamefont {Gamba}}, \ and\
  \bibinfo {author} {\bibfnamefont {S.}~\bibnamefont {Bernuzzi}},\ }\href
  {\doibase 10.1103/PhysRevD.103.024014} {\bibfield  {journal} {\bibinfo
  {journal} {Phys. Rev. D}\ }\textbf {\bibinfo {volume} {103}},\ \bibinfo
  {pages} {024014} (\bibinfo {year} {2021})},\ \Eprint
  {http://arxiv.org/abs/2005.05338} {arXiv:2005.05338 [gr-qc]} \BibitemShut
  {NoStop}%
\bibitem [{\citenamefont {Gamba}\ \emph {et~al.}(2022)\citenamefont {Gamba},
  \citenamefont {Akcay}, \citenamefont {Bernuzzi},\ and\ \citenamefont
  {Williams}}]{Gamba:2021ydi}%
  \BibitemOpen
  \bibfield  {author} {\bibinfo {author} {\bibfnamefont {R.}~\bibnamefont
  {Gamba}}, \bibinfo {author} {\bibfnamefont {S.}~\bibnamefont {Akcay}},
  \bibinfo {author} {\bibfnamefont {S.}~\bibnamefont {Bernuzzi}}, \ and\
  \bibinfo {author} {\bibfnamefont {J.}~\bibnamefont {Williams}},\ }\href
  {\doibase 10.1103/PhysRevD.106.024020} {\bibfield  {journal} {\bibinfo
  {journal} {Phys. Rev. D}\ }\textbf {\bibinfo {volume} {106}},\ \bibinfo
  {pages} {024020} (\bibinfo {year} {2022})},\ \Eprint
  {http://arxiv.org/abs/2111.03675} {arXiv:2111.03675 [gr-qc]} \BibitemShut
  {NoStop}%
\bibitem [{\citenamefont {Ajith}\ \emph {et~al.}(2007)\citenamefont {Ajith}
  \emph {et~al.}}]{Ajith:2007qp}%
  \BibitemOpen
  \bibfield  {author} {\bibinfo {author} {\bibfnamefont {P.}~\bibnamefont
  {Ajith}} \emph {et~al.},\ }\href {\doibase 10.1088/0264-9381/24/19/S31}
  {\bibfield  {journal} {\bibinfo  {journal} {Class. Quant. Grav.}\ }\textbf
  {\bibinfo {volume} {24}},\ \bibinfo {pages} {S689} (\bibinfo {year}
  {2007})},\ \Eprint {http://arxiv.org/abs/0704.3764} {arXiv:0704.3764 [gr-qc]}
  \BibitemShut {NoStop}%
\bibitem [{\citenamefont {Ajith}\ \emph {et~al.}(2008)\citenamefont {Ajith}
  \emph {et~al.}}]{Ajith:2007kx}%
  \BibitemOpen
  \bibfield  {author} {\bibinfo {author} {\bibfnamefont {P.}~\bibnamefont
  {Ajith}} \emph {et~al.},\ }\href {\doibase 10.1103/PhysRevD.77.104017}
  {\bibfield  {journal} {\bibinfo  {journal} {Phys. Rev. D}\ }\textbf {\bibinfo
  {volume} {77}},\ \bibinfo {pages} {104017} (\bibinfo {year} {2008})},\
  \bibinfo {note} {[Erratum: Phys.Rev.D 79, 129901 (2009)]},\ \Eprint
  {http://arxiv.org/abs/0710.2335} {arXiv:0710.2335 [gr-qc]} \BibitemShut
  {NoStop}%
\bibitem [{\citenamefont {Pratten}\ \emph {et~al.}(2021)\citenamefont {Pratten}
  \emph {et~al.}}]{Pratten:2020ceb}%
  \BibitemOpen
  \bibfield  {author} {\bibinfo {author} {\bibfnamefont {G.}~\bibnamefont
  {Pratten}} \emph {et~al.},\ }\href {\doibase 10.1103/PhysRevD.103.104056}
  {\bibfield  {journal} {\bibinfo  {journal} {Phys. Rev. D}\ }\textbf {\bibinfo
  {volume} {103}},\ \bibinfo {pages} {104056} (\bibinfo {year} {2021})},\
  \Eprint {http://arxiv.org/abs/2004.06503} {arXiv:2004.06503 [gr-qc]}
  \BibitemShut {NoStop}%
\bibitem [{\citenamefont {Hamilton}\ \emph {et~al.}(2021)\citenamefont
  {Hamilton}, \citenamefont {London}, \citenamefont {Thompson}, \citenamefont
  {Fauchon-Jones}, \citenamefont {Hannam}, \citenamefont {Kalaghatgi},
  \citenamefont {Khan}, \citenamefont {Pannarale},\ and\ \citenamefont
  {Vano-Vinuales}}]{Hamilton:2021pkf}%
  \BibitemOpen
  \bibfield  {author} {\bibinfo {author} {\bibfnamefont {E.}~\bibnamefont
  {Hamilton}}, \bibinfo {author} {\bibfnamefont {L.}~\bibnamefont {London}},
  \bibinfo {author} {\bibfnamefont {J.~E.}\ \bibnamefont {Thompson}}, \bibinfo
  {author} {\bibfnamefont {E.}~\bibnamefont {Fauchon-Jones}}, \bibinfo {author}
  {\bibfnamefont {M.}~\bibnamefont {Hannam}}, \bibinfo {author} {\bibfnamefont
  {C.}~\bibnamefont {Kalaghatgi}}, \bibinfo {author} {\bibfnamefont
  {S.}~\bibnamefont {Khan}}, \bibinfo {author} {\bibfnamefont {F.}~\bibnamefont
  {Pannarale}}, \ and\ \bibinfo {author} {\bibfnamefont {A.}~\bibnamefont
  {Vano-Vinuales}},\ }\href {\doibase 10.1103/PhysRevD.104.124027} {\bibfield
  {journal} {\bibinfo  {journal} {Phys. Rev. D}\ }\textbf {\bibinfo {volume}
  {104}},\ \bibinfo {pages} {124027} (\bibinfo {year} {2021})},\ \Eprint
  {http://arxiv.org/abs/2107.08876} {arXiv:2107.08876 [gr-qc]} \BibitemShut
  {NoStop}%
\bibitem [{\citenamefont {Hamilton}\ \emph {et~al.}(2023)\citenamefont
  {Hamilton}, \citenamefont {London},\ and\ \citenamefont
  {Hannam}}]{Hamilton:2023znn}%
  \BibitemOpen
  \bibfield  {author} {\bibinfo {author} {\bibfnamefont {E.}~\bibnamefont
  {Hamilton}}, \bibinfo {author} {\bibfnamefont {L.}~\bibnamefont {London}}, \
  and\ \bibinfo {author} {\bibfnamefont {M.}~\bibnamefont {Hannam}},\ }\href
  {\doibase 10.1103/PhysRevD.107.104035} {\bibfield  {journal} {\bibinfo
  {journal} {Phys. Rev. D}\ }\textbf {\bibinfo {volume} {107}},\ \bibinfo
  {pages} {104035} (\bibinfo {year} {2023})},\ \Eprint
  {http://arxiv.org/abs/2301.06558} {arXiv:2301.06558 [gr-qc]} \BibitemShut
  {NoStop}%
\bibitem [{\citenamefont {Ghosh}\ \emph {et~al.}(2024)\citenamefont {Ghosh},
  \citenamefont {Kolitsidou},\ and\ \citenamefont {Hannam}}]{Ghosh:2023mhc}%
  \BibitemOpen
  \bibfield  {author} {\bibinfo {author} {\bibfnamefont {S.}~\bibnamefont
  {Ghosh}}, \bibinfo {author} {\bibfnamefont {P.}~\bibnamefont {Kolitsidou}}, \
  and\ \bibinfo {author} {\bibfnamefont {M.}~\bibnamefont {Hannam}},\ }\href
  {\doibase 10.1103/PhysRevD.109.024061} {\bibfield  {journal} {\bibinfo
  {journal} {Phys. Rev. D}\ }\textbf {\bibinfo {volume} {109}},\ \bibinfo
  {pages} {024061} (\bibinfo {year} {2024})},\ \Eprint
  {http://arxiv.org/abs/2310.16980} {arXiv:2310.16980 [gr-qc]} \BibitemShut
  {NoStop}%
\bibitem [{\citenamefont {Thompson}\ \emph {et~al.}(2024)\citenamefont
  {Thompson}, \citenamefont {Hamilton}, \citenamefont {London}, \citenamefont
  {Ghosh}, \citenamefont {Kolitsidou}, \citenamefont {Hoy},\ and\ \citenamefont
  {Hannam}}]{Thompson:2023ase}%
  \BibitemOpen
  \bibfield  {author} {\bibinfo {author} {\bibfnamefont {J.~E.}\ \bibnamefont
  {Thompson}}, \bibinfo {author} {\bibfnamefont {E.}~\bibnamefont {Hamilton}},
  \bibinfo {author} {\bibfnamefont {L.}~\bibnamefont {London}}, \bibinfo
  {author} {\bibfnamefont {S.}~\bibnamefont {Ghosh}}, \bibinfo {author}
  {\bibfnamefont {P.}~\bibnamefont {Kolitsidou}}, \bibinfo {author}
  {\bibfnamefont {C.}~\bibnamefont {Hoy}}, \ and\ \bibinfo {author}
  {\bibfnamefont {M.}~\bibnamefont {Hannam}},\ }\href {\doibase
  10.1103/PhysRevD.109.063012} {\bibfield  {journal} {\bibinfo  {journal}
  {Phys. Rev. D}\ }\textbf {\bibinfo {volume} {109}},\ \bibinfo {pages}
  {063012} (\bibinfo {year} {2024})},\ \Eprint
  {http://arxiv.org/abs/2312.10025} {arXiv:2312.10025 [gr-qc]} \BibitemShut
  {NoStop}%
\bibitem [{\citenamefont {Schmidt}\ \emph {et~al.}(2015)\citenamefont
  {Schmidt}, \citenamefont {Ohme},\ and\ \citenamefont
  {Hannam}}]{Schmidt:2014iyl}%
  \BibitemOpen
  \bibfield  {author} {\bibinfo {author} {\bibfnamefont {P.}~\bibnamefont
  {Schmidt}}, \bibinfo {author} {\bibfnamefont {F.}~\bibnamefont {Ohme}}, \
  and\ \bibinfo {author} {\bibfnamefont {M.}~\bibnamefont {Hannam}},\ }\href
  {\doibase 10.1103/PhysRevD.91.024043} {\bibfield  {journal} {\bibinfo
  {journal} {Phys. Rev. D}\ }\textbf {\bibinfo {volume} {91}},\ \bibinfo
  {pages} {024043} (\bibinfo {year} {2015})},\ \Eprint
  {http://arxiv.org/abs/1408.1810} {arXiv:1408.1810 [gr-qc]} \BibitemShut
  {NoStop}%
\bibitem [{\citenamefont {Khan}\ \emph {et~al.}(2019)\citenamefont {Khan},
  \citenamefont {Chatziioannou}, \citenamefont {Hannam},\ and\ \citenamefont
  {Ohme}}]{Khan:2018fmp}%
  \BibitemOpen
  \bibfield  {author} {\bibinfo {author} {\bibfnamefont {S.}~\bibnamefont
  {Khan}}, \bibinfo {author} {\bibfnamefont {K.}~\bibnamefont {Chatziioannou}},
  \bibinfo {author} {\bibfnamefont {M.}~\bibnamefont {Hannam}}, \ and\ \bibinfo
  {author} {\bibfnamefont {F.}~\bibnamefont {Ohme}},\ }\href {\doibase
  10.1103/PhysRevD.100.024059} {\bibfield  {journal} {\bibinfo  {journal}
  {Phys. Rev. D}\ }\textbf {\bibinfo {volume} {100}},\ \bibinfo {pages}
  {024059} (\bibinfo {year} {2019})},\ \Eprint
  {http://arxiv.org/abs/1809.10113} {arXiv:1809.10113 [gr-qc]} \BibitemShut
  {NoStop}%
\bibitem [{\citenamefont {Khan}\ \emph {et~al.}(2020)\citenamefont {Khan},
  \citenamefont {Ohme}, \citenamefont {Chatziioannou},\ and\ \citenamefont
  {Hannam}}]{Khan:2019kot}%
  \BibitemOpen
  \bibfield  {author} {\bibinfo {author} {\bibfnamefont {S.}~\bibnamefont
  {Khan}}, \bibinfo {author} {\bibfnamefont {F.}~\bibnamefont {Ohme}}, \bibinfo
  {author} {\bibfnamefont {K.}~\bibnamefont {Chatziioannou}}, \ and\ \bibinfo
  {author} {\bibfnamefont {M.}~\bibnamefont {Hannam}},\ }\href {\doibase
  10.1103/PhysRevD.101.024056} {\bibfield  {journal} {\bibinfo  {journal}
  {Phys. Rev. D}\ }\textbf {\bibinfo {volume} {101}},\ \bibinfo {pages}
  {024056} (\bibinfo {year} {2020})},\ \Eprint
  {http://arxiv.org/abs/1911.06050} {arXiv:1911.06050 [gr-qc]} \BibitemShut
  {NoStop}%
\bibitem [{\citenamefont {Estell\'es}\ \emph {et~al.}(2021)\citenamefont
  {Estell\'es}, \citenamefont {Ramos-Buades}, \citenamefont {Husa},
  \citenamefont {Garc\'{i}a-Quir\'os}, \citenamefont {Colleoni}, \citenamefont
  {Haegel},\ and\ \citenamefont {Jaume}}]{Estelles:2020osj}%
  \BibitemOpen
  \bibfield  {author} {\bibinfo {author} {\bibfnamefont {H.}~\bibnamefont
  {Estell\'es}}, \bibinfo {author} {\bibfnamefont {A.}~\bibnamefont
  {Ramos-Buades}}, \bibinfo {author} {\bibfnamefont {S.}~\bibnamefont {Husa}},
  \bibinfo {author} {\bibfnamefont {C.}~\bibnamefont {Garc\'{i}a-Quir\'os}},
  \bibinfo {author} {\bibfnamefont {M.}~\bibnamefont {Colleoni}}, \bibinfo
  {author} {\bibfnamefont {L.}~\bibnamefont {Haegel}}, \ and\ \bibinfo {author}
  {\bibfnamefont {R.}~\bibnamefont {Jaume}},\ }\href {\doibase
  10.1103/PhysRevD.103.124060} {\bibfield  {journal} {\bibinfo  {journal}
  {Phys. Rev. D}\ }\textbf {\bibinfo {volume} {103}},\ \bibinfo {pages}
  {124060} (\bibinfo {year} {2021})},\ \Eprint
  {http://arxiv.org/abs/2004.08302} {arXiv:2004.08302 [gr-qc]} \BibitemShut
  {NoStop}%
\bibitem [{\citenamefont {Estell\'es}\ \emph
  {et~al.}(2022{\natexlab{a}})\citenamefont {Estell\'es}, \citenamefont {Husa},
  \citenamefont {Colleoni}, \citenamefont {Keitel}, \citenamefont
  {Mateu-Lucena}, \citenamefont {Garc\'{i}a-Quir\'os}, \citenamefont
  {Ramos-Buades},\ and\ \citenamefont {Borchers}}]{Estelles:2020twz}%
  \BibitemOpen
  \bibfield  {author} {\bibinfo {author} {\bibfnamefont {H.}~\bibnamefont
  {Estell\'es}}, \bibinfo {author} {\bibfnamefont {S.}~\bibnamefont {Husa}},
  \bibinfo {author} {\bibfnamefont {M.}~\bibnamefont {Colleoni}}, \bibinfo
  {author} {\bibfnamefont {D.}~\bibnamefont {Keitel}}, \bibinfo {author}
  {\bibfnamefont {M.}~\bibnamefont {Mateu-Lucena}}, \bibinfo {author}
  {\bibfnamefont {C.}~\bibnamefont {Garc\'{i}a-Quir\'os}}, \bibinfo {author}
  {\bibfnamefont {A.}~\bibnamefont {Ramos-Buades}}, \ and\ \bibinfo {author}
  {\bibfnamefont {A.}~\bibnamefont {Borchers}},\ }\href {\doibase
  10.1103/PhysRevD.105.084039} {\bibfield  {journal} {\bibinfo  {journal}
  {Phys. Rev. D}\ }\textbf {\bibinfo {volume} {105}},\ \bibinfo {pages}
  {084039} (\bibinfo {year} {2022}{\natexlab{a}})},\ \Eprint
  {http://arxiv.org/abs/2012.11923} {arXiv:2012.11923 [gr-qc]} \BibitemShut
  {NoStop}%
\bibitem [{\citenamefont {Estell\'es}\ \emph
  {et~al.}(2022{\natexlab{b}})\citenamefont {Estell\'es}, \citenamefont
  {Colleoni}, \citenamefont {Garc\'{i}a-Quir\'os}, \citenamefont {Husa},
  \citenamefont {Keitel}, \citenamefont {Mateu-Lucena}, \citenamefont
  {Planas},\ and\ \citenamefont {Ramos-Buades}}]{Estelles:2021gvs}%
  \BibitemOpen
  \bibfield  {author} {\bibinfo {author} {\bibfnamefont {H.}~\bibnamefont
  {Estell\'es}}, \bibinfo {author} {\bibfnamefont {M.}~\bibnamefont
  {Colleoni}}, \bibinfo {author} {\bibfnamefont {C.}~\bibnamefont
  {Garc\'{i}a-Quir\'os}}, \bibinfo {author} {\bibfnamefont {S.}~\bibnamefont
  {Husa}}, \bibinfo {author} {\bibfnamefont {D.}~\bibnamefont {Keitel}},
  \bibinfo {author} {\bibfnamefont {M.}~\bibnamefont {Mateu-Lucena}}, \bibinfo
  {author} {\bibfnamefont {M.~d.~L.}\ \bibnamefont {Planas}}, \ and\ \bibinfo
  {author} {\bibfnamefont {A.}~\bibnamefont {Ramos-Buades}},\ }\href {\doibase
  10.1103/PhysRevD.105.084040} {\bibfield  {journal} {\bibinfo  {journal}
  {Phys. Rev. D}\ }\textbf {\bibinfo {volume} {105}},\ \bibinfo {pages}
  {084040} (\bibinfo {year} {2022}{\natexlab{b}})},\ \Eprint
  {http://arxiv.org/abs/2105.05872} {arXiv:2105.05872 [gr-qc]} \BibitemShut
  {NoStop}%
\bibitem [{\citenamefont {Mac~Uilliam}\ \emph {et~al.}(2024)\citenamefont
  {Mac~Uilliam}, \citenamefont {Akcay},\ and\ \citenamefont
  {Thompson}}]{MacUilliam:2024oif}%
  \BibitemOpen
  \bibfield  {author} {\bibinfo {author} {\bibfnamefont {J.}~\bibnamefont
  {Mac~Uilliam}}, \bibinfo {author} {\bibfnamefont {S.}~\bibnamefont {Akcay}},
  \ and\ \bibinfo {author} {\bibfnamefont {J.~E.}\ \bibnamefont {Thompson}},\
  }\href {\doibase 10.1103/PhysRevD.109.084077} {\bibfield  {journal} {\bibinfo
   {journal} {Phys. Rev. D}\ }\textbf {\bibinfo {volume} {109}},\ \bibinfo
  {pages} {084077} (\bibinfo {year} {2024})},\ \Eprint
  {http://arxiv.org/abs/2402.06781} {arXiv:2402.06781 [gr-qc]} \BibitemShut
  {NoStop}%
\bibitem [{\citenamefont {Colleoni}\ \emph {et~al.}(2023)\citenamefont
  {Colleoni}, \citenamefont {Vidal}, \citenamefont {Johnson-McDaniel},
  \citenamefont {Dietrich}, \citenamefont {Haney},\ and\ \citenamefont
  {Pratten}}]{Colleoni:2023czp}%
  \BibitemOpen
  \bibfield  {author} {\bibinfo {author} {\bibfnamefont {M.}~\bibnamefont
  {Colleoni}}, \bibinfo {author} {\bibfnamefont {F.~A.~R.}\ \bibnamefont
  {Vidal}}, \bibinfo {author} {\bibfnamefont {N.~K.}\ \bibnamefont
  {Johnson-McDaniel}}, \bibinfo {author} {\bibfnamefont {T.}~\bibnamefont
  {Dietrich}}, \bibinfo {author} {\bibfnamefont {M.}~\bibnamefont {Haney}}, \
  and\ \bibinfo {author} {\bibfnamefont {G.}~\bibnamefont {Pratten}},\
  }\href@noop {} {\  (\bibinfo {year} {2023})},\ \Eprint
  {http://arxiv.org/abs/2311.15978} {arXiv:2311.15978 [gr-qc]} \BibitemShut
  {NoStop}%
\bibitem [{\citenamefont {Abac}\ \emph {et~al.}(2024)\citenamefont {Abac},
  \citenamefont {Dietrich}, \citenamefont {Buonanno}, \citenamefont
  {Steinhoff},\ and\ \citenamefont {Ujevic}}]{Abac:2023ujg}%
  \BibitemOpen
  \bibfield  {author} {\bibinfo {author} {\bibfnamefont {A.}~\bibnamefont
  {Abac}}, \bibinfo {author} {\bibfnamefont {T.}~\bibnamefont {Dietrich}},
  \bibinfo {author} {\bibfnamefont {A.}~\bibnamefont {Buonanno}}, \bibinfo
  {author} {\bibfnamefont {J.}~\bibnamefont {Steinhoff}}, \ and\ \bibinfo
  {author} {\bibfnamefont {M.}~\bibnamefont {Ujevic}},\ }\href {\doibase
  10.1103/PhysRevD.109.024062} {\bibfield  {journal} {\bibinfo  {journal}
  {Phys. Rev. D}\ }\textbf {\bibinfo {volume} {109}},\ \bibinfo {pages}
  {024062} (\bibinfo {year} {2024})},\ \Eprint
  {http://arxiv.org/abs/2311.07456} {arXiv:2311.07456 [gr-qc]} \BibitemShut
  {NoStop}%
\bibitem [{\citenamefont {Roy}\ and\ \citenamefont
  {Vicente}(2024)}]{Roy:2024rhe}%
  \BibitemOpen
  \bibfield  {author} {\bibinfo {author} {\bibfnamefont {S.}~\bibnamefont
  {Roy}}\ and\ \bibinfo {author} {\bibfnamefont {R.}~\bibnamefont {Vicente}},\
  }\href@noop {} {\  (\bibinfo {year} {2024})},\ \Eprint
  {http://arxiv.org/abs/2410.16388} {arXiv:2410.16388 [gr-qc]} \BibitemShut
  {NoStop}%
\bibitem [{\citenamefont {{LIGO Scientific Collaboration, Virgo Collaboration,
  and KAGRA Collaboration}}(2024)}]{lalsuite}%
  \BibitemOpen
  \bibfield  {author} {\bibinfo {author} {\bibnamefont {{LIGO Scientific
  Collaboration, Virgo Collaboration, and KAGRA Collaboration}}},\ }\href
  {\doibase 10.7935/GT1W-FZ16} {\enquote {\bibinfo {title} {{LVK} {A}lgorithm
  {L}ibrary - {LALS}uite},}\ }\bibinfo {howpublished} {free software (GPL)}
  (\bibinfo {year} {2024})\BibitemShut {NoStop}%
\bibitem [{\citenamefont {Kesden}\ \emph {et~al.}(2015)\citenamefont {Kesden},
  \citenamefont {Gerosa}, \citenamefont {O'Shaughnessy}, \citenamefont
  {Berti},\ and\ \citenamefont {Sperhake}}]{Kesden:2014sla}%
  \BibitemOpen
  \bibfield  {author} {\bibinfo {author} {\bibfnamefont {M.}~\bibnamefont
  {Kesden}}, \bibinfo {author} {\bibfnamefont {D.}~\bibnamefont {Gerosa}},
  \bibinfo {author} {\bibfnamefont {R.}~\bibnamefont {O'Shaughnessy}}, \bibinfo
  {author} {\bibfnamefont {E.}~\bibnamefont {Berti}}, \ and\ \bibinfo {author}
  {\bibfnamefont {U.}~\bibnamefont {Sperhake}},\ }\href {\doibase
  10.1103/PhysRevLett.114.081103} {\bibfield  {journal} {\bibinfo  {journal}
  {Phys. Rev. Lett.}\ }\textbf {\bibinfo {volume} {114}},\ \bibinfo {pages}
  {081103} (\bibinfo {year} {2015})},\ \Eprint {http://arxiv.org/abs/1411.0674}
  {arXiv:1411.0674 [gr-qc]} \BibitemShut {NoStop}%
\bibitem [{\citenamefont {Gerosa}\ \emph {et~al.}(2015)\citenamefont {Gerosa},
  \citenamefont {Kesden}, \citenamefont {Sperhake}, \citenamefont {Berti},\
  and\ \citenamefont {O'Shaughnessy}}]{Gerosa:2015tea}%
  \BibitemOpen
  \bibfield  {author} {\bibinfo {author} {\bibfnamefont {D.}~\bibnamefont
  {Gerosa}}, \bibinfo {author} {\bibfnamefont {M.}~\bibnamefont {Kesden}},
  \bibinfo {author} {\bibfnamefont {U.}~\bibnamefont {Sperhake}}, \bibinfo
  {author} {\bibfnamefont {E.}~\bibnamefont {Berti}}, \ and\ \bibinfo {author}
  {\bibfnamefont {R.}~\bibnamefont {O'Shaughnessy}},\ }\href {\doibase
  10.1103/PhysRevD.92.064016} {\bibfield  {journal} {\bibinfo  {journal} {Phys.
  Rev. D}\ }\textbf {\bibinfo {volume} {92}},\ \bibinfo {pages} {064016}
  (\bibinfo {year} {2015})},\ \Eprint {http://arxiv.org/abs/1506.03492}
  {arXiv:1506.03492 [gr-qc]} \BibitemShut {NoStop}%
\bibitem [{\citenamefont {Chatziioannou}\ \emph {et~al.}(2017)\citenamefont
  {Chatziioannou}, \citenamefont {Klein}, \citenamefont {Yunes},\ and\
  \citenamefont {Cornish}}]{Chatziioannou:2017tdw}%
  \BibitemOpen
  \bibfield  {author} {\bibinfo {author} {\bibfnamefont {K.}~\bibnamefont
  {Chatziioannou}}, \bibinfo {author} {\bibfnamefont {A.}~\bibnamefont
  {Klein}}, \bibinfo {author} {\bibfnamefont {N.}~\bibnamefont {Yunes}}, \ and\
  \bibinfo {author} {\bibfnamefont {N.}~\bibnamefont {Cornish}},\ }\href
  {\doibase 10.1103/PhysRevD.95.104004} {\bibfield  {journal} {\bibinfo
  {journal} {Phys. Rev. D}\ }\textbf {\bibinfo {volume} {95}},\ \bibinfo
  {pages} {104004} (\bibinfo {year} {2017})},\ \Eprint
  {http://arxiv.org/abs/1703.03967} {arXiv:1703.03967 [gr-qc]} \BibitemShut
  {NoStop}%
\bibitem [{\citenamefont {Garc\'{i}a-Quir\'os}\ \emph
  {et~al.}(2020)\citenamefont {Garc\'{i}a-Quir\'os}, \citenamefont {Colleoni},
  \citenamefont {Husa}, \citenamefont {Estell\'es}, \citenamefont {Pratten},
  \citenamefont {Ramos-Buades}, \citenamefont {Mateu-Lucena},\ and\
  \citenamefont {Jaume}}]{Garcia-Quiros:2020qpx}%
  \BibitemOpen
  \bibfield  {author} {\bibinfo {author} {\bibfnamefont {C.}~\bibnamefont
  {Garc\'{i}a-Quir\'os}}, \bibinfo {author} {\bibfnamefont {M.}~\bibnamefont
  {Colleoni}}, \bibinfo {author} {\bibfnamefont {S.}~\bibnamefont {Husa}},
  \bibinfo {author} {\bibfnamefont {H.}~\bibnamefont {Estell\'es}}, \bibinfo
  {author} {\bibfnamefont {G.}~\bibnamefont {Pratten}}, \bibinfo {author}
  {\bibfnamefont {A.}~\bibnamefont {Ramos-Buades}}, \bibinfo {author}
  {\bibfnamefont {M.}~\bibnamefont {Mateu-Lucena}}, \ and\ \bibinfo {author}
  {\bibfnamefont {R.}~\bibnamefont {Jaume}},\ }\href {\doibase
  10.1103/PhysRevD.102.064002} {\bibfield  {journal} {\bibinfo  {journal}
  {Phys. Rev. D}\ }\textbf {\bibinfo {volume} {102}},\ \bibinfo {pages}
  {064002} (\bibinfo {year} {2020})},\ \Eprint
  {http://arxiv.org/abs/2001.10914} {arXiv:2001.10914 [gr-qc]} \BibitemShut
  {NoStop}%
\bibitem [{\citenamefont {Sturani}()}]{Sturani_note}%
  \BibitemOpen
  \bibfield  {author} {\bibinfo {author} {\bibfnamefont {R.}~\bibnamefont
  {Sturani}},\ }\href@noop {} {}\bibinfo {howpublished}
  {\url{https://dcc.ligo.org/DocDB/0122/T1500554/023/dLdS.pdf}}\BibitemShut
  {NoStop}%
\bibitem [{\citenamefont {Buonanno}\ \emph {et~al.}(2009)\citenamefont
  {Buonanno}, \citenamefont {Iyer}, \citenamefont {Ochsner}, \citenamefont
  {Pan},\ and\ \citenamefont {Sathyaprakash}}]{Buonanno:2009zt}%
  \BibitemOpen
  \bibfield  {author} {\bibinfo {author} {\bibfnamefont {A.}~\bibnamefont
  {Buonanno}}, \bibinfo {author} {\bibfnamefont {B.}~\bibnamefont {Iyer}},
  \bibinfo {author} {\bibfnamefont {E.}~\bibnamefont {Ochsner}}, \bibinfo
  {author} {\bibfnamefont {Y.}~\bibnamefont {Pan}}, \ and\ \bibinfo {author}
  {\bibfnamefont {B.~S.}\ \bibnamefont {Sathyaprakash}},\ }\href {\doibase
  10.1103/PhysRevD.80.084043} {\bibfield  {journal} {\bibinfo  {journal} {Phys.
  Rev. D}\ }\textbf {\bibinfo {volume} {80}},\ \bibinfo {pages} {084043}
  (\bibinfo {year} {2009})},\ \Eprint {http://arxiv.org/abs/0907.0700}
  {arXiv:0907.0700 [gr-qc]} \BibitemShut {NoStop}%
\bibitem [{\citenamefont {Yu}\ \emph {et~al.}(2023)\citenamefont {Yu},
  \citenamefont {Roulet}, \citenamefont {Venumadhav}, \citenamefont {Zackay},\
  and\ \citenamefont {Zaldarriaga}}]{Yu:2023lml}%
  \BibitemOpen
  \bibfield  {author} {\bibinfo {author} {\bibfnamefont {H.}~\bibnamefont
  {Yu}}, \bibinfo {author} {\bibfnamefont {J.}~\bibnamefont {Roulet}}, \bibinfo
  {author} {\bibfnamefont {T.}~\bibnamefont {Venumadhav}}, \bibinfo {author}
  {\bibfnamefont {B.}~\bibnamefont {Zackay}}, \ and\ \bibinfo {author}
  {\bibfnamefont {M.}~\bibnamefont {Zaldarriaga}},\ }\href {\doibase
  10.1103/PhysRevD.108.064059} {\bibfield  {journal} {\bibinfo  {journal}
  {Phys. Rev. D}\ }\textbf {\bibinfo {volume} {108}},\ \bibinfo {pages}
  {064059} (\bibinfo {year} {2023})},\ \Eprint
  {http://arxiv.org/abs/2306.08774} {arXiv:2306.08774 [gr-qc]} \BibitemShut
  {NoStop}%
\bibitem [{\citenamefont {Bohe}\ \emph {et~al.}(2013)\citenamefont {Bohe},
  \citenamefont {Marsat}, \citenamefont {Faye},\ and\ \citenamefont
  {Blanchet}}]{Bohe:2012mr}%
  \BibitemOpen
  \bibfield  {author} {\bibinfo {author} {\bibfnamefont {A.}~\bibnamefont
  {Bohe}}, \bibinfo {author} {\bibfnamefont {S.}~\bibnamefont {Marsat}},
  \bibinfo {author} {\bibfnamefont {G.}~\bibnamefont {Faye}}, \ and\ \bibinfo
  {author} {\bibfnamefont {L.}~\bibnamefont {Blanchet}},\ }\href {\doibase
  10.1088/0264-9381/30/7/075017} {\bibfield  {journal} {\bibinfo  {journal}
  {Class. Quant. Grav.}\ }\textbf {\bibinfo {volume} {30}},\ \bibinfo {pages}
  {075017} (\bibinfo {year} {2013})},\ \Eprint {http://arxiv.org/abs/1212.5520}
  {arXiv:1212.5520 [gr-qc]} \BibitemShut {NoStop}%
\bibitem [{\citenamefont {Boyle}\ \emph {et~al.}(2011)\citenamefont {Boyle},
  \citenamefont {Owen},\ and\ \citenamefont {Pfeiffer}}]{Boyle:2011gg}%
  \BibitemOpen
  \bibfield  {author} {\bibinfo {author} {\bibfnamefont {M.}~\bibnamefont
  {Boyle}}, \bibinfo {author} {\bibfnamefont {R.}~\bibnamefont {Owen}}, \ and\
  \bibinfo {author} {\bibfnamefont {H.~P.}\ \bibnamefont {Pfeiffer}},\ }\href
  {\doibase 10.1103/PhysRevD.84.124011} {\bibfield  {journal} {\bibinfo
  {journal} {Phys. Rev. D}\ }\textbf {\bibinfo {volume} {84}},\ \bibinfo
  {pages} {124011} (\bibinfo {year} {2011})},\ \Eprint
  {http://arxiv.org/abs/1110.2965} {arXiv:1110.2965 [gr-qc]} \BibitemShut
  {NoStop}%
\bibitem [{\citenamefont {O'Shaughnessy}\ \emph {et~al.}(2013)\citenamefont
  {O'Shaughnessy}, \citenamefont {London}, \citenamefont {Healy},\ and\
  \citenamefont {Shoemaker}}]{OShaughnessy:2012iol}%
  \BibitemOpen
  \bibfield  {author} {\bibinfo {author} {\bibfnamefont {R.}~\bibnamefont
  {O'Shaughnessy}}, \bibinfo {author} {\bibfnamefont {L.}~\bibnamefont
  {London}}, \bibinfo {author} {\bibfnamefont {J.}~\bibnamefont {Healy}}, \
  and\ \bibinfo {author} {\bibfnamefont {D.}~\bibnamefont {Shoemaker}},\ }\href
  {\doibase 10.1103/PhysRevD.87.044038} {\bibfield  {journal} {\bibinfo
  {journal} {Phys. Rev. D}\ }\textbf {\bibinfo {volume} {87}},\ \bibinfo
  {pages} {044038} (\bibinfo {year} {2013})},\ \Eprint
  {http://arxiv.org/abs/1209.3712} {arXiv:1209.3712 [gr-qc]} \BibitemShut
  {NoStop}%
\bibitem [{\citenamefont {O'Shaughnessy}\ \emph {et~al.}(2011)\citenamefont
  {O'Shaughnessy}, \citenamefont {Vaishnav}, \citenamefont {Healy},
  \citenamefont {Meeks},\ and\ \citenamefont
  {Shoemaker}}]{OShaughnessy:2011pmr}%
  \BibitemOpen
  \bibfield  {author} {\bibinfo {author} {\bibfnamefont {R.}~\bibnamefont
  {O'Shaughnessy}}, \bibinfo {author} {\bibfnamefont {B.}~\bibnamefont
  {Vaishnav}}, \bibinfo {author} {\bibfnamefont {J.}~\bibnamefont {Healy}},
  \bibinfo {author} {\bibfnamefont {Z.}~\bibnamefont {Meeks}}, \ and\ \bibinfo
  {author} {\bibfnamefont {D.}~\bibnamefont {Shoemaker}},\ }\href {\doibase
  10.1103/PhysRevD.84.124002} {\bibfield  {journal} {\bibinfo  {journal} {Phys.
  Rev. D}\ }\textbf {\bibinfo {volume} {84}},\ \bibinfo {pages} {124002}
  (\bibinfo {year} {2011})},\ \Eprint {http://arxiv.org/abs/1109.5224}
  {arXiv:1109.5224 [gr-qc]} \BibitemShut {NoStop}%
\bibitem [{\citenamefont {Boyle}\ \emph {et~al.}(2019)\citenamefont {Boyle}
  \emph {et~al.}}]{Boyle:2019kee}%
  \BibitemOpen
  \bibfield  {author} {\bibinfo {author} {\bibfnamefont {M.}~\bibnamefont
  {Boyle}} \emph {et~al.},\ }\href {\doibase 10.1088/1361-6382/ab34e2}
  {\bibfield  {journal} {\bibinfo  {journal} {Class. Quant. Grav.}\ }\textbf
  {\bibinfo {volume} {36}},\ \bibinfo {pages} {195006} (\bibinfo {year}
  {2019})},\ \Eprint {http://arxiv.org/abs/1904.04831} {arXiv:1904.04831
  [gr-qc]} \BibitemShut {NoStop}%
\bibitem [{sxs()}]{sxsdb}%
  \BibitemOpen
  \href@noop {} {\enquote {\bibinfo {title} {{SXS Gravitational Waveform
  Database}},}\ }\bibinfo {howpublished}
  {\url{https://data.black-holes.org/waveforms/index.html}}\BibitemShut
  {NoStop}%
\bibitem [{\citenamefont {Harry}\ \emph {et~al.}(2016)\citenamefont {Harry},
  \citenamefont {Privitera}, \citenamefont {Boh\'e},\ and\ \citenamefont
  {Buonanno}}]{Harry:2016ijz}%
  \BibitemOpen
  \bibfield  {author} {\bibinfo {author} {\bibfnamefont {I.}~\bibnamefont
  {Harry}}, \bibinfo {author} {\bibfnamefont {S.}~\bibnamefont {Privitera}},
  \bibinfo {author} {\bibfnamefont {A.}~\bibnamefont {Boh\'e}}, \ and\ \bibinfo
  {author} {\bibfnamefont {A.}~\bibnamefont {Buonanno}},\ }\href {\doibase
  10.1103/PhysRevD.94.024012} {\bibfield  {journal} {\bibinfo  {journal} {Phys.
  Rev. D}\ }\textbf {\bibinfo {volume} {94}},\ \bibinfo {pages} {024012}
  (\bibinfo {year} {2016})},\ \Eprint {http://arxiv.org/abs/1603.02444}
  {arXiv:1603.02444 [gr-qc]} \BibitemShut {NoStop}%
\bibitem [{\citenamefont {Barsotti}\ \emph {et~al.}(2018)\citenamefont
  {Barsotti}, \citenamefont {Fritschel}, \citenamefont {Evans},\ and\
  \citenamefont {Gras}}]{adligopsd}%
  \BibitemOpen
  \bibfield  {author} {\bibinfo {author} {\bibfnamefont {L.}~\bibnamefont
  {Barsotti}}, \bibinfo {author} {\bibfnamefont {P.}~\bibnamefont {Fritschel}},
  \bibinfo {author} {\bibfnamefont {M.}~\bibnamefont {Evans}}, \ and\ \bibinfo
  {author} {\bibfnamefont {S.}~\bibnamefont {Gras}},\ }\href
  {https://dcc.ligo.org/T1800044/public} {\emph {\bibinfo {title} {{The updated
  Advanced LIGO design curve}}}},\ \bibinfo {type} {Tech. Rep.}\ \bibinfo
  {number} {LIGO-T1800044}\ (\bibinfo  {institution} {{LIGO Project}},\
  \bibinfo {year} {2018})\BibitemShut {NoStop}%
\bibitem [{\citenamefont {Abbott}\ \emph
  {et~al.}(2021{\natexlab{c}})\citenamefont {Abbott} \emph
  {et~al.}}]{LIGOScientific:2019lzm}%
  \BibitemOpen
  \bibfield  {author} {\bibinfo {author} {\bibfnamefont {R.}~\bibnamefont
  {Abbott}} \emph {et~al.} (\bibinfo {collaboration} {LIGO Scientific,
  Virgo}),\ }\href {\doibase 10.1016/j.softx.2021.100658} {\bibfield  {journal}
  {\bibinfo  {journal} {SoftwareX}\ }\textbf {\bibinfo {volume} {13}},\
  \bibinfo {pages} {100658} (\bibinfo {year} {2021}{\natexlab{c}})},\ \Eprint
  {http://arxiv.org/abs/1912.11716} {arXiv:1912.11716 [gr-qc]} \BibitemShut
  {NoStop}%
\bibitem [{\citenamefont {Ashton}\ \emph {et~al.}(2019)\citenamefont {Ashton}
  \emph {et~al.}}]{Ashton:2018jfp}%
  \BibitemOpen
  \bibfield  {author} {\bibinfo {author} {\bibfnamefont {G.}~\bibnamefont
  {Ashton}} \emph {et~al.},\ }\href {\doibase 10.3847/1538-4365/ab06fc}
  {\bibfield  {journal} {\bibinfo  {journal} {Astrophys. J. Suppl.}\ }\textbf
  {\bibinfo {volume} {241}},\ \bibinfo {pages} {27} (\bibinfo {year} {2019})},\
  \Eprint {http://arxiv.org/abs/1811.02042} {arXiv:1811.02042 [astro-ph.IM]}
  \BibitemShut {NoStop}%
\bibitem [{\citenamefont {Collaboration}\ \emph {et~al.}(2021)\citenamefont
  {Collaboration}, \citenamefont {Collaboration},\ and\ \citenamefont
  {Collaboration}}]{gwtc3-zenodo}%
  \BibitemOpen
  \bibfield  {author} {\bibinfo {author} {\bibfnamefont {L.~S.}\ \bibnamefont
  {Collaboration}}, \bibinfo {author} {\bibfnamefont {V.}~\bibnamefont
  {Collaboration}}, \ and\ \bibinfo {author} {\bibfnamefont {K.}~\bibnamefont
  {Collaboration}},\ }\href {\doibase 10.5281/zenodo.5546663} {\enquote
  {\bibinfo {title} {{GWTC-3: Compact Binary Coalescences Observed by LIGO and
  Virgo During the Second Part of the Third Observing Run — Parameter
  estimation data release}},}\ } (\bibinfo {year} {2021})\BibitemShut {NoStop}%
\bibitem [{\citenamefont {Zhang}\ \emph {et~al.}(2023)\citenamefont {Zhang},
  \citenamefont {Fragione}, \citenamefont {Kimball},\ and\ \citenamefont
  {Kalogera}}]{Zhang:2023fpp}%
  \BibitemOpen
  \bibfield  {author} {\bibinfo {author} {\bibfnamefont {R.~C.}\ \bibnamefont
  {Zhang}}, \bibinfo {author} {\bibfnamefont {G.}~\bibnamefont {Fragione}},
  \bibinfo {author} {\bibfnamefont {C.}~\bibnamefont {Kimball}}, \ and\
  \bibinfo {author} {\bibfnamefont {V.}~\bibnamefont {Kalogera}},\ }\href
  {\doibase 10.3847/1538-4357/ace4c1} {\bibfield  {journal} {\bibinfo
  {journal} {Astrophys. J.}\ }\textbf {\bibinfo {volume} {954}},\ \bibinfo
  {pages} {23} (\bibinfo {year} {2023})},\ \Eprint
  {http://arxiv.org/abs/2302.07284} {arXiv:2302.07284 [astro-ph.HE]}
  \BibitemShut {NoStop}%
\bibitem [{\citenamefont {Fishbach}\ \emph {et~al.}(2022)\citenamefont
  {Fishbach}, \citenamefont {Kimball},\ and\ \citenamefont
  {Kalogera}}]{Fishbach:2022lzq}%
  \BibitemOpen
  \bibfield  {author} {\bibinfo {author} {\bibfnamefont {M.}~\bibnamefont
  {Fishbach}}, \bibinfo {author} {\bibfnamefont {C.}~\bibnamefont {Kimball}}, \
  and\ \bibinfo {author} {\bibfnamefont {V.}~\bibnamefont {Kalogera}},\ }\href
  {\doibase 10.3847/2041-8213/ac86c4} {\bibfield  {journal} {\bibinfo
  {journal} {Astrophys. J. Lett.}\ }\textbf {\bibinfo {volume} {935}},\
  \bibinfo {pages} {L26} (\bibinfo {year} {2022})},\ \Eprint
  {http://arxiv.org/abs/2207.02924} {arXiv:2207.02924 [astro-ph.HE]}
  \BibitemShut {NoStop}%
\bibitem [{\citenamefont {Udall}\ \emph {et~al.}(2024)\citenamefont {Udall},
  \citenamefont {Hourihane}, \citenamefont {Miller}, \citenamefont {Davis},
  \citenamefont {Chatziioannou}, \citenamefont {Isi},\ and\ \citenamefont
  {Deshong}}]{Udall:2024ovp}%
  \BibitemOpen
  \bibfield  {author} {\bibinfo {author} {\bibfnamefont {R.}~\bibnamefont
  {Udall}}, \bibinfo {author} {\bibfnamefont {S.}~\bibnamefont {Hourihane}},
  \bibinfo {author} {\bibfnamefont {S.}~\bibnamefont {Miller}}, \bibinfo
  {author} {\bibfnamefont {D.}~\bibnamefont {Davis}}, \bibinfo {author}
  {\bibfnamefont {K.}~\bibnamefont {Chatziioannou}}, \bibinfo {author}
  {\bibfnamefont {M.}~\bibnamefont {Isi}}, \ and\ \bibinfo {author}
  {\bibfnamefont {H.}~\bibnamefont {Deshong}},\ }\href@noop {} {\  (\bibinfo
  {year} {2024})},\ \Eprint {http://arxiv.org/abs/2409.03912} {arXiv:2409.03912
  [gr-qc]} \BibitemShut {NoStop}%
\bibitem [{\citenamefont {Hannam}\ \emph
  {et~al.}(2022{\natexlab{b}})\citenamefont {Hannam}, \citenamefont {Hoy},
  \citenamefont {Thompson}, \citenamefont {Fairhurst}, \citenamefont {Raymond},
  , \citenamefont {members of~the LIGO},\ and\ \citenamefont
  {collaborations}}]{hannam_mark_2022_6672460}%
  \BibitemOpen
  \bibfield  {author} {\bibinfo {author} {\bibfnamefont {M.}~\bibnamefont
  {Hannam}}, \bibinfo {author} {\bibfnamefont {C.}~\bibnamefont {Hoy}},
  \bibinfo {author} {\bibfnamefont {J.~E.}\ \bibnamefont {Thompson}}, \bibinfo
  {author} {\bibfnamefont {S.}~\bibnamefont {Fairhurst}}, \bibinfo {author}
  {\bibfnamefont {V.}~\bibnamefont {Raymond}}, , \bibinfo {author}
  {\bibnamefont {members of~the LIGO}}, \ and\ \bibinfo {author} {\bibfnamefont
  {V.}~\bibnamefont {collaborations}},\ }\href {\doibase
  10.5281/zenodo.6672460} {\enquote {\bibinfo {title} {{Measurement of
  general-relativistic precession in a black-hole binary - data release}},}\ }
  (\bibinfo {year} {2022}{\natexlab{b}})\BibitemShut {NoStop}%
\bibitem [{noi()}]{noise}%
  \BibitemOpen
  \href@noop {} {\enquote {\bibinfo {title} {{Noise curves used for Simulations
  in the update of the Observing Scenarios Paper}},}\ }\bibinfo {howpublished}
  {\url{https://dcc.ligo.org/LIGO-T2000012/public}}\BibitemShut {NoStop}%
\bibitem [{\citenamefont {Liu}\ \emph {et~al.}(2024)\citenamefont {Liu},
  \citenamefont {Cao},\ and\ \citenamefont {Zhu}}]{Liu:2023ldr}%
  \BibitemOpen
  \bibfield  {author} {\bibinfo {author} {\bibfnamefont {X.}~\bibnamefont
  {Liu}}, \bibinfo {author} {\bibfnamefont {Z.}~\bibnamefont {Cao}}, \ and\
  \bibinfo {author} {\bibfnamefont {Z.-H.}\ \bibnamefont {Zhu}},\ }\href
  {\doibase 10.1088/1361-6382/ad72ca} {\bibfield  {journal} {\bibinfo
  {journal} {Class. Quant. Grav.}\ }\textbf {\bibinfo {volume} {41}},\ \bibinfo
  {pages} {195019} (\bibinfo {year} {2024})},\ \Eprint
  {http://arxiv.org/abs/2310.04552} {arXiv:2310.04552 [gr-qc]} \BibitemShut
  {NoStop}%
\bibitem [{\citenamefont {Gamba}\ \emph {et~al.}(2024)\citenamefont {Gamba},
  \citenamefont {Chiaramello},\ and\ \citenamefont {Neogi}}]{Gamba:2024cvy}%
  \BibitemOpen
  \bibfield  {author} {\bibinfo {author} {\bibfnamefont {R.}~\bibnamefont
  {Gamba}}, \bibinfo {author} {\bibfnamefont {D.}~\bibnamefont {Chiaramello}},
  \ and\ \bibinfo {author} {\bibfnamefont {S.}~\bibnamefont {Neogi}},\ }\href
  {\doibase 10.1103/PhysRevD.110.024031} {\bibfield  {journal} {\bibinfo
  {journal} {Phys. Rev. D}\ }\textbf {\bibinfo {volume} {110}},\ \bibinfo
  {pages} {024031} (\bibinfo {year} {2024})},\ \Eprint
  {http://arxiv.org/abs/2404.15408} {arXiv:2404.15408 [gr-qc]} \BibitemShut
  {NoStop}%
\bibitem [{\citenamefont {Speagle}(2020)}]{Speagle_2020}%
  \BibitemOpen
  \bibfield  {author} {\bibinfo {author} {\bibfnamefont {J.~S.}\ \bibnamefont
  {Speagle}},\ }\href {\doibase 10.1093/mnras/staa278} {\bibfield  {journal}
  {\bibinfo  {journal} {Mon. Not. R. Astron. Soc.}\ }\textbf {\bibinfo {volume}
  {493}},\ \bibinfo {pages} {3132} (\bibinfo {year} {2020})},\ \Eprint
  {http://arxiv.org/abs/1904.02180} {arXiv:1904.02180 [astro-ph.IM]}
  \BibitemShut {NoStop}%
\bibitem [{\citenamefont {Hunter}(2007)}]{Hunter:2007ouj}%
  \BibitemOpen
  \bibfield  {author} {\bibinfo {author} {\bibfnamefont {J.~D.}\ \bibnamefont
  {Hunter}},\ }\href {\doibase 10.1109/MCSE.2007.55} {\bibfield  {journal}
  {\bibinfo  {journal} {Comput. Sci. Eng.}\ }\textbf {\bibinfo {volume} {9}},\
  \bibinfo {pages} {90} (\bibinfo {year} {2007})}\BibitemShut {NoStop}%
\bibitem [{\citenamefont {Harris}\ \emph {et~al.}(2020)\citenamefont {Harris}
  \emph {et~al.}}]{Harris:2020xlr}%
  \BibitemOpen
  \bibfield  {author} {\bibinfo {author} {\bibfnamefont {C.~R.}\ \bibnamefont
  {Harris}} \emph {et~al.},\ }\href {\doibase 10.1038/s41586-020-2649-2}
  {\bibfield  {journal} {\bibinfo  {journal} {Nature (London)}\ }\textbf
  {\bibinfo {volume} {585}},\ \bibinfo {pages} {357} (\bibinfo {year}
  {2020})},\ \Eprint {http://arxiv.org/abs/2006.10256} {arXiv:2006.10256
  [cs.MS]} \BibitemShut {NoStop}%
\bibitem [{\citenamefont {Nitz}\ \emph {et~al.}()\citenamefont {Nitz} \emph
  {et~al.}}]{PyCBC}%
  \BibitemOpen
  \bibfield  {author} {\bibinfo {author} {\bibfnamefont {A.~H.}\ \bibnamefont
  {Nitz}} \emph {et~al.},\ }\href {\doibase 10.5281/zenodo.596388} {\enquote
  {\bibinfo {title} {{PyCBC software}},}\ }\bibinfo {note}
  {\url{https://doi.org/10.5281/zenodo.596388}}\BibitemShut {NoStop}%
\bibitem [{\citenamefont {Virtanen}\ \emph {et~al.}(2020)\citenamefont
  {Virtanen} \emph {et~al.}}]{Virtanen:2019joe}%
  \BibitemOpen
  \bibfield  {author} {\bibinfo {author} {\bibfnamefont {P.}~\bibnamefont
  {Virtanen}} \emph {et~al.},\ }\href {\doibase 10.1038/s41592-019-0686-2}
  {\bibfield  {journal} {\bibinfo  {journal} {Nat. Methods}\ }\textbf {\bibinfo
  {volume} {17}},\ \bibinfo {pages} {261} (\bibinfo {year} {2020})},\ \Eprint
  {http://arxiv.org/abs/1907.10121} {arXiv:1907.10121 [cs.MS]} \BibitemShut
  {NoStop}%
\end{thebibliography}%


\end{document}